\documentclass[british,english,aps,aip,jcp,english,dvipsnames,preprint]{revtex4-1}
\usepackage[T1]{fontenc}
\usepackage[latin9]{inputenc}
\usepackage{amsmath}
\usepackage{amssymb}
\usepackage{graphicx}
\usepackage{esint}

\makeatletter

\providecommand{\tabularnewline}{\\}

\makeatother

\usepackage{babel}
\begin{document}

\title{The Importance of the Pre-exponential Factor in Semiclassical Molecular
Dynamics}

\author{Giovanni \surname{Di Liberto}}

\affiliation{Dipartimento di Chimica, Università degli Studi di Milano, via Golgi
19, 20133 Milano, Italy}

\author{Michele \surname{Ceotto}}

\affiliation{Dipartimento di Chimica, Università degli Studi di Milano, via Golgi
19, 20133 Milano, Italy}

\email{michele.ceotto@unimi.it}

\begin{abstract}
This paper deals with the critical issue of approximating the pre-exponential
factor in semiclassical molecular dynamics. The pre-exponential factor
is important because it accounts for the quantum contribution to the
semiclassical propagator of the classical Feynman path fluctuations.
Pre-exponential factor approximations are necessary when chaotic or
complex systems are simulated. We introduced pre-exponential factor
approximations based either on analytical considerations or numerical
regularization. The approximations are tested for power spectrum calculations
of more and more chaotic model systems and on several molecules, for
which exact quantum mechanical values are available. The results show
that the pre-exponential factor approximations introduced are accurate
enough to be safely employed for semiclassical simulations of complex
systems.
\end{abstract}
\maketitle

\section{Introduction\label{sec:Introduction}}

Semiclassical (SC) molecular dynamics is a well established molecular
dynamics approach for including all quantum effects starting from
classical trajectories.\cite{Miller_avd_74,Miller_PNAS,Kay_review}
Since its introduction,\cite{Miller_IVR} the Semiclassical Initial
Value Representation (SC-IVR) formulation of the semiclassical propagator
in the coherent state representations\cite{Heller_frozengaussian,Herman_Kluk_2,Kay_100}
has become a molecular dynamics tool that embodies accuracy and, at
the same time, practicability.\cite{Miller_includingQuantumEffects,Kay_HKderivation_06,Kay_IVRcorrelationfunctions_04,Kay_collinearHe_07,Kay_tunnellingHigerorderHK,Herman_review,Makri_review,Pollak_prefactorfree_04,Pollak_gaussianpropagator_06,Pollak_autocorrelation,Manolopoulos,Coker,Coker2,Grossmann,DeAguiar_SC_IVR_11,Roy,Charu_electrontranssferSCIVR_11,Nakamura_PCCP,Nanbu_nonadaFrozenGauss,Ceotto_mixedSC,Grossmann_SCBoseHubbardmodel_16}
SC-IVR depends only on local potential and it is very promising for
the future, since it has been implemented with ``on the fly'' direct
molecular dynamics approaches,\cite{Ceotto_MCSCIVR,Ceotto_1traj,Ceotto_cursofdimensionality_11,Ceotto_acceleratedSCIVR,Ceotto_NH3,Jorge,Pollak_CH2OInternalconversion_13,Roy_AbinitioSCIVR}
allowing for calculations when an analytical fitting of the Potential
Energy Surface (PES) is not possible. This aspect is fundamental when
pursuing the simulation of complex systems, where the high number
of degrees of freedom does not allow for a compact analytical PES
formulation.\cite{Hase_abinitioMDformaldehyde_94,Hase_BOMDHessianIntegrator_99,Hase_reviewBODirectDynamics_03,Jiri_intjquantchem,Martinez_AIMS,Martins_AIMD_91,Marx_book,Hase_ScienceSN2} 

The main stumbling block of the SC-IVR propagator is represented by
the pre-exponential factor, which we will describe below. Several
approximations has been employed in the past to obviate this limitation.
Analytical considerations includes the linearization of the propagator
(LSC-IVR) that can be derived also using Wigner's transform of the
quantum operators involved,\cite{Coker,Miller308_Wigner_98,Miller312_FaradayDisc_98,Miller320_LSCIVRgeneralization_99,Miller_Liu,Geva,Koda_IVRWigner,Koda_Mix_SC_Wigner},
the interaction picture,\cite{Jiri_intjquantchem,Pollak_Petersen_interaction}
or the Forward-Backward FB SC-IVR approximation, which is suitable
for correlation function calculations.\cite{Pollak_autocorrelation,MillerFBSCIVR,Miller340_GeneralizedFilinov_01,Makri_FBSCIVR}
Also, the pre-exponential factor can be partially suppressed in a
series expansion of the propagator,\cite{Pollak_prefactorfree_04,Conte_Pollak10,Conte_Pollak12}
or totally suppressed in the amplitude-free quasicorrelation function.\cite{Takatsuka_eigenstates_05}
Numerical considerations lead to the introduction of filtering techniques,
such as the one by Filinov\cite{Miller340_GeneralizedFilinov_01,Filinov_filter}
or the time averaging one in the instance of spectroscopic calculations.\cite{Ceotto_MCSCIVR,Ceotto_1traj,Ceotto_acceleratedSCIVR,Ceotto_cursofdimensionality_11,Ceotto_NH3,Alex_Mik,Ceotto_Zhang_JCTC,Ceotto_GPU,Ceotto_david,Ceotto_eigenfunctions,Ceotto_mixedSC}
Considering that during ``on the fly'' direct dynamics semiclassical
simulations, the calculation of the Hessian, necessary at each time-step
for the pre-exponential factor calculation, is the computational time
bottleneck, a compact finite difference (CFD) numerical approximation
for the Hessian has also been implemented.\cite{Ceotto_acceleratedSCIVR,Ceotto_Zhang_JCTC}

In this paper, after introducing the origin and the physical importance
of the semiclassical pre-exponential factor, we extensively test different
approximations to the pre-exponential factor and introduce new ones.
The tests are performed on both artificial chaotic systems and real
molecules, in order to give a complete overview of the range of applicability
of the approximations and provide a reliable tool for complex system
simulations. The following Section presents the motivations of this
work and Section \ref{sec:SC-IVR_power_spectrum} recalls the SC-IVR
expression for power spectra calculations. Section \ref{sec:adiabatic}
illustrates the adiabatic approximation of the pre-exponential prefactor,
which still implies the numerical integration of the pre-exponential
factor components. Section \ref{sec:The-poor-person's} recalls the
``poor person's'' approximation. Section \ref{sec:logderivative}
formulates the log-derivative representation of the pre-exponential
factor which leads to a set of approximations like the harmonic approximation
(Section \ref{sec:harmonic}), Johnson's approximation (Section \ref{sec:Johnson}),
one approximation designed by Miller (Section \ref{sec:ourapprox})
and, eventually, our new approximations at the end of the same Section.
Numerical approximation of the pre-exponential factor are presented
in Section \ref{sec:Numerical-approaches} and numerical tests follow
in Section \ref{sec:numerical_tests}, both for model chaotic systems
(Sections \ref{sec:Henon_Helies} and \ref{sec:quartic_potential})
and molecules (Sections \ref{sec:H2O}, \ref{sec:CO2}, \ref{sec:CH2O},
\ref{sec:CH4_CH2D2}). Section \ref{sec:Conclusions} concludes the
paper.

\section{Motivation\label{sec:Motivation}}

In the Feynman's path integral representation,\cite{Feynman_Hibbs}
the quantum propagator going from the starting point $\mathbf{q}_{0}$
to the final one $\mathbf{q}_{t}$ is formulated as a collection of
paths
\begin{equation}
\left\langle \mathbf{q}_{t}\left|e^{-i\hat{H}t/\hbar}\right|\mathbf{q}_{0}\right\rangle =\int_{\mathbf{q}_{0}}^{\mathbf{q}_{t}}\mathcal{D}\left[\mathbf{q}\left(t\right)\right]e^{iS_{t}\left[\mathbf{q}_{0},\mathbf{q}_{t}\right]/\hbar}\label{eq:FeynamnPaths}
\end{equation}
where $S_{t}\left[\mathbf{q}_{0},\mathbf{q}_{t}\right]$ is the action
functional for time $t$ and $\mathcal{D}\left[\mathbf{q}\left(t\right)\right]$
is the differential over all possible paths (even the infinity length
ones!). The main obstacle to the numerical integration of Eq.(\ref{eq:FeynamnPaths})
is given by the oscillatory integrand. A common strategy is to approximate
the integral (\ref{eq:FeynamnPaths}) to the contribution that comes
from the paths where the phase is stationary, i.e. $\delta S_{t}\left[\mathbf{q}\left(t\right)\right]=0$,
provided that starting and ending points are fixed. In this case,
Eq.(\ref{eq:FeynamnPaths}) becomes
\begin{equation}
\left\langle \mathbf{q}_{t}\left|e^{-i\hat{H}t/\hbar}\right|\mathbf{q}_{0}\right\rangle \approx\int_{\mathbf{q}_{0}}^{\mathbf{q}_{t}}\mathcal{D}\left[\mathbf{q}\left(t\right)\right]\mbox{exp}\left[\frac{i}{\hbar}\left(S_{t}^{cl}\left(\mathbf{q}_{0},\mathbf{q}_{t}\right)+\frac{1}{2}\frac{\delta^{2}S_{t}^{cl}\left(\mathbf{q}_{0},\mathbf{q}_{t}\right)}{\delta\mathbf{q}\left(t\right)^{2}}\delta\mathbf{q}\left(t\right)^{2}\right)\right]\label{eq:FeynmanPathsApprox}
\end{equation}
where the sum is now restricted to the classical paths from $\mathbf{q}_{0}$
to $\mathbf{q}_{t}$ and $S_{t}^{cl}\left(\mathbf{q}_{0},\mathbf{q}_{t}\right)$
is the action of the classical paths. It is important to stress that
Eq.(\ref{eq:FeynmanPathsApprox}) is the embrio of several semiclassical
approximations and it accounts not only for the classical paths contributions,
but also for the vicinity of each path via second order path fluctuations.
The goal of this paper is to determine how important these fluctuations
are to ``sew quantum mechanical flash onto classical bones''\cite{Feynman_Hibbs,Berry_Mount}
and, thus, for an accurate quantum mechanics description of molecular
vibrations and molecular dynamics in general. 

By performing the integration in Eq.(\ref{eq:FeynmanPathsApprox}),
the van Vleck propagator is derived
\begin{eqnarray}
\left\langle \mathbf{q}_{t}\left|e^{-i\hat{H}t/\hbar}\right|\mathbf{q}_{0}\right\rangle  & \approx & \sum_{\begin{array}{c}
classical\\
paths
\end{array}}\sqrt{\frac{1}{\left(2\pi i\hbar\right)^{F}}\left|-\frac{\partial^{2}S_{t}^{cl}\left(\mathbf{q}_{0},\mathbf{q}_{t}\right)}{\partial\mathbf{q}_{t}\partial\mathbf{q}_{0}}\right|}e^{iS_{t}^{cl}\left(\mathbf{q}_{0},\mathbf{q}_{t}\right)/\hbar-i\nu\pi/2}\label{eq:vanVleck}\\
 & = & \sum_{\begin{array}{c}
classical\\
paths
\end{array}}\sqrt{\frac{1}{\left(2\pi i\hbar\right)^{F}}\left|\frac{\partial\mathbf{q}_{t}}{\partial\mathbf{p}_{0}}\right|^{-1}}e^{iS_{t}^{cl}\left(\mathbf{q}_{0},\mathbf{q}_{t}\right)/\hbar-i\nu\pi/2}
\end{eqnarray}
where the integral is now a sum over all classical trajectories going
from $\mathbf{q}_{0}$ with initial momentum $\mathbf{p}_{0}$ to
$\mathbf{q}_{t}$ in an amount of time $t$ for $F$ degrees of freedom.
$\nu$ is the Maslov or Morse index and it takes into account the
number of times along each trajectory that the determinant in Eq.
(\ref{eq:vanVleck}) diverges. The squared root in Eq.(\ref{eq:vanVleck})
is usually termed as the ``semiclassical pre-exponential factor''
and it embodies the second order path-fluctuations of Eq.(\ref{eq:FeynmanPathsApprox}).
Unfortunately Eq.(\ref{eq:vanVleck}) is plagued by the improbable
task of finding classical trajectories with fixed boundary values
and the integrand diverges whenever the determinant is zero. The semiclassical
``Initial Value Representation'' (SC-IVR) trick introduced by Miller\cite{Miller_IVR}
avoids these issues by writing the wavefunction evolution in terms
of the classical paths and the sum over the classical paths as a phase
space integration which includes the Jacobian accounting for the change
of variable
\begin{equation}
\left\langle \chi\left|e^{-i\hat{H}t/\hbar}\right|\chi\right\rangle \approx\int\int d\mathbf{p}_{0}d\mathbf{q}_{0}\sqrt{\frac{1}{\left(2\pi i\hbar\right)^{F}}\left|\frac{\partial\mathbf{q}_{t}}{\partial\mathbf{p}_{0}}\right|}\chi^{*}\left(\mathbf{q}_{t}\right)\chi\left(\mathbf{q}_{0}\right)e^{iS_{t}^{cl}\left(\mathbf{q}_{0},\mathbf{p}_{0}\right)/\hbar-i\nu\pi/2}\label{eq:Miller_IVR}
\end{equation}
In Eq.(\ref{eq:Miller_IVR}), no root search is required and the zero
of the determinant at caustics is not a numerical issue anymore. The
second order path-fluctuations are now represented by the square root
term in Eq.(\ref{eq:Miller_IVR}), which quantifies how much the final
position depends on the initial momentum.

A natural representation of the wavefunction in Eq.(\ref{eq:Miller_IVR})
is given by coherent states of the type
\begin{equation}
\left\langle \mathbf{x}\left|\mathbf{p}_{t}\mathbf{q}_{t}\right.\right\rangle =\left(\frac{det(\mathbf{\gamma})}{\pi^{F}}\right)^{\frac{1}{4}}e^{-\frac{1}{2}\left(\mathbf{x}-\mathbf{q}_{t}\right)^{T}\mathbf{\gamma}\left(\mathbf{x}-\mathbf{q}_{t}\right)+\frac{i}{\hbar}\mathbf{p}_{t}^{T}\left(\mathbf{x}-\mathbf{q}_{t}\right)}\label{eq:coherent_state}
\end{equation}
where \textbf{$\mathbf{\gamma}$} is the coherent state width diagonal
matrix containing time-independent coefficients. This frozen Gaussian-dressed
semiclassical dynamics idea was introduced by Heller\cite{Heller_frozengaussian}
and later implemented by Herman and Kluk\cite{Herman_Kluk_2} and
Kay,\cite{Kay_100} in the case of the SC-IVR propagator of Eq.(\ref{eq:Miller_IVR}).
The final expression for the quantum propagator is
\begin{equation}
\left\langle \chi\left|e^{-i\hat{H}t/\hbar}\right|\chi\right\rangle \approx\left(\frac{1}{2\pi\hbar}\right)^{F}\iintop d\mathbf{p}_{0}d\mathbf{q}_{0}C_{t}\left(\mathbf{p}_{0},\mathbf{q}_{0}\right)e^{\frac{i}{\hbar}S_{t}\left(\mathbf{p}_{0},\mathbf{q}_{0}\right)}\left\langle \chi\right.\left|\mathbf{p}_{t}\mathbf{q}_{t}\left\rangle \right\langle \mathbf{p}_{0}\mathbf{q}_{0}\right|\left.\chi\right\rangle \label{eq:HHKK_prop}
\end{equation}
where we have dropped ``cl'' for the classical action $S_{t}\left(\mathbf{p}_{0},\mathbf{q}_{0}\right)$
and the original second order path-fluctuation of Eq.(\ref{eq:FeynmanPathsApprox})
is now equal to
\begin{equation}
C_{t}\left(\mathbf{p}_{0},\mathbf{q}_{0}\right)=\sqrt{\mbox{det}\left[\frac{1}{2}\left(\mathbf{M_{qq}}+\frac{1}{\mathbf{\gamma}}\mathbf{M_{pp}}\mathbf{\gamma}+\frac{i}{\hbar\boldsymbol{\gamma}}\mathbf{M_{pq}}+\frac{\hbar}{i}\mathbf{M_{qp}}\boldsymbol{\gamma}\right)\right]}\label{eq:C_t_prefactor}
\end{equation}
where \textbf{$\mathbf{M_{qq}}$}, etc., are elements of the $F\times F$
monodromy (or stability) matrix\cite{Goldstein}
\begin{equation}
\mathbf{M}\left(t\right)\equiv\left(\begin{array}{cc}
\mathbf{M_{pp}} & \mathbf{M_{pq}}\\
\mathbf{M_{qp}} & \mathbf{M_{qq}}
\end{array}\right)=\left(\begin{array}{cc}
\partial\mathbf{p}_{t}/\partial\mathbf{p}_{0} & \partial\mathbf{p}_{t}/\partial\mathbf{q}_{0}\\
\partial\mathbf{q}_{t}/\partial\mathbf{p}_{0} & \partial\mathbf{q}_{t}/\partial\mathbf{q}_{0}
\end{array}\right).\label{eq:Monodromy_Matrix}
\end{equation}
In a system following the classical Hamilton equations of motion for
$\left(\mathbf{p}_{t},\mathbf{q}_{t}\right)$, as enforced by the
stationary condition of the action $S_{t}\left(\mathbf{p}_{0},\mathbf{q}_{0}\right)$
of Eq.(\ref{eq:FeynmanPathsApprox}), the evolution of the monodromy
matrix in Eq.(\ref{eq:Monodromy_Matrix}) is
\begin{equation}
\frac{d}{dt}\mathbf{M}\left(t\right)=\left(\begin{array}{cc}
\mathbf{0} & -\mathbf{K}_{t}\\
\mathbf{m}^{-1} & \mathbf{0}
\end{array}\right)\mathbf{M}\left(t\right)\label{eq:Monodromy_evolution}
\end{equation}
where $\mathbf{K}_{t}=\partial^{2}V\left(\mathbf{q}_{t}\right)/\partial\mathbf{q}_{t}^{2}$
is the local Hessian, $V\left(\mathbf{q}_{t}\right)$ is the potential
of the system, and $\mathbf{m}^{-1}$ is the inverse of the mass tensor
and it is equal to the identity in mass-scaled coordinates. The SC-IVR
of Eq.(\ref{eq:HHKK_prop}) has been successfully employed in many
fields using several variants. It provides a globally uniform asymptotic
approximation to the quantum propagator. Each monodromy matrix element
describes the dependency of the phase space trajectory $\left(\mathbf{p}_{t},\mathbf{q}_{t}\right)$
with respect to its initial conditions $\left(\mathbf{p}_{0},\mathbf{q}_{0}\right)$.
Thus, the matrix $\mathbf{M}$ is the classical representation of
the quantum fluctuations about a classical trajectory.\cite{WangThossReviewSCIVR}
Unfortunately, the semiclassical pre-exponential factor poses two
main serious issues for the application of the SC-IVR propagator to
complex systems. First, the calculation of $C_{t}\left(\mathbf{p}_{0},\mathbf{q}_{0}\right)$
represents the bottleneck as the dimensionality of the problem increases,
because the numerical effort per trajectory has an unfavorable scaling
with respect to the number of degrees of freedom. Then, for chaotic
dynamics, the monodromy matrix elements become exponentially large,
with the exponent being the Lyapunov number, which is needed to properly
account for the strong dependency on the initial conditions. This
amplifies the oscillatory behavior of the phase space integrand and
undermines the accuracy and feasibility of any numerical approaches
to evaluate the phase space integration necessary to obtain the semiclassical
propagator.\cite{signproblem} The only way out rather than exponentially
improving the number of trajectories to have the chaotic trajectories
contribution mutually cancelled,\cite{Kay_101} it is to find reasonable
approximations for the calculation of $C_{t}\left(\mathbf{p}_{0},\mathbf{q}_{0}\right)$.
The goal of this paper is to provide suitable approximations to avoid
the pre-exponential factor to become huge. However, such an approximation
cannot simply consist in the complete neglect of the pre-exponential
factor, which would generally be a very rough and so not desirable
approximation. In fact, in the $\hbar$ expansion of the Schroedinger
equation solution given by Miller and Kay,\cite{Kay_HKderivation_06}
the semiclassical propagator (and thus the pre-exponential factor)
appears at zero-th order. Furthermore, also in the perturbation approach
of Pollak and co-workers, $C_{t}\left(\mathbf{p}_{0},\mathbf{q}_{0}\right)$
turns out already in the unperturbated zero order term.\cite{Pollak_pertrubation}

Kay\cite{Kay_101} has proposed to simply remove the trajectories
that are unstable and that cause the trouble, whenever along the evolution
\begin{equation}
\left|C_{t}\left(\mathbf{p}_{0},\mathbf{q}_{0}\right)\right|^{2}\geq\mbox{D}_{t}\label{eq:Kay_criterium}
\end{equation}
where $\mbox{D}_{t}$ is a time dependent or independent quantity.
One can choose $\mbox{D}_{t}$ according to the target value. In our
cases, the target values are the vibrational energy levels and a $\mbox{D}_{t}$
equals to the number of trajectories does not perturb our results.
In this procedure, discarded trajectories still contribute to the
Monte Carlo phase space integration at times preceding the rejection.
Thus, also chaotic trajectories contributes to the propagator, but
at shorter times. When rejecting trajectories in the Monte Carlo integration
of Eq.(\ref{eq:HHKK_prop}), one should ask himself if enough trajectories
would survive the removal process to provide any useful semiclassical
information. Miller and coworkers\cite{Miller327_logderivative_00}
came up with a numerical approach borrowed from quantum scattering
calculations. They formulate the pre-exponential factor evolution
in terms of log-derivative quantities. On one hand, this approach
avoids the branch cut problem which has hampered other formulations.
On the other, the numerical issues induced by the chaotic dynamics
still remains. Another possible solution is the ``poor person's''
approximation.\cite{Pollak_renormalizationFrozenGaussian_11} Here,
the pre-exponential factor is taken to be constant with respect to
the phase space Monte Carlo integration and approximated to the one
of the most probable trajectory, according to the Husimi distribution
of the integrand in Eq.(\ref{eq:HHKK_prop}).

Unfortunately, none of these procedures completely eliminate the problems
arising from chaotic trajectories and practical schemes need to be
developed in order to adopt the semiclassical propagator for obtaining
quantum information of complex systems. The present work tests previous
approximations of the pre-exponential factor $C_{t}\left(\mathbf{p}_{0},\mathbf{q}_{0}\right)$
and proposes new and more efficient ones, and shows advantages with
regard to previous approximations.

\section{SC-IVR expression for power spectrum calculations\label{sec:SC-IVR_power_spectrum}}

In this paper, the accuracy of the pre-exponential factor approximations
will be tested by looking at the power spectrum $I\left(E\right)$
of several models and molecular systems. $I\left(E\right)$ is defined
as
\begin{eqnarray}
I\left(E\right) & \equiv & \left\langle \chi\left|\delta\left(\hat{H}-E\right)\right|\chi\right\rangle \nonumber \\
 & =\frac{1}{2\pi\hbar} & \int_{-\infty}^{+\infty}\left\langle \chi\left|e^{-i\hat{H}t/\hbar}\right|\chi\right\rangle e^{iEt/\hbar}dt\label{eq:power_spectrum}
\end{eqnarray}
where $\left|\boldsymbol{\mathbf{\chi}}\right\rangle $ is a reference
state of the type $\left|\mathbf{p}_{eq}\mathbf{q}_{eq}\right\rangle $
and $\hat{H}$ is the Hamiltonian of the system. We choose $\mathbf{q}_{eq}$
to be the global minimum position vector with respect to the potential
energy of $\hat{H}$ and $\mathbf{p}_{eq}$ is taken such that $p_{eq,j}^{2}/2m=\hbar\omega_{j}\left(n+1/2\right)$,
where $\omega_{j}$ is the frequency of the $j-th$ normal mode. The
semiclassical expression of $\left\langle \chi\left|e^{-i\hat{H}t/\hbar}\right|\chi\right\rangle $
is reported in Eq.(\ref{eq:HHKK_prop}) and the matrix $\gamma$ of
(\ref{eq:coherent_state}) is taken to be diagonal and constant in
time, with $\gamma_{j}=m\omega_{j}/\hbar$ for the $j-th$ mode. The
SC-IVR expression for the power spectrum calculations is obtained
by substituting Eq.(\ref{eq:HHKK_prop}) into Eq.(\ref{eq:power_spectrum})
to obtain
\begin{eqnarray}
I\left(E\right) & = & \frac{1}{2\pi\hbar}\int_{-\infty}^{+\infty}dt\:\iintop d\mathbf{p}_{0}d\mathbf{q}_{0}\:e^{iEt/\hbar}\left(\frac{1}{2\pi\hbar}\right)^{F}\nonumber \\
 & \times & C_{t}\left(\mathbf{p}_{0},\mathbf{q}_{0}\right)e^{\frac{i}{\hbar}S_{t}\left(\mathbf{p}_{0},\mathbf{q}_{0}\right)}\left\langle \chi\right.\left|\mathbf{p}_{t}\mathbf{q}_{t}\left\rangle \right\langle \mathbf{p}_{0}\mathbf{q}_{0}\right|\left.\chi\right\rangle .\label{eq:HK_powerspectrum}
\end{eqnarray}

Several approaches has been introduced to speed up the phase space
integration of Eq.(\ref{eq:HK_powerspectrum}).\cite{MillerFBSCIVR,Alex_Mik,Ceotto_MCSCIVR,Wang_Dmatrix}
Here we employ the time-averaging filter to reduce the number of phase
space trajectories needed for the convergence of Monte Carlo integration.
An additional time integration is inserted in Eq.(\ref{eq:HK_powerspectrum}),
and the phase space average is performed for a time-averaged integrand.
After approximating the pre-exponential factor as $C_{t}\left(\mathbf{p}_{0},\mathbf{q}_{0}\right)=\mbox{exp}\left[i\phi\left(t\right)/\hbar\right]$,
the following time averaged semiclassical expression for the power
spectrum of Eq.(\ref{eq:HK_powerspectrum}) can be obtained 
\begin{equation}
I\left(E\right)=\left(\frac{1}{2\pi\hbar}\right)^{F}\iintop d\mathbf{p}_{0}d\mathbf{q}_{0}\frac{1}{2\pi\hbar T}\left|\intop_{0}^{T}dte^{\frac{i}{\hbar}\left[S_{t}\left(\mathbf{p}_{0},\mathbf{q}_{0}\right)+Et+\phi\left(t\right)\right]}\left\langle \boldsymbol{\chi}\left|\mathbf{p}_{t}\mathbf{q}_{t}\right.\right\rangle \right|^{2}.\label{eq:separable}
\end{equation}
Clearly, the longer the time-averaging $T$ is, the greater is the
advantage of the time filter.

\section{The Adiabatic pre-exponential factor approximation\label{sec:adiabatic}}

The idea of the adiabatic approximation of the pre-exponential factor
$C_{t}\left(\mathbf{p}_{0},\mathbf{q}_{0}\right)$ by Miller and coworkers\cite{Miller318_prefactoradiabaticGuallar_99,Miller328_prefactorapprox_00}
is to assume that the monodromy matrix elements are adiabatic with
respect to each other. The instantaneous normal mode framework is
enforced by the diagonalization of the Hessian at each time-step.
First, the auxiliary variables
\begin{eqnarray}
\mathbf{Q}_{t} & = & \mathbf{M_{qq}}-i\hbar\mathbf{M_{qp}}\gamma\label{eq:Q}\\
\mathbf{P}_{t} & = & \mathbf{M_{pq}}-i\hbar\mathbf{M_{pp}}\gamma\label{eq:P}
\end{eqnarray}
are introduced, and the equations of motion of $\mathbf{P}_{t}$ and
$\mathbf{Q}_{t}$ are 
\begin{equation}
\begin{cases}
\dot{\mathbf{Q}}_{t}= & \mathbf{P}_{t}\\
\dot{\mathbf{P}}_{t}= & -\mathbf{K}_{t}\mathbf{Q}_{t}
\end{cases}\label{eq:Qdot_Pdot}
\end{equation}
where initial conditions $\mathbf{Q}_{0}=\mathbf{1}$ and $\mathbf{P}_{0}=-i\hbar$\textbf{$\gamma$}
can be obtained from Eqs. (\ref{eq:Monodromy_Matrix}) and (\ref{eq:Monodromy_evolution}).
Then, the pre-exponential factor of Eq.(\ref{eq:C_t_prefactor}) becomes
\begin{equation}
C_{t}\left(\mathbf{p}_{0},\mathbf{q}_{0}\right)=\sqrt{\frac{1}{2^{F}}\mbox{det}\left[\mathbf{Q}_{t}+\frac{i}{\hbar\gamma}\mathbf{P}_{t}\right]}.\label{eq:C_t_PQ}
\end{equation}
This formulation is still exact. The set of instantaneous mass-scaled
normal mode coordinates is calculated at each time step by the matrix
$\mathbf{U}_{t}$ such that 
\begin{equation}
\mathbf{U}_{t}^{\dagger}\mathbf{K}_{t}\mathbf{U}_{t}\equiv\mathbf{\omega}_{t}^{2}\label{eq:diagonal_Kt}
\end{equation}
where $\mathbf{\omega}_{t}^{2}$ is the instantaneous diagonal Hessian
matrix. In the adiabatic approximation the time derivatives of $\mathbf{U}_{t}$
are neglected and the new transformed matrices

\begin{eqnarray}
\tilde{\mathbf{Q}}_{t} & \equiv & \mathbf{U}_{t}^{\dagger}\mathbf{Q}_{t}\mathbf{U}_{t}\label{eq:Qtilde}\\
\tilde{\mathbf{P}}_{t} & \equiv & \mathbf{U}_{t}^{\dagger}\mathbf{P}_{t}\mathbf{U}_{t}\label{eq:Ptilde}
\end{eqnarray}
remain diagonal at all times $t$. The system of equations (\ref{eq:Qdot_Pdot})
for the new variables of Eqs. (\ref{eq:Qtilde}) and (\ref{eq:Ptilde})
becomes a set of $F-$independent one-dimensional second-order differential
equations. Finally, the expression of the pre-exponential factor in
the adiabatic approximation is
\begin{equation}
C_{t}\left(\mathbf{p}_{0},\mathbf{q}_{0}\right)\approx\sqrt{\prod_{j}^{F}\frac{1}{2}\left(\tilde{Q_{t}}\left(j,j\right)+\frac{i}{\hbar\gamma}\tilde{P_{t}}\left(j,j\right)\right)}\label{eq:C_t_adiabatic}
\end{equation}
where $\tilde{Q_{t}}\left(j,j\right)$ and $\tilde{P_{t}}\left(j,j\right)$
are the diagonal elements of the matrices respectively defined in
Eqs. (\ref{eq:Qtilde}) and (\ref{eq:Ptilde}) and evolved according
to Eq.(\ref{eq:Qdot_Pdot}). This approximation should be good as
far as each frequency $\omega_{j,t}$ of the $j-th$ mode is well
separated and modes are not strongly coupled, i.e. adiabatic with
respect to each other. The opposite situation, the diabatic limit,
when frequencies are in resonance, is also favorable to the adiabatic
approximation, since the instantaneous normal mode diagonalization
can fit a local adiabatic representation. The intermediate cases,
where coupling cannot be removed, are the worse case scenario for
the adiabatic approximation.

The basic advantages of this approximation is to reduce the computational
cost. However, integration of Eq.(\ref{eq:Qdot_Pdot}) is still sensitive
to the initial conditions and problems related to chaotic dynamics
will hinder a straightforward application of Eq.(\ref{eq:C_t_adiabatic}).

\section{The ``poor person's'' approximation\label{sec:The-poor-person's}}

A more drastic approximation is the ``poor person's'' one, that
we will abbreviate as ``PPs''.\cite{Pollak_renormalizationFrozenGaussian_11}
This approximation is motivated by the observation that the approximated
propagator should (i) be exact for harmonic systems, (ii) be not very
sensitive to the choice of the coherent states width parameter, (ii)
be local in the potential, and (iv) retains normalization. Given the
conditions (i)-(iv), the approximation should also save computational
time, making complex systems simulations possible. The PPs formulation
approximates Eq.(\ref{eq:HHKK_prop}) as 
\begin{equation}
\left\langle \chi\left|e^{-i\hat{H}t/\hbar}\right|\chi\right\rangle \approx\left(\frac{1}{2\pi\hbar}\right)^{F}C_{t}\left(\mathbf{p}_{eq},\mathbf{q}_{eq}\right)\iintop d\mathbf{p}_{0}d\mathbf{q}_{0}e^{\frac{i}{\hbar}S_{t}\left(\mathbf{p}_{0},\mathbf{q}_{0}\right)}\left\langle \chi\left(\mathbf{p}_{eq},\mathbf{q}_{eq}\right)\right.\left|\mathbf{p}_{t}\mathbf{q}_{t}\left\rangle \right\langle \mathbf{p}_{0}\mathbf{q}_{0}\right|\left.\chi\left(\mathbf{p}_{eq},\mathbf{q}_{eq}\right)\right\rangle \label{eq:PPs}
\end{equation}
where the phase point $\left(\mathbf{p}_{eq},\mathbf{q}_{eq}\right)$
is the location of the coherent reference state $\left|\chi\right\rangle $
and the center of the Husimi distribution employed for the Monte Carlo
phase space sampling. In this way, the pre-exponential factor $C_{t}$
is calculated for a single (and the most probable) trajectory and
enforced to all the others. Eq.(\ref{eq:PPs}) is exact for the harmonic
oscillator, where $C_{t}$ does not depend on the phase space initial
coordinates. The monodromy matrix still needs to be calculated for
the trajectory starting at $\left(\mathbf{p}_{eq},\mathbf{q}_{eq}\right)$
and the approximation can not be applied when the system is so chaotic
that the monodromy matrix of that single trajectory can not be evolved.
The PPs approximation is particularly advantageous for ``on the fly''
simulations, where the Hessian calculation is very demanding.

\section{The log-derivative formulation of the pre-exponential factor and
its approximations\label{sec:logderivative}}

To overcome the numerical issues of the monodromy matrix evolution
described above, Miller and coworkers wrote the evolution of the pre-exponential
factor $C_{t}\left(\mathbf{p}_{0},\mathbf{q}_{0}\right)$ using the
log-derivative formulation.\cite{Miller327_logderivative_00} The
log-derivative matrix $\mathbf{R}_{t}$ is defined by
\begin{equation}
\boldsymbol{R}_{t}=\frac{\dot{\mathbf{Q}}_{t}}{\mathbf{Q}_{t}}=\frac{\mathbf{P}_{t}}{\mathbf{Q}_{t}}\label{eq:logderivative}
\end{equation}
and it is properly defined since $\mbox{det}\left(\mathbf{Q}_{t}\right)$
is never zero.\cite{Kay_100,Kay_101} The pre-exponential factor can
now be written as
\begin{equation}
C_{t}\left(\mathbf{p}_{0},\mathbf{q}_{0}\right)=\sqrt{\mbox{det}\left[\frac{1}{2}\left(I+\frac{i}{\hbar\gamma}\mathbf{R}_{t}\right)\right]}e^{\frac{1}{2}\intop_{0}^{t}d\tau\mbox{Tr}\left[\mathbf{R}_{\tau}\right]}\label{eq:prefactor_logdev}
\end{equation}
and one is left with the calculation of the matrix $\mathbf{R}_{t}$
at each time step. By deriving Eq.(\ref{eq:logderivative}) on both
sides with respect to time and using Eq.(\ref{eq:Qdot_Pdot}), the
equation of motion
\begin{equation}
\dot{\mathbf{R}}_{t}=-\mathbf{K}_{t}-\mathbf{R}_{t}^{2}\label{eq:equation_Rt}
\end{equation}
is what must be solved for the calculation of the pre-exponential
factor. No approximation has been introduced so far and Eq.(\ref{eq:prefactor_logdev})
is an exact formulation of the pre-exponential factor. The issues
related to the stability matrix for chaotic systems are hidden inside
the integration of the Riccati's equation (\ref{eq:equation_Rt}).
A possible simplification is to assume that the force constant matrix
$\mathbf{K}_{t}$ is slowly varying and one can set the squared root
in Eq.(\ref{eq:prefactor_logdev}) equal to unity. However, this approximation
does not remove the numerical issues related to chaotic motion. For
these reasons, one should better employ the following approximations.

\subsection{The Harmonic approximation\label{sec:harmonic}}

This is a crude approximation which is equivalent to take at any time
in Eq. (\ref{eq:equation_Rt})
\begin{equation}
\mathbf{K}_{t}\approx\mathbf{K}_{0}=\mathbf{\omega}_{0}^{2}\label{eq:Harmonic_Kt}
\end{equation}
where $\mathbf{\omega}_{o}^{2}$ are the diagonal Hessian matrix elements
at equilibrium position. Since, for harmonic oscillators, the coherent
state width matrix $\mathbf{\gamma}$ is constant and equal to $m\mathbf{\omega}_{0}/\hbar$,
the solution of Eq.(\ref{eq:equation_Rt}) is analytical
\begin{equation}
\mathbf{R}_{t}=-\hbar\mathbf{\boldsymbol{\gamma}}\frac{i+tan\left(\hbar\boldsymbol{\gamma}t\right)}{1-itan\left(\hbar\boldsymbol{\gamma}t\right)}=-i\hbar\boldsymbol{\gamma}\label{eq:Rt_harmonic}
\end{equation}
and the pre-exponential factor is approximated as
\begin{equation}
C_{t}\left(\mathbf{p}_{0},\mathbf{q}_{0}\right)=e^{-i\hbar\sum_{j=1}^{F}\gamma_{j}t/2}=e^{-i\sum_{j=1}^{F}\omega_{0,j}t/2}\label{eq:prefactor_Harmonic}
\end{equation}
where $\omega_{0,j}$ is the harmonic frequency of the $j-th$ mode.
The same result can be obtained by inserting $\mathbf{K}_{0}$ into
Eq.(\ref{eq:Monodromy_evolution}) and solving the set of differential
equations.

\subsection{The Johnson Multichannel approximation\label{sec:Johnson}}

To improve the accuracy of the harmonic approximation, one can naively
replace in Eq.(\ref{eq:prefactor_Harmonic}) $\omega_{0,j}t$ with
$\int_{0}^{t}\omega_{\tau,j}d\tau$, i.e. the initial harmonic frequencies
with instantaneous ones and consider the integral over time. A more
elegant way to reach the same conclusion is to assume that the term
$\mathbf{\dot{R}}_{t}$ in Eq.(\ref{eq:equation_Rt}) can be disregarded
since the log-derivative matrix $\mathbf{R}_{t}$ is much more slowly
variant than $\mathbf{Q}_{t}$. The equation solution of Eq.(\ref{eq:equation_Rt})
becomes 
\begin{equation}
\mathbf{R}_{t}=-i\sqrt{\mathbf{K}_{t}}\label{eq:Rt_Johnson}
\end{equation}
where the minus sign has been chosen to satisfy the initial conditions
$\mathbf{R}_{0}=-i\hbar$$\gamma$. By inserting Eq.(\ref{eq:Rt_Johnson})
into Eq.(\ref{eq:prefactor_logdev}), the following approximation
is obtained
\begin{equation}
C_{t}\left(\mathbf{p}_{0},\mathbf{q}_{0}\right)=\sqrt{\mbox{det}\left[\frac{1}{2}\left(\mathbf{I}+\frac{\sqrt{\mathbf{K}_{t}}}{\hbar\boldsymbol{\gamma}}\right)\right]}e^{-i\intop_{0}^{t}\mbox{Tr}\left(\sqrt{\mathbf{K}_{\tau}}\right)d\tau/2}.\label{eq:Johnson_1}
\end{equation}
The pre-exponential term in Eq.(\ref{eq:Johnson_1}) is also slowly
variant and by approximating each matrix element ratio
\begin{equation}
\frac{\omega_{t,j}}{\hbar\gamma_{j}}\approx1\label{eq:approx_ratio}
\end{equation}
the Johnson's ``multichannel WKB'' approximation of the semiclassical
pre-exponential factor is derived
\begin{equation}
C_{t}\left(\mathbf{p}_{0},\mathbf{q}_{0}\right)\approx\mbox{exp}\left[-\frac{i}{\hbar}\int_{0}^{t}\sum_{j=1}^{F}\left(\frac{\hbar}{2}\omega_{\tau,j}\right)d\tau\right].\label{eq:Johnson_2}
\end{equation}
Eq.(\ref{eq:Johnson_2}) approximates the pre-exponential factor as
the phase arising from the local zero-point energy along the trajectory.
This approximation has already been employed in the past.\cite{Roy_AbinitioSCIVR,Roy_ZPEaccurate_05,Roy_singeltrajectory_07,Roy_excitedstatesconstrained_07,Roy_hydrogenbonded_07}

\subsection{A recursive perturbative approach\label{sec:ourapprox}}

A possible accuracy improvement of the Sec.\ref{sec:harmonic} is
the following perturbative approach. We initially follow Miller and
coworkers,\cite{Miller327_logderivative_00} and we assume that $\mathbf{R}_{t}$
is given by the harmonic value in Eq.(\ref{eq:Rt_harmonic}) corrected
by a perturbation term $\varepsilon$
\begin{equation}
\mathbf{R}_{t}=-i\hbar\boldsymbol{\gamma}+\boldsymbol{\varepsilon}.\label{eq:Miller1}
\end{equation}
 By inserting (\ref{eq:Miller1}) into the Riccati's equation (\ref{eq:equation_Rt}),
and assuming the perturbation constant in time, i.e. $\dot{\varepsilon}\approx0$,
\begin{equation}
-\mathbf{K}_{t}+\hbar^{2}\boldsymbol{\gamma}^{2}=\boldsymbol{\varepsilon}^{2}-2i\hbar\boldsymbol{\gamma\varepsilon}\label{eq:Miller2}
\end{equation}
and neglecting the higher order terms in $\varepsilon$, the following
expression for the perturbation term is obtained
\begin{equation}
\boldsymbol{\varepsilon}=-\frac{i}{2}\left(\frac{\mathbf{K}_{t}}{\hbar\mathbf{\gamma}}-\hbar\mathbf{\gamma}\right).\label{eq:Miller_eps}
\end{equation}
The resulting approximation of the log-derivative matrix (\ref{eq:logderivative})
is 
\begin{equation}
\mathbf{R}_{t}^{(1)}=-\frac{i}{2}\left(\hbar\boldsymbol{\gamma}+\frac{\mathbf{K}_{t}}{\hbar\boldsymbol{\gamma}}\right)\label{eq:Miller_approx}
\end{equation}
as previously suggested by Miller.\cite{Miller327_logderivative_00}
Eq.(\ref{eq:Miller_approx}) will provide the approximate pre-exponential
factor once inserted into Eq.(\ref{eq:prefactor_logdev}). Since the
Hessian $\mathbf{K}_{t}$ is always real, the expression of $\mathbf{R}_{t}^{(1)}$
in Eq.(\ref{eq:Miller_approx}) is purely imaginary. This pre-exponential
factor approximation mainly differs from the harmonic (\ref{eq:prefactor_Harmonic})
and Johnson's (\ref{eq:Johnson_2}) ones in the exponential term,
which is linearly dependent on the Hessian.

We now want to systematically improve the approximation (\ref{eq:Miller_approx}).
The idea is to use Eq.(\ref{eq:Miller_approx}) as a more accurate
solution than the harmonic one (\ref{eq:Rt_harmonic}), insert it
into the Riccati equation and obtain a new perturbative correction.
A new solution will be obtained by iteratively using the new correction
as an initial guess. We start by inserting 
\begin{equation}
\mathbf{R}_{t}^{(2)}=\mathbf{R}_{t}^{(1)}+\boldsymbol{\varepsilon}=-\frac{i}{2}\left[\frac{\mathbf{K}_{t}}{\hbar\boldsymbol{\gamma}}+\hbar\boldsymbol{\gamma}\right]+\boldsymbol{\varepsilon}\label{eq:our_ansatz}
\end{equation}
into (\ref{eq:equation_Rt}), and disregard higher order and time-derivative
terms of $\epsilon$ and Hessian time-derivatives. We obtain the following
equation
\begin{equation}
0=\frac{1}{4}\left(\hbar^{2}\boldsymbol{\gamma}^{2}+\frac{\mathbf{K}_{t}^{2}}{\hbar^{2}\boldsymbol{\gamma}^{2}}+2\mathbf{K}_{t}\right)+i\boldsymbol{\varepsilon}\left(\frac{\mathbf{K}_{t}}{\hbar\boldsymbol{\gamma}}+\hbar\boldsymbol{\gamma}\right)-\mathbf{K}_{t}\label{eq:eps_ansatz}
\end{equation}
which brings 
\begin{equation}
\boldsymbol{\varepsilon}=\frac{i}{4}\frac{\left(\hbar\boldsymbol{\gamma}-\frac{\mathbf{K}_{t}}{\hbar\boldsymbol{\gamma}}\right)^{2}}{\frac{\mathbf{K}_{t}}{\hbar\boldsymbol{\gamma}}+\hbar\boldsymbol{\gamma}}.\label{eq:eps_ansatz2}
\end{equation}
Then, the substitution of Eq.(\ref{eq:eps_ansatz2}) into Eq.(\ref{eq:our_ansatz})
provides the expression
\begin{equation}
\mathbf{R}_{t}^{(2)}=-\frac{i}{2}\left[\frac{\mathbf{K}_{t}}{\hbar\boldsymbol{\gamma}}+\hbar\boldsymbol{\gamma}\right]+\frac{i}{4}\frac{\left(\hbar\boldsymbol{\gamma}-\frac{\mathbf{K}_{t}}{\hbar\boldsymbol{\gamma}}\right)^{2}}{\left(\hbar\boldsymbol{\gamma}+\frac{\mathbf{K}_{t}}{\hbar\boldsymbol{\gamma}}\right)}.\label{eq:Rt2}
\end{equation}
Again, this solution is purely imaginary and the dependence on the
Hessian matrix is more complex than previous ones. Eq.(\ref{eq:Rt2})
is better written in terms of $\mathbf{R}_{t}^{(1)}$ as
\begin{equation}
\mathbf{R}_{t}^{(2)}=\mathbf{R}_{t}^{(1)}+\frac{1}{2^{3}}\frac{\left(\hbar\boldsymbol{\gamma}-\frac{\mathbf{K}_{t}}{\hbar\boldsymbol{\gamma}}\right)^{2}}{\mathbf{R}_{t}^{(1)}}.\label{eq:Rt2_Rt1}
\end{equation}
We can now look for the next order $\mathbf{R}_{t}^{(3)}=\mathbf{R}_{t}^{(2)}+\boldsymbol{\varepsilon}$
by inserting this guess into the Riccati's equation, take zero time
derivative for $\mathbf{K}_{t}$ and $\varepsilon$ as usual, and
disregarding the higher order perturbation terms, we obtain 
\begin{equation}
\mathbf{R}_{t}^{(3)}=\mathbf{R}_{t}^{(2)}-\frac{1}{2^{7}}\frac{\left(\hbar\boldsymbol{\gamma}-\frac{\mathbf{K}_{t}}{\hbar\boldsymbol{\gamma}}\right)^{4}}{\mathbf{R}_{t}^{(1)^{2}}\mathbf{R}_{t}^{(2)}}\label{eq:Rt3}
\end{equation}
and, in the same fashion, one can find 
\begin{equation}
\mathbf{R}_{t}^{(4)}=\mathbf{R}_{t}^{(3)}+\frac{1}{2^{15}}\frac{\left(\hbar\boldsymbol{\gamma}-\frac{\mathbf{K}_{t}}{\hbar\boldsymbol{\gamma}}\right)^{8}}{\mathbf{R}_{t}^{(1)^{4}}\mathbf{R}_{t}^{(2)^{2}}\mathbf{R}_{t}^{(3)}}.\label{eq:Rt3-1}
\end{equation}
By induction, the final $n-order$ correction of the harmonic log-derivative
matrix is in closed form equal to
\begin{equation}
\mathbf{R}_{t}^{(n)}=\mathbf{R}_{t}^{(n-1)}+\frac{\left(-\right)^{n}}{2^{\left(2^{n}-1\right)}}\frac{\left(\hbar\boldsymbol{\gamma}-\frac{\mathbf{K}_{t}}{\hbar\boldsymbol{\gamma}}\right)^{2^{\left(n-1\right)}}}{\Pi_{j=0}^{n-2}\left(\mathbf{R}_{t}^{(n-1-j)}\right)^{2^{j}}}.\label{eq:Rt_n}
\end{equation}
We stress that Eq.(\ref{eq:Rt_n}) is not the formal solution of the
Riccati equation (\ref{eq:equation_Rt}), even if it is a closed form
for an $n-th$ order perturbation correction, because it has assumed
that the Hessian is constant, i.e. $\dot{\mathbf{K}}_{t}\approx0$,
throughout the derivation.

\section{Numerical approximations\label{sec:Numerical-approaches}}

An alternative route with respect to the analytical approximations
presented in the previous Sections, is to perform numerical approximations.
We consider two possibilities, the Log-derivative symplectic integrator
and the monodromy matrix regularization. We employ either one of these
numerical approximations as an alternative to the analytical approximations.

\subsection{Log-derivative symplectic integration\label{sub:Log-derivative-symplectic-integr}}

Another approach to solve the evolution of the monodromy matrix elements
in presence of chaos is to employ high order numerical algorithms.
We usually employ the \emph{4th} order symplectic algorithm described
in Appendix of Ref.\cite{Manolopoulos}(c), and originally due to
Calvo \emph{et al}.,\cite{Calvo} to solve Eq.(\ref{eq:Monodromy_evolution}).
One can similarly use such an accurate algorithm to solve the Riccati
equation instead. Manolopoulos and Gray\cite{Gray_Manolopoulos} showed
that the system of equations 
\begin{equation}
\begin{cases}
\mathbf{X}_{k} & =\mathbf{R}_{k-1}+b_{k}\mathbf{K}_{k}\Delta t\\
\mathbf{R}_{k} & =\left[\mathbf{I}+a_{k}\mathbf{X}_{k}\Delta t\right]^{-1}\mathbf{X}_{k}
\end{cases}\label{eq:integrator}
\end{equation}
does this task when suitable coefficients $a_{k}$ and $b_{k}$\cite{Gray_Manolopoulos}
are employed ($\mathbf{X}$ is an auxiliary variable). We implemented
Eq.(\ref{eq:integrator}) in our calculations. The results indicate
that when the trajectory is experiencing a chaotic potential, the
numerical calculation of the log-derivative $\mathbf{R}_{t}$ cannot
be managed, similarly to the case of the monodromy matrix elements.

\subsection{Monodromy Matrix regularization\label{sub:Monodromy-Matrix-regularization}}

Another route to deal with chaotic potentials is to introduce an artificial
and \emph{ad hoc} numerical method to tame the exponentially growing
value of the monodromy matrix elements. A possible procedure is to
monitor the monodromy elements at each time step. After the diagonalization
of the monodromy matrix, the degrees of freedom mostly responsible
for the chaotic behaviour can be identified by looking at their complex
eigenvalues. More specifically, each element of the monodromy matrix
can be written as 
\begin{equation}
m_{ij}=u_{ik}\lambda_{k}u_{kj}^{-1}\label{eq:monodromy_element}
\end{equation}
where $u_{ik}$ and $u_{ik}^{-1}$ are the elements of the $\mathbf{U}$
orthogonal matrix that diagonalizes the monodromy matrix and the sum
over $k$ is implied. The greater the modulus of an eigenvalue $\lambda_{s}$
is, the more sensitive to the initial conditions and chaotic the $s-$degree
of freedom is. Then, a brute force regularization approach consists
in setting either the most chaotic eigenvector or eigenvalue or both
equal to zero in the following way 
\begin{equation}
\mathbf{\tilde{U}}^{-1}=\left(\begin{array}{ccc}
... & ... & ...\\
0 & 0 & 0\\
... & ... & ...
\end{array}\right);\;\;\mathbf{\tilde{U}}=\left(\begin{array}{ccc}
... & 0 & ...\\
... & 0 & ...\\
... & 0 & ...
\end{array}\right)\label{eq:modified_U}
\end{equation}
where the $s-th$ column and row is set to zero and a modified diagonal
matrix is obtained
\begin{equation}
\tilde{\mathbf{\Lambda}}=\left(\begin{array}{ccc}
...\\
 & 0\\
 &  & ...
\end{array}\right)\label{eq:modified_Lambda}
\end{equation}
by setting to zero the $s-th$ diagonal element of the $\mathbf{\Lambda}$
eigenvalues matrix. The criterion for setting the eigenvector or the
eigenvalue equal to zero is when $\left|\lambda_{s}\right|\geq\epsilon_{thr}$,
where $\epsilon_{thr}$ is an arbitrary positive number. Considering
that for unstable manifolds monodromy matrix eigenvalues are real,
this criterion can be directly applied by checking the absolute value
of the real eigenvalues. A tamed monodromy matrix $\tilde{\mathbf{M}}$
suitable for time evolution is then obtained by transforming back
the modified eigenvalues matrix $\tilde{\mathbf{\Lambda}}$ using
the modified orthogonal matrices $\tilde{\mathbf{U}}$
\begin{equation}
\mathbf{\tilde{M}}=\mathbf{\tilde{U}}\tilde{\mathbf{\Lambda}}\mathbf{\tilde{U}}^{-1}.\label{eq:tamed_monodromy_matrix}
\end{equation}
A possible procedure for applying Eq.(\ref{eq:tamed_monodromy_matrix})
is to monitor the larger real eigenvalues and apply either Eqs. (\ref{eq:modified_U})
or (\ref{eq:modified_Lambda}) or both whenever this is above $\epsilon_{thr}$.
Numerical tests showed either choice is equivalent. However, it may
be necessary to apply the regularization to more than a single degree
of freedom, when the system is very chaotic. We applied multiple regularizations
when a single one failed to limit numerical divergence.

\section{Numerical tests\label{sec:numerical_tests}}

To assess the accuracy of the pre-exponential factor $C_{t}\left(\mathbf{p}_{0},\mathbf{q}_{0}\right)$
approximations introduced above, we consider both chaotic model potentials,
as well as real molecular systems. The chaotic potentials are the
bidimensional Henon-Heiles potential\cite{Brewer_99} and a bidimensional
quartic potential.\cite{Kay_101,Pollak_chaoticquartic_89} These examples
are famously chaotic systems and their accurate spectrum calculation
represents a tough challenge for semiclassical dynamics. Spectra have
been calculated using both Eq.(\ref{eq:HK_powerspectrum}), and the
time-averaged expression of Eq.(\ref{eq:separable}). The second set
of systems is represented by molecules of growing dimensionality and
complexity, i.e. $\mbox{H}_{2}$, $\mbox{H}_{2}\mbox{O}$, $\mbox{C}\mbox{O}_{2}$,
$\mbox{H}_{2}\mbox{CO}$, $\mbox{CH}_{4}$, and $\mbox{CH}_{2}\mbox{D}_{2}$,
and the spectra has been calculated using Eq.(\ref{eq:separable}).
When the pre-exponential factor is not approximated, the semiclassical
trajectories are rejected either if $1-\mbox{det}\left|\mathbf{M}^{T}\left(t\right)\mathbf{M}\left(t\right)\right|>10^{-5}$,
which is a quite strict criteria for the accuracy of the monodromy
matrix $\mathbf{M}\left(t\right)$ evolution, or using Kay's ad hoc
method of Eq.(\ref{eq:Kay_criterium}). In alternative, when using
the numerical regularization of Subsection \ref{sub:Monodromy-Matrix-regularization},
we tested different threshold values for the highest monodromy matrix
eigenvalue, and we found out that $\epsilon_{thr}=1.15\times10^{3}$
is high enough to not perturb vibrational spectra for both model and
molecular systems. This set of examples will allow the reader to fully
appreciate the accuracy of the approximations for future applications,
not only for models but also for real molecular systems. In the following,
unless specified, atomic units $\left(\hbar=1\right)$ are adopted.

\subsection{Bidimensional Henon-Heiles potential\label{sec:Henon_Helies}}

Our first example of a model chaotic potential is the bidimensional
Henon-Heiles potential
\begin{equation}
V\left(x,y\right)=\frac{1}{2}\left(x^{2}+y^{2}\right)+\lambda x^{2}y-\frac{\lambda}{3}y^{3}\label{eq:HH_pot}
\end{equation}
where the mass and the harmonic frequencies are taken to be equal
to unit. The $\lambda$ parameter modulates the amount of chaos added
to the otherwise harmonic motion. There are four stationary points
for this potential. The minimum is at the origin and the others are
saddle points. We choose to look at the power spectrum for two values
of $\lambda$. One is $\lambda=0.11803$, which is the same employed
by others,\cite{Miller340_GeneralizedFilinov_01,Brewer_99} and it
represents a soft chaos motion. The other is $\lambda=0.400$ and
it reproduces a quite strongly chaotic motion, as far as we are aware
never considered before in semiclassical dynamics. For case 1 and
2 below, the length of a typical semiclassical trajectory with an
approximated pre-exponential factor is $5000$ time steps of $0.1$
a.u. each. Semiclassical results are compared with exact quantum mechanical
Discrete Variable Representation (DVR) calculations.\cite{Colbert_DVR}

\paragraph*{Case 1: Soft chaos}

\begin{figure}
\begin{centering}
\includegraphics[scale=0.4]{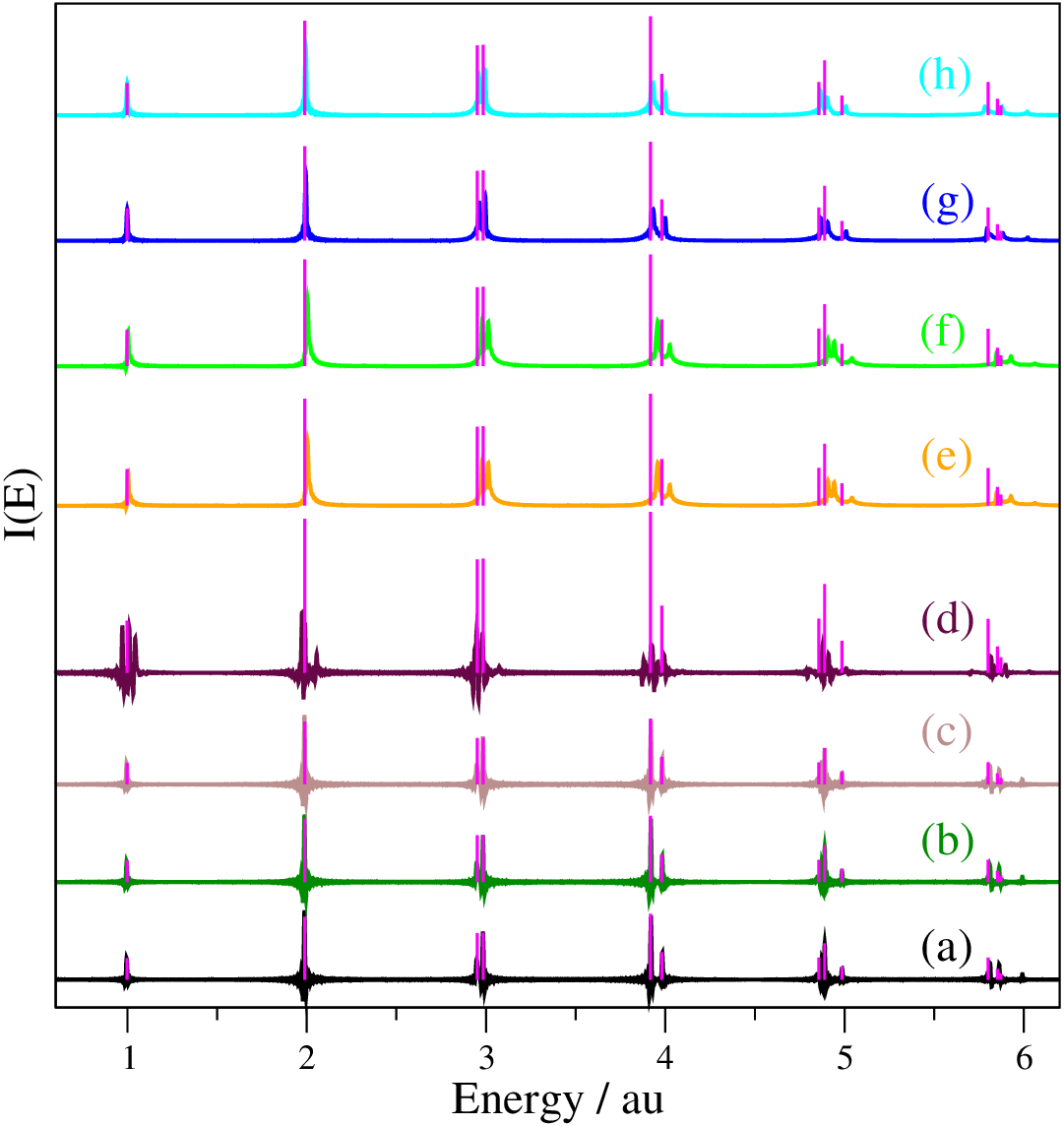}
\par\end{centering}

\caption{\label{fig:spettri_Henon_Helies_soft_HK}SC-IVR spectra of a bidimensional
Henon-Heiles potential with $\lambda=0.11803$ using Eq.(\ref{eq:HK_powerspectrum}).
(a) Black continuous lines are for the rejection criterium $1-\mbox{det}\left|\mathbf{M}^{T}\left(t\right)\mathbf{M}\left(t\right)\right|>10^{-5}$;
(b) dark green continuous lines for the rejection method of Kay (\ref{eq:Kay_criterium});
(c) brown for the regularization of the monodromy matrix (\ref{eq:tamed_monodromy_matrix});
(d) maroon for the PPs approximation; (e) orange for the harmonic
pre-exponential factor approximation of Eq.(\ref{eq:prefactor_Harmonic});
(f) light green spectrum for the approximation in Eq.(\ref{eq:Miller_approx})
$\mathbf{R}_{t}^{(1)}$; (g) blue for the pre-exponential factor reported
in Eq.(\ref{eq:Rt2}) $\mathbf{R}_{t}^{(2)}$ and (h) cyan for Eq.(\ref{eq:Rt3})
$\mathbf{R}_{t}^{(3)}$. Exact quantum mechanical values are indicated
by the vertical magenta lines with an height which is equal to the
square of the overlap between the SC reference state and the exact
eigenstate calculated by DVR.}
\end{figure}

The power spectrum is calculated employing Eq.(\ref{eq:HK_powerspectrum})
and sampling $10^{7}$ trajectories for the Monte Carlo integration,
which is already enough for convergence. The sampling is performed
such that the position center is set equal to the equilibrium positions
and the momentum center is located at the first harmonic vibrational
level, i.e. $p_{j}=\sqrt{3\hbar\omega_{j}}$ in mass-scaled coordinates,
where $\omega_{j}$ is the harmonic frequency of the $j-th$ mode.
This choice is evident when observing that the second and third peaks
in Fig.(\ref{fig:spettri_Henon_Helies_soft_HK}) are the most intense
ones. The coupling $\lambda=0.11803$ is small and only $28\%$ of
the trajectories are rejected using $1-\mbox{det}\left|\mathbf{M}^{T}\left(t\right)\mathbf{M}\left(t\right)\right|>10^{-5}$,
while 26\% using Kay's criterium of Eq.(\ref{eq:Kay_criterium}),
as it should be for a soft chaotic regime. We find the two rejection
criteria to be very similar in terms of accuracy, shape of the spectra
and number of rejected trajectories. Instead, $10^{6}$ trajectories
are more than enough to converge the Monte Carlo integration for the
calculation of the spectra using Eq.(\ref{eq:HK_powerspectrum}) in
conjuction with the analytical and the numerical pre-exponential factor
approximations described above.
Fig.(\ref{fig:spettri_Henon_Helies_soft_HK}) reports the power spectra
at the level of Eq.(\ref{eq:HK_powerspectrum}). The bottom spectra
(a) is calculated using the $\mbox{det}\left|\mathbf{M}^{T}\left(t\right)\mathbf{M}\left(t\right)\right|$
rejection criterium, while (b) using Eq.(\ref{eq:Kay_criterium}).
The two spectra are almost identical. As far as the numerical regularization
of Eq.(\ref{eq:tamed_monodromy_matrix}) reported at spectrum (c),
the results are in very good agreement with (a) and (b). Only $28\%$
(the same percent of the determinat rejection criterium) trajectories
have been regularized and the most choatic one was tamed for $278$
times out of $5000$ steps. Spectrum (d) is computed with the PPs
pre-exponential factor approximation of section \ref{sec:The-poor-person's},
while spectrum (e) refers to the harmonic pre-exponential factor of
section \ref{sec:harmonic}. Spectrum (f) is obtained by using $\mathbf{R}_{t}^{(1)}$
approximation of Eq.(\ref{eq:Miller_approx}), while (g) and (h) derive
from our ansatzs presented in Section \ref{sec:ourapprox} and formulated
in Eq.(\ref{eq:Rt2}) and Eq.(\ref{eq:Rt3}) respectively. Fig.(\ref{fig:spettri_Henon_Helies_soft_HK})
shows quite a good agreement, both in peak position and intensity
between all approximations and the SC-IVR results. The prefactor approximations
formulated in Eq.(\ref{eq:Rt2}) and Eq.(\ref{eq:Rt3}) works better
than the harmonic and $\mathbf{R}_{t}^{(1)}$ approximations. The
Johnson approximation of sec.\ref{sec:Johnson} cannot be applied
for the Henon-Heiles potential because $\omega_{j,t}$ in Eq.(\ref{eq:Johnson_2})
is often imaginary, making the exponential term too big to be calculated
(overflowing code error). The adiabatic approximation couldn't be
applied, since Eq.s (\ref{eq:Qtilde}) and (\ref{eq:Ptilde}) are
too chaotic and cannot be integrated numerically. The computed energy
levels are reported in Table (\ref{tab:Henon-Helies-soft_HK}). 
\begin{table}
\centering{}\caption{\label{tab:Henon-Helies-soft_HK}Power spectrum of the Henon-Heiles
potential with $\lambda=0.11803.$ Comparison between results (in
Atomic Units) obtained using Eq.(\ref{eq:HK_powerspectrum}) at different
level of approximation. From left to right: Exact DVR values, SC-IVR
values using the rejection criterium $1-\mbox{det}\left|\mathbf{M}^{T}\left(t\right)\mathbf{M}\left(t\right)\right|>10^{-5}$,
SC-IVR calculation using the ad hoc Kay's rejection method of Eq.(\ref{eq:Kay_criterium}),
SC-IVR calculation using the monodromy matrix regularization (\ref{eq:tamed_monodromy_matrix}),
the PPs approximation (\ref{eq:PPs}), the harmonic approximation
(\ref{eq:prefactor_Harmonic}), $\mathbf{R}_{t}^{(1)}$ approximation
(\ref{eq:Miller_approx}), and our approximations of Eqs. (\ref{eq:Rt2})
and (\ref{eq:Rt3}). In the last row the Mean Average Errors (MAE)
are reported.}
\begin{tabular}{cccccccccc}
 & {\footnotesize{}Exact} & {\footnotesize{}SC-IVR} & {\footnotesize{}Kay's method} & {\footnotesize{}Regularization} & {\footnotesize{}PPs} & {\footnotesize{}HO} & {\footnotesize{}$\mathbf{R}_{t}^{(1)}$} & {\footnotesize{}$\mathbf{R}_{t}^{(2)}$} & {\footnotesize{}$\mathbf{R}_{t}^{(3)}$}\tabularnewline
\cline{2-10} 
 & {\footnotesize{}0.998} & {\footnotesize{}0.995} & {\footnotesize{}0.995} & {\footnotesize{}0.995} & {\footnotesize{}0.971} & {\footnotesize{}1.003} & {\footnotesize{}1.003} & {\footnotesize{}0.998} & {\footnotesize{}0.998}\tabularnewline
\cline{2-10} 
 & {\footnotesize{}1.989} & {\footnotesize{}1.987} & {\footnotesize{}1.987} & {\footnotesize{}1.987} & {\footnotesize{}1.974} & {\footnotesize{}2.004} & {\footnotesize{}2.004} & {\footnotesize{}1.994} & {\footnotesize{}1.994}\tabularnewline
\cline{2-10} 
 & {\footnotesize{}1.989} & {\footnotesize{}1.987} & {\footnotesize{}1.987} & {\footnotesize{}1.987} & {\footnotesize{}1.974} & {\footnotesize{}2.004} & {\footnotesize{}2.004} & {\footnotesize{}1.994} & {\footnotesize{}1.994}\tabularnewline
\cline{2-10} 
 & {\footnotesize{}2.951} & {\footnotesize{}2.947} & {\footnotesize{}2.948} & {\footnotesize{}2.948} & {\footnotesize{}2.948} & {\footnotesize{}2.979} & {\footnotesize{}2.979} & {\footnotesize{}2.962} & {\footnotesize{}2.961}\tabularnewline
\cline{2-10} 
 & {\footnotesize{}2.984} & {\footnotesize{}2.983} & {\footnotesize{}2.983} & {\footnotesize{}2.983} & {\footnotesize{}2.980} & {\footnotesize{}3.012} & {\footnotesize{}3.012} & {\footnotesize{}2.995} & {\footnotesize{}2.994}\tabularnewline
\cline{2-10} 
 & {\footnotesize{}2.984} & {\footnotesize{}2.983} & {\footnotesize{}2.983} & {\footnotesize{}2.983} & {\footnotesize{}2.980} & {\footnotesize{}3.012} & {\footnotesize{}3.012} & {\footnotesize{}2.995} & {\footnotesize{}2.994}\tabularnewline
\cline{2-10} 
 & {\footnotesize{}3.917} & {\footnotesize{}3.92} & {\footnotesize{}3.920} & {\footnotesize{}3.920} & {\footnotesize{}3.920} & {\footnotesize{}3.958} & {\footnotesize{}3.958} & {\footnotesize{}3.931} & {\footnotesize{}3.931}\tabularnewline
\cline{2-10} 
 & {\footnotesize{}3.918} & {\footnotesize{}3.92} & {\footnotesize{}3.920} & {\footnotesize{}3.920} & {\footnotesize{}3.920} & {\footnotesize{}3.958} & {\footnotesize{}3.958} & {\footnotesize{}3.931} & {\footnotesize{}3.931}\tabularnewline
\cline{2-10} 
 & {\footnotesize{}3.980} & {\footnotesize{}3.982} & {\footnotesize{}3.982} & {\footnotesize{}3.983} & {\footnotesize{}3.995} & {\footnotesize{}4.025} & {\footnotesize{}4.025} & {\footnotesize{}4.000} & {\footnotesize{}3.999}\tabularnewline
\cline{2-10} 
 & {\footnotesize{}3.984} & {\footnotesize{}3.982} & {\footnotesize{}3.982} & {\footnotesize{}3.983} & {\footnotesize{}3.995} & {\footnotesize{}4.025} & {\footnotesize{}4.025} & {\footnotesize{}4.000} & {\footnotesize{}3.999}\tabularnewline
\cline{2-10} 
 & {\footnotesize{}4.856} & {\footnotesize{}4.873} & {\footnotesize{}4.873} & {\footnotesize{}4.874} & {\footnotesize{}4.876} & {\footnotesize{}4.907} & {\footnotesize{}4.907} & {\footnotesize{}4.868} & {\footnotesize{}4.864}\tabularnewline
\cline{2-10} 
 & {\footnotesize{}4.888} & {\footnotesize{}4.889} & {\footnotesize{}4.889} & {\footnotesize{}4.889} & {\footnotesize{}4.910} & {\footnotesize{}4.942} & {\footnotesize{}4.942} & {\footnotesize{}4.906} & {\footnotesize{}4.903}\tabularnewline
\cline{2-10} 
 & {\footnotesize{}4.888} & {\footnotesize{}4.889} & {\footnotesize{}4.889} & {\footnotesize{}4.889} & {\footnotesize{}4.910} & {\footnotesize{}4.942} & {\footnotesize{}4.942} & {\footnotesize{}4.906} & {\footnotesize{}4.903}\tabularnewline
\cline{2-10} 
 & {\footnotesize{}4.985} & {\footnotesize{}4.985} & {\footnotesize{}4.985} & {\footnotesize{}4.986} & {\footnotesize{}5.009} & {\footnotesize{}5.041} & {\footnotesize{}5.041} & {\footnotesize{}5.008} & {\footnotesize{}5.007}\tabularnewline
\cline{2-10} 
 & {\footnotesize{}4.985} & {\footnotesize{}4.985} & {\footnotesize{}4.985} & {\footnotesize{}4.986} & {\footnotesize{}5.009} & {\footnotesize{}5.041} & {\footnotesize{}5.041} & {\footnotesize{}5.008} & {\footnotesize{}5.007}\tabularnewline
\cline{2-10} 
 & {\footnotesize{}5.800} & {\footnotesize{}5.812} & {\footnotesize{}5.811} & {\footnotesize{}5.811} & {\footnotesize{}5.818} & {\footnotesize{}5.849} & {\footnotesize{}5.849} & {\footnotesize{}5.795} & {\footnotesize{}5.783}\tabularnewline
\cline{2-10} 
 & {\footnotesize{}5.800} & {\footnotesize{}5.812} & {\footnotesize{}5.811} & {\footnotesize{}5.811} & {\footnotesize{}5.818} & {\footnotesize{}5.849} & {\footnotesize{}5.849} & {\footnotesize{}5.795} & {\footnotesize{}5.783}\tabularnewline
\cline{2-10} 
 & {\footnotesize{}5.853} & {\footnotesize{}5.862} & {\footnotesize{}5.862} & {\footnotesize{}5.862} & {\footnotesize{}5.833} & {\footnotesize{}5.863} & {\footnotesize{}5.863} & {\footnotesize{}5.882} & {\footnotesize{}5.878}\tabularnewline
\cline{2-10} 
 & {\footnotesize{}5.872} & {\footnotesize{}5.878} & {\footnotesize{}5.878} & {\footnotesize{}5.878} & {\footnotesize{}5.898} & {\footnotesize{}5.928} & {\footnotesize{}5.928} & {\footnotesize{}5.882} & {\footnotesize{}5.878}\tabularnewline
\hline 
{\footnotesize{}MAE} &  & {\footnotesize{}0.004} & {\footnotesize{}0.004} & {\footnotesize{}0.004} & {\footnotesize{}0.015} & {\footnotesize{}0.038} & {\footnotesize{}0.038} & {\footnotesize{}0.013} & {\footnotesize{}0.013}\tabularnewline
\hline 
\end{tabular}
\end{table}

When using the time averaged power spectrum approximation of Eq.(\ref{eq:separable}),
we run only 5000 trajectories after verifying that $10^{3}$ trajectories
are enough to reach numerical convergence. 
\begin{figure}
\begin{centering}
\includegraphics[scale=0.4]{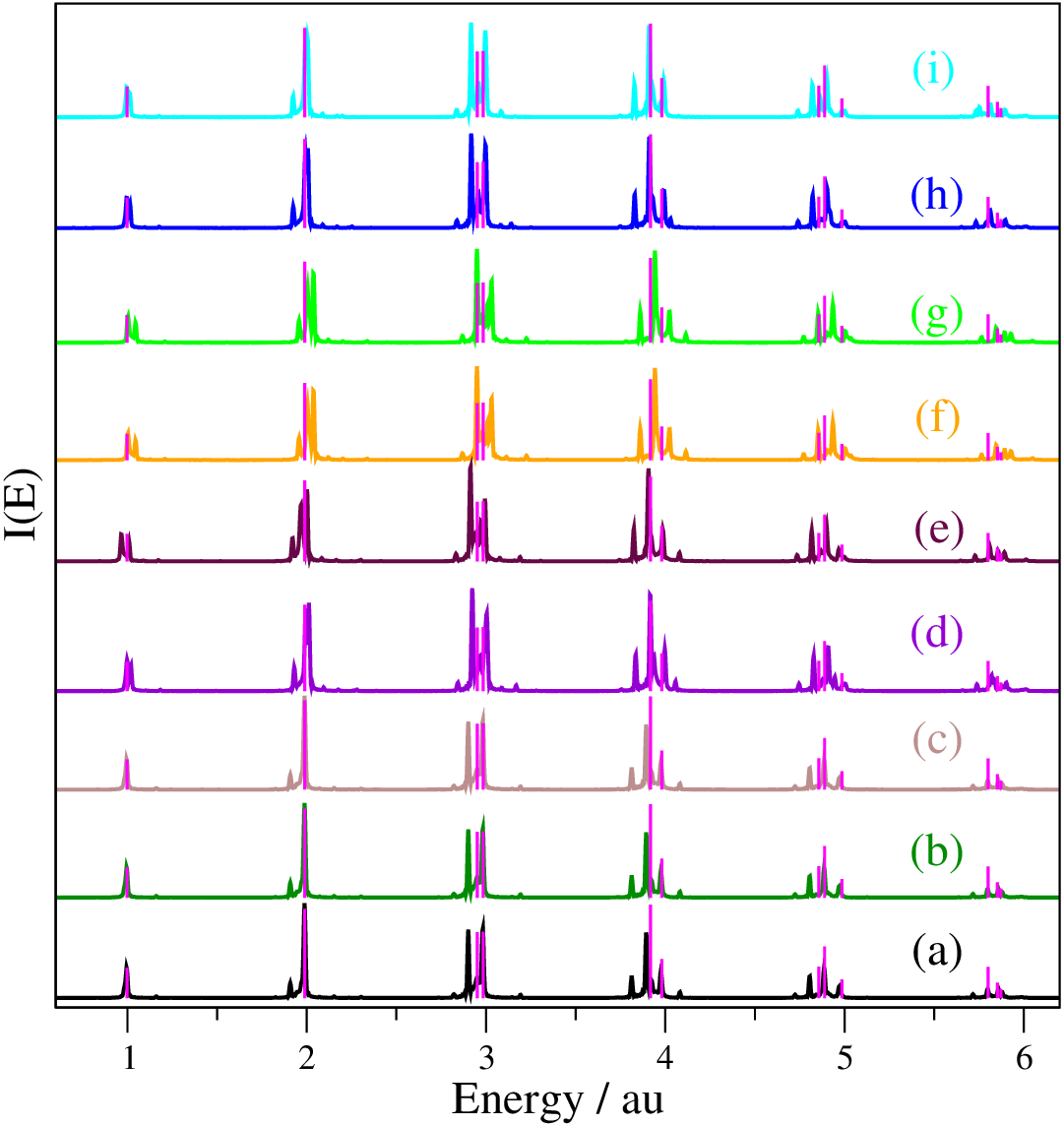}
\par\end{centering}

\caption{\label{fig:spettri_Henon_Helies_soft_TA}TA-SC-IVR (Eq.(\ref{eq:separable}))
spectra of a bidimensional Henon-Heiles potential with $\lambda=0.11803$.
(a) Black continuous lines are for semiclassical spectra (\ref{eq:separable})
using the rejection criterium $1-\mbox{det}\left|\mathbf{M}^{T}\left(t\right)\mathbf{M}\left(t\right)\right|>10^{-5}$;
(b) dark green continuous lines are for the rejection method of Kay
(\ref{eq:Kay_criterium}); (c) brown for the regularization of the
monodromy matrix (\ref{eq:tamed_monodromy_matrix}); (d) violet for
the adiabatic approximation in Eq.(\ref{eq:C_t_adiabatic}); (e) maroon
for the PPs approximation; (f) orange for the harmonic pre-exponential
factor approximation (Eq.(\ref{eq:prefactor_Harmonic})); (g) light
green spectrum for the approximation in Eq.(\ref{eq:Miller_approx})
$\mathbf{R}_{t}^{(1)}$; (h) blue for the pre-exponential factor reported
in Eq.(\ref{eq:Rt2}) $\mathbf{R}_{t}^{(2)}$ and (i) cyan for the
pre-exponential factor reported in Eq.(\ref{eq:Rt3}) $\mathbf{R}_{t}^{(3)}$.
Exact quantum mechanical values are indicated by the vertical magenta
lines with an height which is equal to square of the overlap between
the SC reference state $\left|\chi\right\rangle $ and the exact eigenstate
calculated by DVR.}
\end{figure}
 The results are reported in Fig. (\ref{fig:spettri_Henon_Helies_soft_TA})
at different level of approximation. The bottom spectra (a) and (b)
are calculated by using Eq.(\ref{eq:separable}) and without any of
the pre-exponential factor approximations. Starting from the bottom,
(c) is performed by using Eq.(\ref{eq:tamed_monodromy_matrix}), where
$12\%$ of trajectories have been regularized and for the most chaotic
one Eq.(\ref{eq:tamed_monodromy_matrix}) is employed $99$ times.
Spectrum (d) is at the level of adiabatic approximation (see Section
(\ref{sec:adiabatic})), the spectrum (e) is computed with the PPs
pre-exponential factor approximation of section \ref{sec:The-poor-person's},
(f) refers to the harmonic pre-exponential factor of section \ref{sec:harmonic},
the (g) spectrum is obtained by using $\mathbf{R}_{t}^{(1)}$ approximation
of Eq.(\ref{eq:Miller_approx}), (h) and (i) derive from our ansatzs
presented in Section \ref{sec:ourapprox} and formulated in Eq.(\ref{eq:Rt2})
and Eq.(\ref{eq:Rt3}).We observe a quite good agreement between all
approximations and the original SC-IVR calculations, both in peak
position and intensity. The Johnson approximation can not be applied
also in this case. In addition respect to Fig.(\ref{fig:spettri_Henon_Helies_soft_HK}),
we can apply the adiabatic approximation, since less trajectories
are required for the time averaged spectrum.
\begin{table}
\centering{}\caption{\label{tab:Henon-Helies-soft_TA}Time averaged spectra for the Henon-Heiles
potential with $\lambda=0.11803.$ Comparison between results (in
Atomic Units) obtained with different approximations. From left to
right: Exact values, TA-SC-IVR values (\ref{eq:separable}) using
the rejection criterium $1-\mbox{det}\left|\mathbf{M}^{T}\left(t\right)\mathbf{M}\left(t\right)\right|>10^{-5}$,
TA-SC-IVR calculation using the ad hoc Kay's rejection method of Eq.(\ref{eq:Kay_criterium}),
monodromy matrix regularization (\ref{eq:tamed_monodromy_matrix}),
adiabatic approximation (\ref{eq:C_t_adiabatic}), PPs approximation
(\ref{eq:PPs}), harmonic approximation (\ref{eq:prefactor_Harmonic}),
$\mathbf{R}_{t}^{(1)}$ approximation (\ref{eq:Miller_approx}), and
our approximations of Eqs. (\ref{eq:Rt2}) and (\ref{eq:Rt3}). In
the last row the Mean Average Errors (MAE) are reported.}
\begin{tabular}{ccccccccccc}
 & {\footnotesize{}Exact} & {\footnotesize{}SC-IVR} & {\footnotesize{}Kay's method} & {\footnotesize{}Regularization} & {\footnotesize{}Adiabatic} & {\footnotesize{}PPs} & {\footnotesize{}HO} & {\footnotesize{}$\mathbf{R}_{t}^{(1)}$} & {\footnotesize{}$\mathbf{R}_{t}^{(2)}$} & {\footnotesize{}$\mathbf{R}_{t}^{(3)}$}\tabularnewline
\cline{2-11} 
 & {\footnotesize{}0.998} & {\footnotesize{}0.995} & {\footnotesize{}0.995} & {\footnotesize{}0.995} & {\footnotesize{}0.998} & {\footnotesize{}0.965} & {\footnotesize{}1.003} & {\footnotesize{}1.003} & {\footnotesize{}0.997} & {\footnotesize{}0.997}\tabularnewline
\cline{2-11} 
 & {\footnotesize{}1.989} & {\footnotesize{}1.988} & {\footnotesize{}1.988} & {\footnotesize{}1.988} & {\footnotesize{}1.995} & {\footnotesize{}1.967} & {\footnotesize{}2.004} & {\footnotesize{}2.004} & {\footnotesize{}1.993} & {\footnotesize{}1.993}\tabularnewline
\cline{2-11} 
 & {\footnotesize{}1.989} & {\footnotesize{}1.988} & {\footnotesize{}1.988} & {\footnotesize{}1.988} & {\footnotesize{}2.012} & {\footnotesize{}2.001} & {\footnotesize{}2.038} & {\footnotesize{}2.038} & {\footnotesize{}2.007} & {\footnotesize{}2.005}\tabularnewline
\cline{2-11} 
 & {\footnotesize{}2.951} & {\footnotesize{}2.901} & {\footnotesize{}2.901} & {\footnotesize{}2.901} & {\footnotesize{}2.923} & {\footnotesize{}2.913} & {\footnotesize{}2.950} & {\footnotesize{}2.950} & {\footnotesize{}2.917} & {\footnotesize{}2.917}\tabularnewline
\cline{2-11} 
 & {\footnotesize{}2.984} & {\footnotesize{}2.983} & {\footnotesize{}2.983} & {\footnotesize{}2.982} & {\footnotesize{}3.004} & {\footnotesize{}2.994} & {\footnotesize{}3.031} & {\footnotesize{}3.031} & {\footnotesize{}2.997} & {\footnotesize{}2.996}\tabularnewline
\cline{2-11} 
 & {\footnotesize{}2.984} & {\footnotesize{}2.983} & {\footnotesize{}2.983} & {\footnotesize{}2.982} & {\footnotesize{}3.004} & {\footnotesize{}2.994} & {\footnotesize{}3.031} & {\footnotesize{}3.031} & {\footnotesize{}2.997} & {\footnotesize{}2.996}\tabularnewline
\cline{2-11} 
 & {\footnotesize{}3.917} & {\footnotesize{}3.893} & {\footnotesize{}3.893} & {\footnotesize{}3.893} & {\footnotesize{}3.916} & {\footnotesize{}3.907} & {\footnotesize{}3.943} & {\footnotesize{}3.942} & {\footnotesize{}3.911} & {\footnotesize{}3.910}\tabularnewline
\cline{2-11} 
 & {\footnotesize{}3.918} & {\footnotesize{}3.893} & {\footnotesize{}3.893} & {\footnotesize{}3.893} & {\footnotesize{}3.916} & {\footnotesize{}3.907} & {\footnotesize{}3.943} & {\footnotesize{}3.942} & {\footnotesize{}3.911} & {\footnotesize{}3.910}\tabularnewline
\cline{2-11} 
 & {\footnotesize{}3.980} & {\footnotesize{}3.975} & {\footnotesize{}3.975} & {\footnotesize{}3.975} & {\footnotesize{}3.997} & {\footnotesize{}3.987} & {\footnotesize{}4.024} & {\footnotesize{}4.023} & {\footnotesize{}3.993} & {\footnotesize{}3.992}\tabularnewline
\cline{2-11} 
 & {\footnotesize{}3.984} & {\footnotesize{}3.975} & {\footnotesize{}3.975} & {\footnotesize{}3.975} & {\footnotesize{}3.997} & {\footnotesize{}3.987} & {\footnotesize{}4.024} & {\footnotesize{}4.023} & {\footnotesize{}3.993} & {\footnotesize{}3.992}\tabularnewline
\cline{2-11} 
 & {\footnotesize{}4.856} & {\footnotesize{}4.805} & {\footnotesize{}4.805} & {\footnotesize{}4.805} & {\footnotesize{}4.828} & {\footnotesize{}4.818} & {\footnotesize{}4.854} & {\footnotesize{}4.853} & {\footnotesize{}4.822} & {\footnotesize{}4.821}\tabularnewline
\cline{2-11} 
 & {\footnotesize{}4.888} & {\footnotesize{}4.886} & {\footnotesize{}4.886} & {\footnotesize{}4.886} & {\footnotesize{}4.909} & {\footnotesize{}4.899} & {\footnotesize{}4.935} & {\footnotesize{}4.934} & {\footnotesize{}4.902} & {\footnotesize{}4.912}\tabularnewline
\cline{2-11} 
 & {\footnotesize{}4.888} & {\footnotesize{}4.886} & {\footnotesize{}4.886} & {\footnotesize{}4.886} & {\footnotesize{}4.909} & {\footnotesize{}4.899} & {\footnotesize{}4.935} & {\footnotesize{}4.934} & {\footnotesize{}4.902} & {\footnotesize{}4.912}\tabularnewline
\cline{2-11} 
 & {\footnotesize{}4.985} & {\footnotesize{}4.970} & {\footnotesize{}4.97} & {\footnotesize{}4.97} & {\footnotesize{}4.99} & {\footnotesize{}4.968} & {\footnotesize{}5.005} & {\footnotesize{}5.004} & {\footnotesize{}4.984} & {\footnotesize{}4.984}\tabularnewline
\cline{2-11} 
 & {\footnotesize{}4.985} & {\footnotesize{}4.970} & {\footnotesize{}4.97} & {\footnotesize{}4.97} & {\footnotesize{}5.003} & {\footnotesize{}4.968} & {\footnotesize{}5.005} & {\footnotesize{}5.004} & {\footnotesize{}5.002} & {\footnotesize{}5.000}\tabularnewline
\cline{2-11} 
 & {\footnotesize{}5.800} & {\footnotesize{}5.798} & {\footnotesize{}5.798} & {\footnotesize{}5.798} & {\footnotesize{}5.820} & {\footnotesize{}5.810} & {\footnotesize{}5.846} & {\footnotesize{}5.845} & {\footnotesize{}5.811} & {\footnotesize{}5.812}\tabularnewline
\cline{2-11} 
 & {\footnotesize{}5.800} & {\footnotesize{}5.798} & {\footnotesize{}5.798} & {\footnotesize{}5.798} & {\footnotesize{}5.820} & {\footnotesize{}5.810} & {\footnotesize{}5.846} & {\footnotesize{}5.845} & {\footnotesize{}5.811} & {\footnotesize{}5.812}\tabularnewline
\cline{2-11} 
 & {\footnotesize{}5.853} & {\footnotesize{}5.859} & {\footnotesize{}5.859} & {\footnotesize{}5.859} & {\footnotesize{}5.835} & {\footnotesize{}5.857} & {\footnotesize{}5.894} & {\footnotesize{}5.893} & {\footnotesize{}5.874} & {\footnotesize{}5.870}\tabularnewline
\cline{2-11} 
 & {\footnotesize{}5.872} & {\footnotesize{}5.879} & {\footnotesize{}5.879} & {\footnotesize{}5.879} & {\footnotesize{}5.902} & {\footnotesize{}5.892} & {\footnotesize{}5.929} & {\footnotesize{}5.927} & {\footnotesize{}5.896} & {\footnotesize{}5.894}\tabularnewline
\hline 
{\footnotesize{}MAE} &  & {\footnotesize{}0.011} & {\footnotesize{}0.011} & {\footnotesize{}0.012 } & {\footnotesize{}0.017} & {\footnotesize{}0.016 } & {\footnotesize{}0.033} & {\footnotesize{}0.032} & {\footnotesize{}0.014} & {\footnotesize{}0.015 }\tabularnewline
\hline 
\end{tabular}
\end{table}
Table (\ref{tab:Henon-Helies-soft_TA}) confirms the accuracy of the
separable time-averaging SC-IVR (\ref{eq:separable}) values reported
in the second column with respect to the exact ones in the first column,
calculated by DVR. For the soft chaos Henon-Heiles power spectrum
calculation, SC-IVR displays an energy mean average error (MAE) which
is about 1\% of the zero point value. The ``Regularization'' column
shows that the artificial numerical regularization of Eq.(\ref{eq:tamed_monodromy_matrix})
is not influential again, showing the negligible contribution of the
chaotic trajectories to the spectrum calculation of this system. In
this case the prefactor approximations have been tested on the top
of the separable approximation. All other columns report the results
with different pre-exponential factor approximations and they should
be compared with the SC-IVR ones. $\mathbf{R}_{t}^{(1)}$ approximation
(\ref{eq:Miller_approx}), and the harmonic oscillator one (\ref{eq:prefactor_Harmonic})
are, as before, quite similar and they usually overestimate the exact
and semiclassical results as expected, since they do not properly
account for anharmonicity. Also the PPs overestimates by about the
same amount. The adiabatic approximation (\ref{eq:C_t_adiabatic})
in the fourth column is more accurate than the PPs, the Harmonic and
$\mathbf{R}_{t}^{(1)}$ ones, but still overestimates the original
SC-IVR values. Finally, the ansatzs of Eq.(\ref{eq:Rt2}) and (\ref{eq:Rt3}),
are the better performing pre-exponential factor analytical approximations
and quite similar to the adiabatic one, where no harmonic assumptions
have been introduced.

\subsubsection*{Case 2: Strong chaos}

\begin{figure}
\begin{centering}
\includegraphics[scale=0.4]{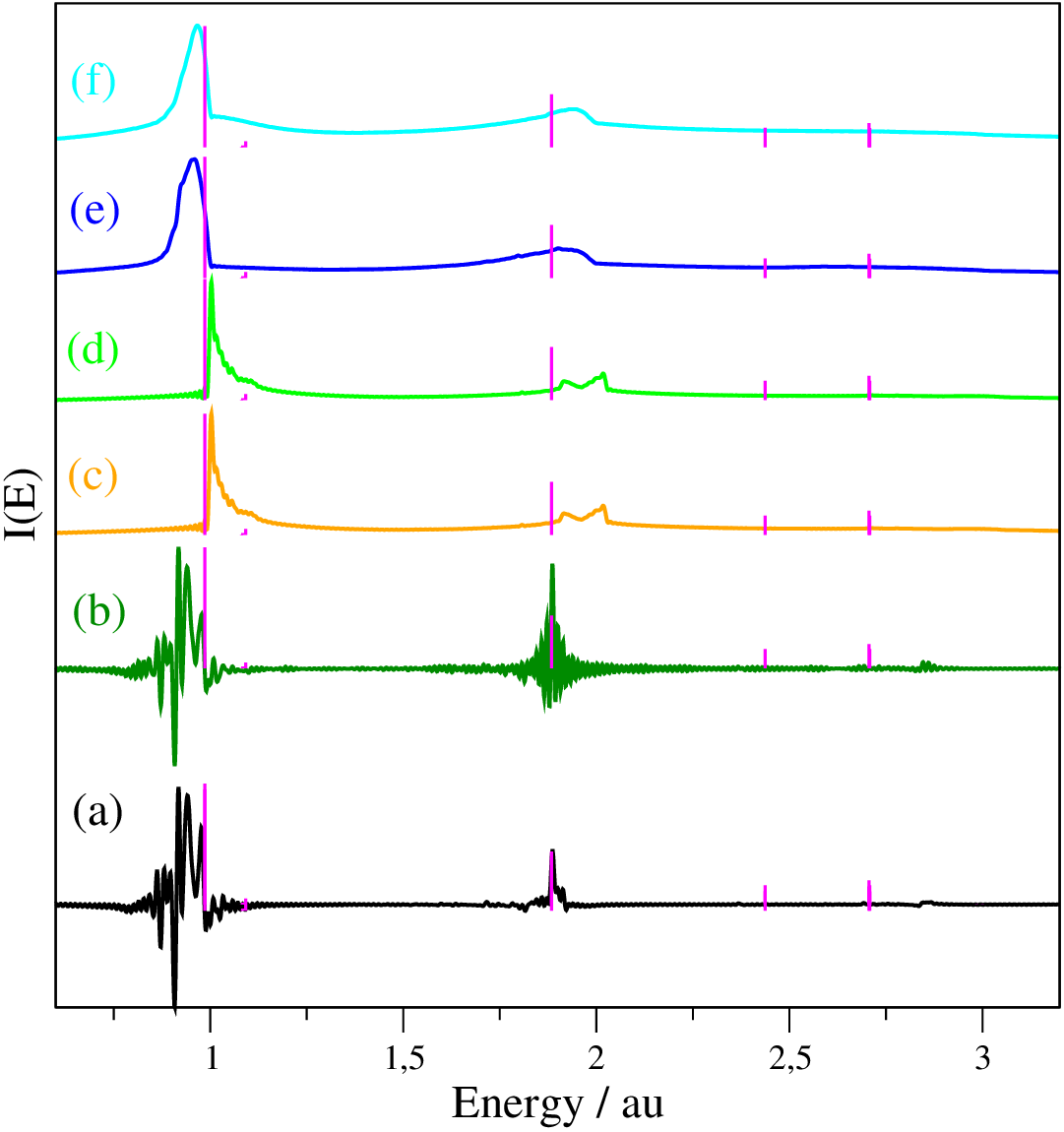}
\par\end{centering}

\caption{\label{fig:spettri_Henon_Helies-strong_HK}SC-IVR spectra of a bidimensional
Henon-Heiles potential with $\lambda=0.4$ using Eq.(\ref{eq:HK_powerspectrum}).
(a) Black continuous lines are for the rejection criterium $1-\mbox{det}\left|\mathbf{M}^{T}\left(t\right)\mathbf{M}\left(t\right)\right|>10^{-5}$;
(b) dark green continuous lines for the rejection method of Kay (\ref{eq:Kay_criterium});
(c) orange for the harmonic pre-exponential factor approximation (Eq.(\ref{eq:prefactor_Harmonic}));
(d) light green spectrum for $\mathbf{R}_{t}^{(1)}$ approximation
in Eq.(\ref{eq:Miller_approx}); (e) blue for the pre-exponential
factor reported in Eq.(\ref{eq:Rt2}) and (f) cyan for the pre-exponential
factor reported in Eq.(\ref{eq:Rt3}). Exact quantum mechanical values
are indicated by the vertical magenta lines with an height which is
equal to square of the overlap between the SC reference state and
the exact eigenstate calculated by DVR.}
\end{figure}

We now look at a strong chaotic motion scenario by increasing the
value of the coupling term to $\lambda=0.4$. For this value of $\lambda$,
states above the ground one are quasi-bound and complex valued. Nevertheless,
the SC-IVR can reproduce the real part of the vibrational eigenvalues.
In the case of Eq.(\ref{eq:HK_powerspectrum}), due to the high rejection
ratio, we sample $10^{8}$ trajectories in conjunction with the $\mbox{det}\left|\mathbf{M}^{T}\left(t\right)\mathbf{M}\left(t\right)\right|$
and Kay's criterium, while $10^{7}$ trajectories are more than enough
for the prefactor approximated spectra calculation. The system is
so chaotic, that Eq.(\ref{eq:tamed_monodromy_matrix}) could not avoid
the monodromy matrix elements numerical divergence to infinity when
applied either to the modulus of the biggest real eigenvalue or to
the moduli of the real eigenvalues greater than $\epsilon_{thr}$.
The PPs approximation lead to a spectrum which is too noisy to find
peaks, and for this reason we choose to do not report it in Fig. (\ref{fig:spettri_Henon_Helies-strong_HK}).
Each peak value is reported in Table (\ref{tab:Henon-Helies-strong_HK}).
\begin{table}
\centering{}\caption{\label{tab:Henon-Helies-strong_HK}Henon-Heiles potential with $\lambda=0.4.$
Column labels as in Table (\ref{tab:Henon-Helies-soft_HK}).}
\begin{tabular}{cccccccc}
 & {\footnotesize{}Ex.} & {\footnotesize{}SC-IVR} & {\footnotesize{}Kay's method} & {\footnotesize{}HO} & {\footnotesize{}$\mathbf{R}_{t}^{(1)}$} & {\footnotesize{}$\mathbf{R}_{t}^{(2)}$} & {\footnotesize{}$\mathbf{R}_{t}^{(3)}$}\tabularnewline
\hline 
 & {\footnotesize{}0.986} & {\footnotesize{}0.918} & {\footnotesize{}0.918} & {\footnotesize{}1.003} & {\footnotesize{}1.003} & {\footnotesize{}0.953} & {\footnotesize{}0.967}\tabularnewline
\hline 
 & {\footnotesize{}1.081} & {\footnotesize{}1.078} & {\footnotesize{}1.073} & {\footnotesize{}1.106} & {\footnotesize{}1.092} & {\footnotesize{}1.011} & {\footnotesize{}1.01}\tabularnewline
\hline 
 & {\footnotesize{}1.084} & {\footnotesize{}1.078} & {\footnotesize{}1.073} & {\footnotesize{}1.106} & {\footnotesize{}1.092} & {\footnotesize{}1.011} & {\footnotesize{}1.01}\tabularnewline
\hline 
 & {\footnotesize{}1.092} & {\footnotesize{}1.078} & {\footnotesize{}1.073} & {\footnotesize{}1.106} & {\footnotesize{}1.016} & {\footnotesize{}1.011} & {\footnotesize{}1.01}\tabularnewline
\hline 
 & {\footnotesize{}1.883} & {\footnotesize{}1.886} & {\footnotesize{}1.886} & {\footnotesize{}2.018} & {\footnotesize{}2.018} & {\footnotesize{}1.902} & {\footnotesize{}1.932}\tabularnewline
\hline 
 & {\footnotesize{}1.884} & {\footnotesize{}1.886} & {\footnotesize{}1.886} & {\footnotesize{}2.018} & {\footnotesize{}2.018} & {\footnotesize{}1.902} & {\footnotesize{}1.945}\tabularnewline
\hline 
 & {\footnotesize{}2.437} & {\footnotesize{}2.368} & {\footnotesize{}2.367} & {\footnotesize{}2.714} & {\footnotesize{}2.713} & {\footnotesize{}2.517} & {\footnotesize{}2.508}\tabularnewline
\hline 
 & {\footnotesize{}2.706} & {\footnotesize{}2.693} & {\footnotesize{}2.694} & {\footnotesize{}2.779} & {\footnotesize{}2.779} & {\footnotesize{}2.653} & {\footnotesize{}2.647}\tabularnewline
\hline 
 & {\footnotesize{}2.708} & {\footnotesize{}2.693} & {\footnotesize{}2.694} & {\footnotesize{}2.779} & {\footnotesize{}2.779} & {\footnotesize{}2.653} & {\footnotesize{}2.647}\tabularnewline
\hline 
{\footnotesize{}MAE} &  & {\footnotesize{}0.022} & {\footnotesize{}0.023} & {\footnotesize{}0.085} & {\footnotesize{}0.089} & {\footnotesize{}0.054} & {\footnotesize{}0.061}\tabularnewline
\hline 
\end{tabular}
\end{table}

\begin{figure}
\begin{centering}
\includegraphics[scale=0.4]{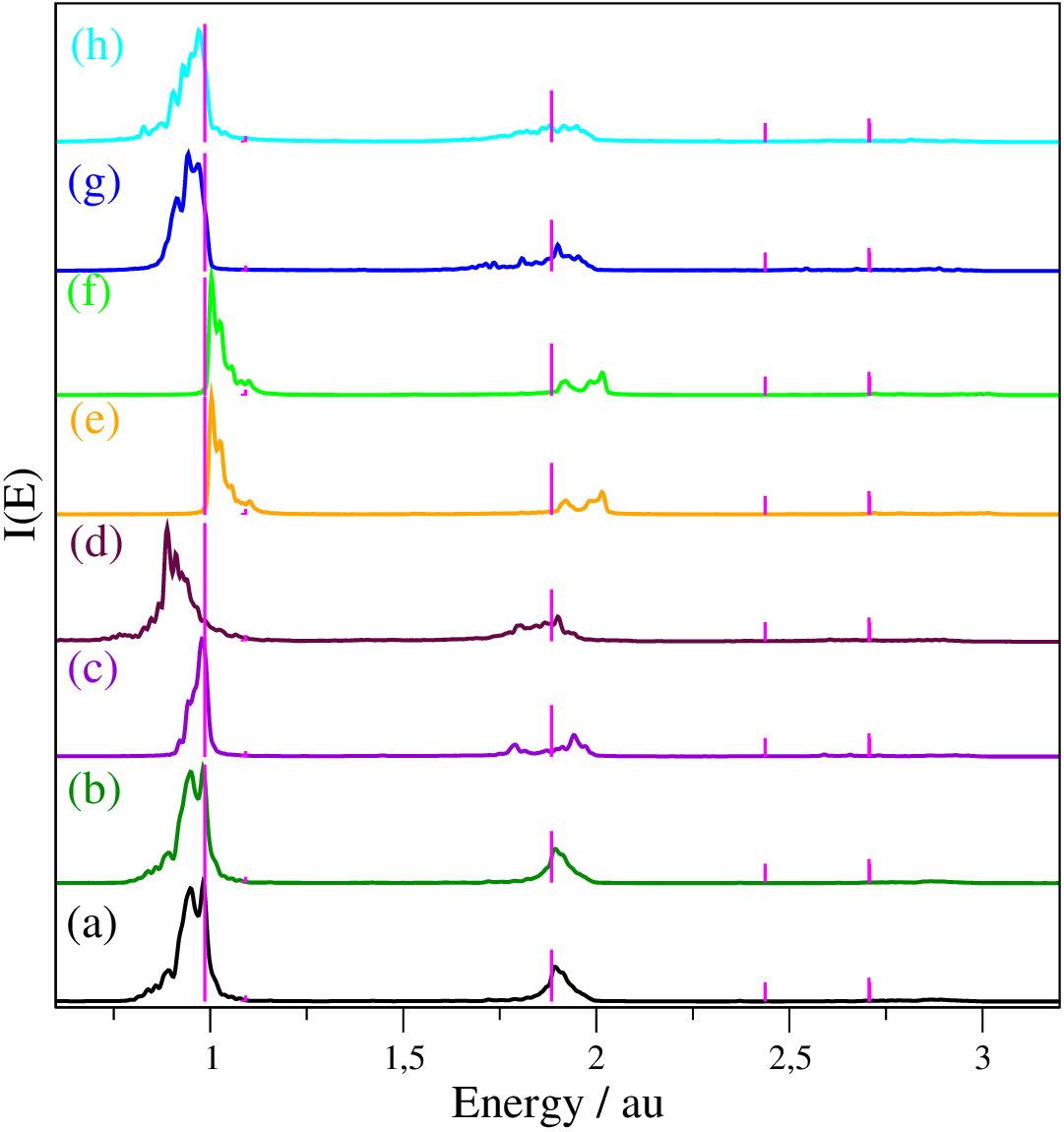}
\par\end{centering}

\caption{\label{fig:spettri_Henon_Helies-strong_TA}TA-SC-IVR spectra of a
bidimensional Henon-Heiles potential with $\lambda=0.4$. a) Black
continuous lines are for semiclassical spectra (\ref{eq:separable})
using the rejection criterium $1-\mbox{det}\left|\mathbf{M}^{T}\left(t\right)\mathbf{M}\left(t\right)\right|>10^{-5}$;
(b) dark green continuous lines are for semiclassical spectra computed
using the rejection method of Kay; (c) violet for the adiabatic approximation
in Eq.(\ref{eq:C_t_adiabatic}); (d) maroon for the PPs approximation;
(e) orange for the harmonic pre-exponential factor approximation (Eq.(\ref{eq:prefactor_Harmonic}));
(f) light green spectrum for $\mathbf{R}_{t}^{(1)}$ approximation
in Eq.(\ref{eq:Miller_approx}); (g) blue for the pre-exponential
factor reported in Eq.(\ref{eq:Rt2}) and (h) cyan for the pre-exponential
factor reported in Eq.(\ref{eq:Rt3}). Exact quantum mechanical values
are indicated by the vertical magenta lines with an height which is
equal to square of the overlap between the SC reference state and
the exact eigenstate calculated by DVR.}
\end{figure}

Again the two rejection criteria seem to lead to very similar spectra.The
present approximations show comparable results and better than the
Harmonic and $\mathbf{R}_{t}^{(1)}$ ones. The spectra are reported
in Fig.(\ref{fig:spettri_Henon_Helies-strong_HK}).
 In the case of TA-SC-IVR calculations we sampled $50000$ trajectories
for the Monte Carlo integration of Eq.(\ref{eq:separable}) rejecting
91\% of the trajectories when using both rejection criteria. Instead,
$5000$ trajectories are enough for the approximated pre-exponential
factor calculations. All power spectra are reported in Fig. (\ref{fig:spettri_Henon_Helies-strong_TA})
and each peak value is reported in Table (\ref{tab:Henon-Helies-strong_TA}).

\begin{table}
\centering{}\caption{\label{tab:Henon-Helies-strong_TA}Henon-Heiles potential with $\lambda=0.4.$
Column labels as in Table (\ref{tab:Henon-Helies-soft_TA}).}
\begin{tabular}{cccccccccc}
 & {\footnotesize{}Ex.} & {\footnotesize{}TA-SC-IVR} & {\footnotesize{}Kay's method} & {\footnotesize{}Adiabatic} & {\footnotesize{}PPs} & {\footnotesize{}HO} & {\footnotesize{}$\mathbf{R}_{t}^{(1)}$} & {\footnotesize{}$\mathbf{R}_{t}^{(2)}$} & {\footnotesize{}$\mathbf{R}_{t}^{(3)}$}\tabularnewline
\cline{2-10} 
 & {\footnotesize{}0.986} & {\footnotesize{}0.949} & {\footnotesize{}0.949} & {\footnotesize{}0.98} & {\footnotesize{}0.889} & {\footnotesize{}1.004} & {\footnotesize{}1.003} & {\footnotesize{}0.945} & {\footnotesize{}0.973}\tabularnewline
\cline{2-10} 
 & {\footnotesize{}1.081} & {\footnotesize{}1.078} & {\footnotesize{}1.077} & {\footnotesize{}1.088} & {\footnotesize{}1.093} & {\footnotesize{}1.102} & {\footnotesize{}1.003} & {\footnotesize{}1.083} & {\footnotesize{}1.083}\tabularnewline
\cline{2-10} 
 & {\footnotesize{}1.084} & {\footnotesize{}1.078} & {\footnotesize{}1.077} & {\footnotesize{}1.088} & {\footnotesize{}1.093} & {\footnotesize{}1.102} & {\footnotesize{}1.023} & {\footnotesize{}1.083} & {\footnotesize{}1.083}\tabularnewline
\cline{2-10} 
 & {\footnotesize{}1.092} & {\footnotesize{}1.078} & {\footnotesize{}1.077} & {\footnotesize{}1.088} & {\footnotesize{}1.093} & {\footnotesize{}1.102} & {\footnotesize{}1.023} & {\footnotesize{}1.102} & {\footnotesize{}1.097}\tabularnewline
\cline{2-10} 
 & {\footnotesize{}1.883} & {\footnotesize{}1.895} & {\footnotesize{}1.895} & {\footnotesize{}1.789} & {\footnotesize{}1.900} & {\footnotesize{}2.015} & {\footnotesize{}2.015} & {\footnotesize{}1.900} & {\footnotesize{}1.881}\tabularnewline
\cline{2-10} 
 & {\footnotesize{}1.884} & {\footnotesize{}1.895} & {\footnotesize{}1.895} & {\footnotesize{}1.942} & {\footnotesize{}1.900} & {\footnotesize{}2.015} & {\footnotesize{}2.015} & {\footnotesize{}1.900} & {\footnotesize{}1.881}\tabularnewline
\cline{2-10} 
 & {\footnotesize{}2.437} & {\footnotesize{}2.373} & {\footnotesize{}2.373} & {\footnotesize{}2.436} & {\footnotesize{}2.402} & {\footnotesize{}2.517} & {\footnotesize{}2.516} & {\footnotesize{}2.544} & {\footnotesize{}2.312}\tabularnewline
\cline{2-10} 
 & {\footnotesize{}2.706} & {\footnotesize{}2.761} & {\footnotesize{}2.761} & {\footnotesize{}2.591} &  & {\footnotesize{}2.722} & {\footnotesize{}2.722} & {\footnotesize{}2.676} & {\footnotesize{}2.687}\tabularnewline
\cline{2-10} 
 & {\footnotesize{}2.708} & {\footnotesize{}2.761} & {\footnotesize{}2.761} & {\footnotesize{}2.659} &  & {\footnotesize{}2.722} & {\footnotesize{}2.722} & {\footnotesize{}2.676} & {\footnotesize{}2.687}\tabularnewline
\cline{2-10} 
{\footnotesize{}MAE} &  & {\footnotesize{}0.028} & {\footnotesize{}0.029} & {\footnotesize{}0.038} & {\footnotesize{}0.027} & {\footnotesize{}0.049} & {\footnotesize{}0.044} & {\footnotesize{}0.028} & {\footnotesize{}0.021}\tabularnewline
\hline 
\end{tabular}
\end{table}
 The original semiclassical values reported in the second column are
less accurate in this case. Nevertheless, the MAE is still about 3\%
the zero point energy value. As in the Herman-Kluk calculation of
Eq.(\ref{eq:HK_powerspectrum}), it is not possible to obtain the
spectrum with the monodromy matrix regularization. Once again, the
harmonic and $\mathbf{R}_{t}^{(1)}$ approximations are quite similar.
The PPs approximation is on average overestimating the peak values.
As stressed above, the pre-exponential factor approximated results
should be compared with the SC-IVR column and the better MAE of the
last PPs approximation is probably due to compensation of errors.
Finally, the strong chaotic regime confirms the better level of accuracy
of the perturbative recursive approximations of Eqs. (\ref{eq:Rt2})
and (\ref{eq:Rt3}).

\subsection{Bidimensional quartic-like potential\label{sec:quartic_potential}}

We now consider an even more severe chaotic model, the bidimensional
potential of two Morse oscillators with a significant quartic potential
contribution of the type
\begin{equation}
V\left(\mathbf{q}\right)=\sum_{i=1}^{2}D\left[1-e^{-\alpha_{i}\left(q_{i}-q_{i}^{eq}\right)}\right]^{2}+\lambda\left[\frac{\beta}{4}\left(\left(q_{1}-q_{1}^{eq}\right)^{4}+\left(q_{2}-q_{2}^{eq}\right)^{4}\right)+\left(q_{1}-q_{1}^{eq}\right)^{2}\left(q_{2}-q_{2}^{eq}\right)^{2}\right]\label{eq:quartic_pot}
\end{equation}
where $\mathbf{q}\equiv\left(q_{1}^{eq},q_{2}^{eq}\right)$ is the
equilibrium position, $D$ and $\alpha_{i}$ are the one-dimensional
unitary mass Morse parameters, $\beta$ tunes the amount of quartic
oscillator contributions and $\lambda$ also the amount of coupling
between the oscillators. The Morse potential parameters are such that
the equilibrium position is at the origin, $D=0.2\:\mbox{a.u.},$
the frequencies $\omega_{1}=3000\:\mbox{cm}^{-1}$ and $\omega_{2}=1700\:\mbox{cm}^{-1}$.
The parameters of the quartic potential are $\beta=0.02\:\mbox{a.u.}$
and $\lambda$ is tuned according to the amount of chaos one wants
to introduce. If we would had taken a pure quartic oscillator which
has been studied in past years,\cite{Kay_101,Pollak_chaoticquartic_89}
on one side, we would have not had any Hessian term in the potential
and the previous approximation could have not been tested. On the
other side, this would not be realistic since \emph{ab initio} calculations
of equilibrium properties of real molecule is such that Hessian and
normal modes can be calculated. As in the case of the Henon-Heiles
potential, we consider two values of coupling $\lambda$, which correspond
to small and strong coupling.

\subsubsection*{Case 1: $\lambda=1\cdot10^{-6}$}

We run $10^{8}$ trajectories to overcome the high rejection rate,
which is 97\% for the $1-\mbox{det}\left|\mathbf{M}^{T}\left(t\right)\mathbf{M}\left(t\right)\right|>10^{-3}$
criterium and 96\% using Eq. (\ref{eq:Kay_criterium}). Instead, for
the approximated prefactor approximations, $10^{7}$ classical trajectories
are enough since there is no rejection in this case. Each trajectory
is $5000$ time-steps long, and each time-step is $10\:\mbox{a.u.}$
long. 
\begin{figure}
\begin{centering}
\includegraphics[scale=0.4]{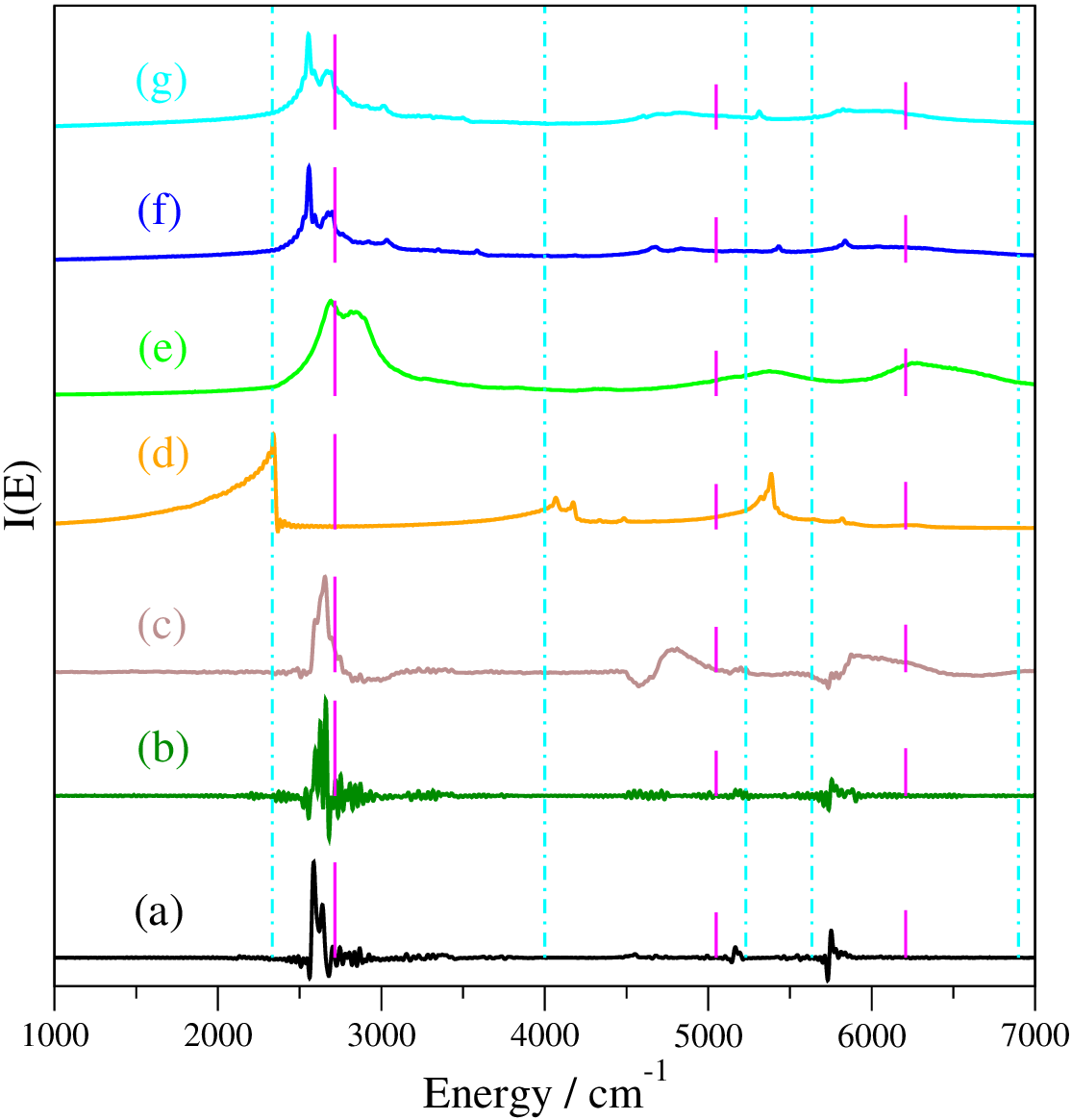}
\par\end{centering}

\caption{\label{fig:2spettri_quartic_soft_HK}Power spectrum of the potential
(\ref{eq:quartic_pot}) with $\lambda=10^{-6}$ using Eq.(\ref{eq:HK_powerspectrum})
and its approximations. (a) Black line for the rejection criterium
$1-\mbox{det}\left|\mathbf{M}^{T}\left(t\right)\mathbf{M}\left(t\right)\right|>10^{-3}$,
(b) dark green line for Kay's rejection method of Eq. (\ref{eq:Kay_criterium}),
(c) brown line for the spectrum computed using the regularization
procedure (\ref{eq:tamed_monodromy_matrix}), (d) orange line for
the HO spectrum, (e) light green line for the $\mathbf{R}_{t}^{(1)}$
approximation spectrum, (f) blue line for the spectrum computed using
Eq.(\ref{eq:Rt2}), and (g) cyan line for the spectrum computed using
Eq.(\ref{eq:Rt3}). The vertical magenta lines represent the exact
energy levels with an intensity equals to square of the overlap between
the SC reference state and the exact eigenstate calculated by DVR.
The vertical cyan dash-dotted lines represents the uncoupled Morse
potential energy levels.}
\end{figure}
The Herman-Kluk spectra of Eq.(\ref{eq:HK_powerspectrum}) reproduce
approximatively the first three energy levels as shown in Figure (\ref{fig:2spettri_quartic_soft_HK}).
From the same Figure, the two rejection criteria lead to very similar
spectra and the regularization procedure provides features quite similar
to the original Herman-Kluk spectrum, in particular for the ZPE peak.
The Johnson, the adiabatic and the PPs approximations of Secs. (\ref{sec:Johnson}),
(\ref{sec:adiabatic}), and (\ref{sec:The-poor-person's}) respectively,
lead to too noisy spectra for energy levels to be detected. The harmonic
approximation results are very similar to the uncoupled energy levels,
while approximation of Eq.(\ref{eq:Miller_approx}) and our proposed
ones of Eqs. (\ref{eq:Rt3}) and (\ref{eq:Rt2}) give quite good results. 

When calculating the spectra using the TA-SC-IVR expression of Eq.(\ref{eq:separable}),
we run 80000 trajectories when the rejection criteria are used, and
5000 trajectories when we use the approximations of the pre-exponential
factor propagators. The numerical taming of Eq.(\ref{eq:tamed_monodromy_matrix})
can not avoid the numerical issues when the cut-off is applied both
to the modulus of the biggest real eigenvalue and to the moduli of
the real eigenvalues greater than $\epsilon_{thr}$. 
\begin{figure}
\begin{centering}
\includegraphics[scale=0.4]{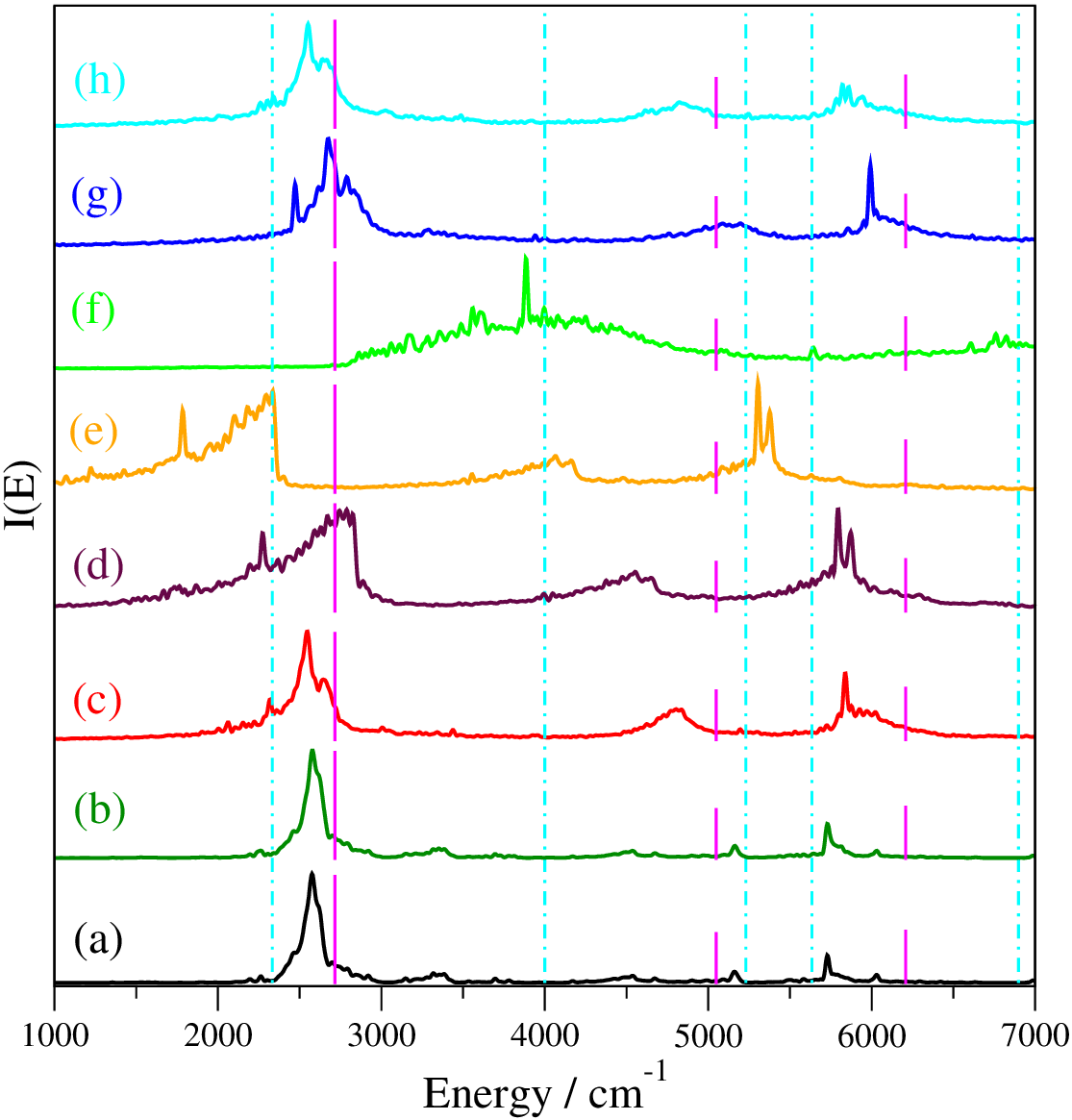}
\par\end{centering}

\caption{\label{fig:2spettri_quartic_soft_TA}Power spectrum of the potential
(\ref{eq:quartic_pot}) with $\lambda=10^{-6}$ using the time averaged
formula of Eq.(\ref{eq:separable}). (a) Black line for the rejection
criterium $1-\mbox{det}\left|\mathbf{M}^{T}\left(t\right)\mathbf{M}\left(t\right)\right|>10^{-3}$,
(b) dark green line for the Kay's rejection method of Eq. (\ref{eq:Kay_criterium}),
(c) red line for the Johnson's approximation spectrum, (d) maroon
line for the spectrum computed using the PPs approximation, (e) orange
line for the HO spectrum, (f) light green line for the $\mathbf{R}_{t}^{(1)}$
approximation spectrum, (g) blue line for the spectrum computed using
Eq.(\ref{eq:Rt2}), and (h) cyan line for the spectrum computed using
Eq.(\ref{eq:Rt3}). The vertical magenta lines represent the exact
energy levels with an intensity equals to the square of the overlap
between the SC reference state $\left|\chi\right\rangle $ and the
exact eigenstate calculated by DVR. The vertical cyan dash-dotted
lines are the uncoupled Morse potential energy levels.}
\end{figure}

Fig.(\ref{fig:2spettri_quartic_soft_TA}) reports the power spectra
at different semiclassical pre-exponential factor level of approximation
using Eq.(\ref{eq:separable}). The (a) spectrum is the original TA-SC-IVR
spectrum of Eq.(\ref{eq:separable}) using $1-\mbox{det}\left|\mathbf{M}^{T}\left(t\right)\mathbf{M}\left(t\right)\right|>10^{-3}$,
while the spectrum (b) is obtained employing the ad-hoc method of
Kay (\ref{eq:Kay_criterium}). The (c) spectrum is obtained using
the Johnson's approximation (\ref{eq:Johnson_2}), the (d) spectrum
is computed using the PPs approximation (\ref{eq:PPs}), the (e) spectrum
the harmonic approximation (\ref{eq:prefactor_Harmonic}), the (f)
spectrum using $\mathbf{R}_{t}^{(1)}$ approximation (\ref{eq:Miller_approx}),
the (g) spectrum using $\mathbf{R}_{t}^{(2)}$, and, finally, the
(h) spectrum using $\mathbf{R}_{t}^{(3)}$. The exact values are indicated
as vertical magenta lines with intensity equals to the overlap between
the SC reference state $\left|\chi\right\rangle $ and the DVR eigenvector,
while the uncoupled Morse oscillators values are the vertical dot-dashed
cyan lines. The adiabatic approximation couldn't be applied, since
Eq.s (\ref{eq:Qtilde}) and (\ref{eq:Ptilde}) are too chaotic and
cannot be integrated numerically.

The TA-SC-IVR is quite approximated in this case and it approximately
reproduces the first three peaks. It presents a ghost peak at about
$3400\:\mbox{cm}^{-1}$ and the highest peak is significantly shifted
toward the uncoupled Morse value. Johnson's approximation is mimicking
quite well the sequence of exact peaks, while the PPs is mainly reproducing
the ground energy peak. The $\mathbf{R}_{t}^{(1)}$ approximation
spectrum is too noisy to judge. The harmonic approximation is definitely
shifted toward the uncoupled Morse oscillators values, while the present
approximations of Eqs. (\ref{eq:Rt2}) and (\ref{eq:Rt3}) are well
reproducing the exact values. In particular, the higher order correction
of Eq.(\ref{eq:Rt3}) is more accurate with respect to the (a) TA-SC-IVR
spectrum. This extreme example tells us that when the system is strongly
chaotic, the semiclassical separable time-averaging SC-IVR is not
very accurate and the approximated pre-exponential factors can better
mimic the exact spectroscopic sequence.

\subsubsection*{Case 2: $\lambda=2.5\cdot10^{-6}$}

Since we want to test the pre-exponential factor approximations to
even more extreme (and probably unrealistic) cases, we consider an
even bigger coupling value between the Morse and the quartic part
of the potential. We run the same number of trajectories of the previous
case for the Herman-Kluk expression of Eq.(\ref{eq:HK_powerspectrum}).
With these values of $\lambda$, the regularization method fails because
of the highly chaotic regime of the potential. This is proved by the
high ratio of rejected trajectories, 99.1\% and 98.6\% found when
the alternative rejection criteria $1-\mbox{det}\left|\mathbf{M}^{T}\left(t\right)\mathbf{M}\left(t\right)\right|>10^{-3}$
and Eq. (\ref{eq:Kay_criterium}) are employed. Again, the two spectra
are quite similar, while the harmonic approximation is more similar
to the uncoupled eigenvalues than the coupled ones. The $\mathbf{R}_{t}^{(1)}$
approximation seems to work very well, while the approximations of
Eqs. (\ref{eq:Rt2}) and (\ref{eq:Rt3}), follow the original SC-IVR
spectrum.
\begin{figure}
\begin{centering}
\includegraphics[scale=0.45]{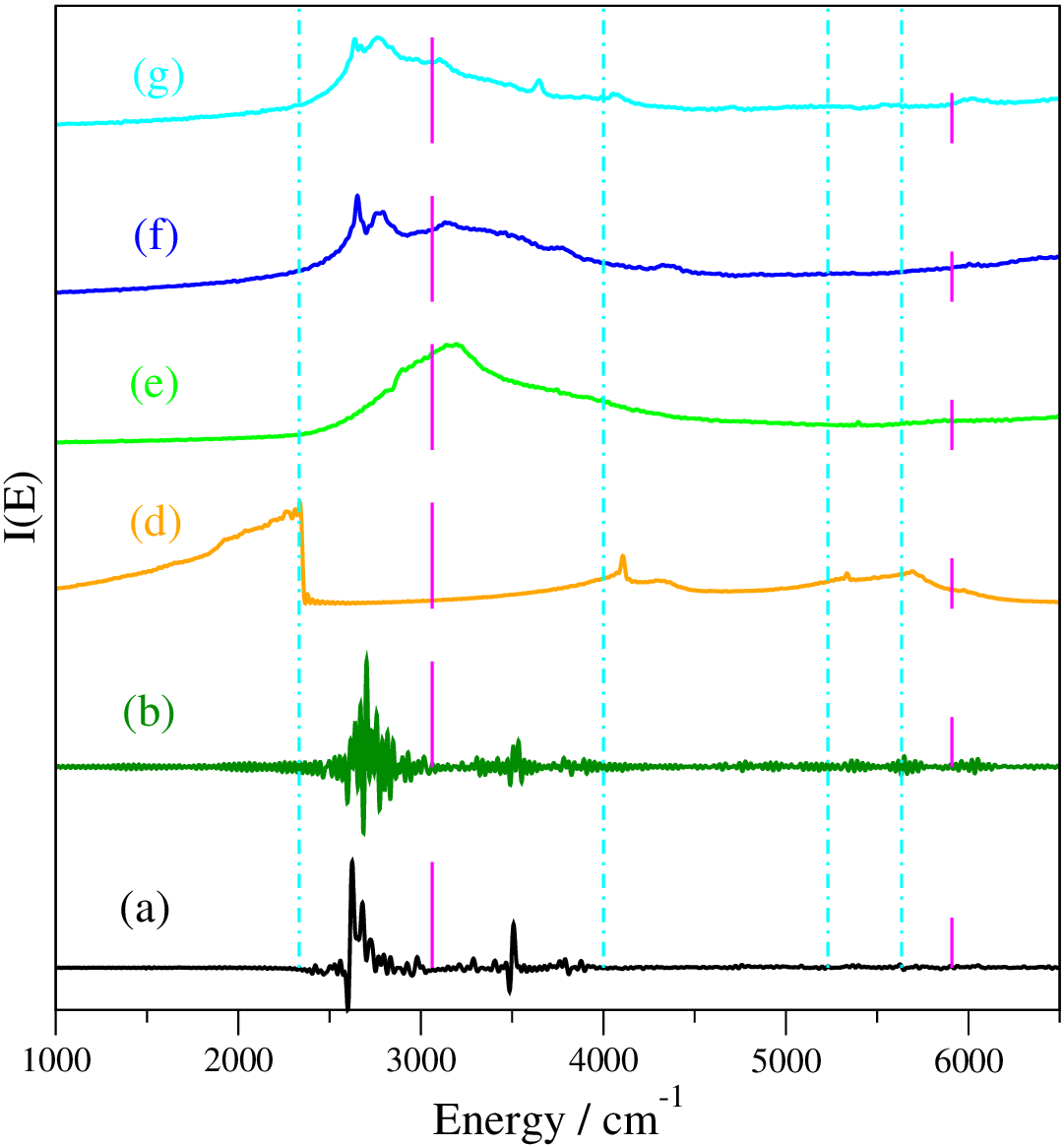}
\par\end{centering}

\caption{\label{fig:2spettri_quartic-strong_HK} The same as in Fig.(\ref{fig:2spettri_quartic_soft_HK})
but with $\lambda$ equals to $2.5\cdot10^{-6}$.}
\end{figure}
 When TA-SC-IVR calculations are employed, we run $250000$ trajectories
for $5000$ time-steps of $10\:\mbox{a.u.}$ each, of which $98.3\%$
are rejected using $1-\mbox{det}\left|\mathbf{M}^{T}\left(t\right)\mathbf{M}\left(t\right)\right|>10^{-3}$and
97.5\% using the method of Kay of Eq. (\ref{eq:Kay_criterium}). The
approximated pre-exponential factor calculations are performed as
above, i.e. with $5000$ trajectories. The monodromy matrix regularization
fails as in the previous case. Instead, the Johnson approximation
lead to a resolute spectrum. 
\begin{figure}
\begin{centering}
\includegraphics[scale=0.45]{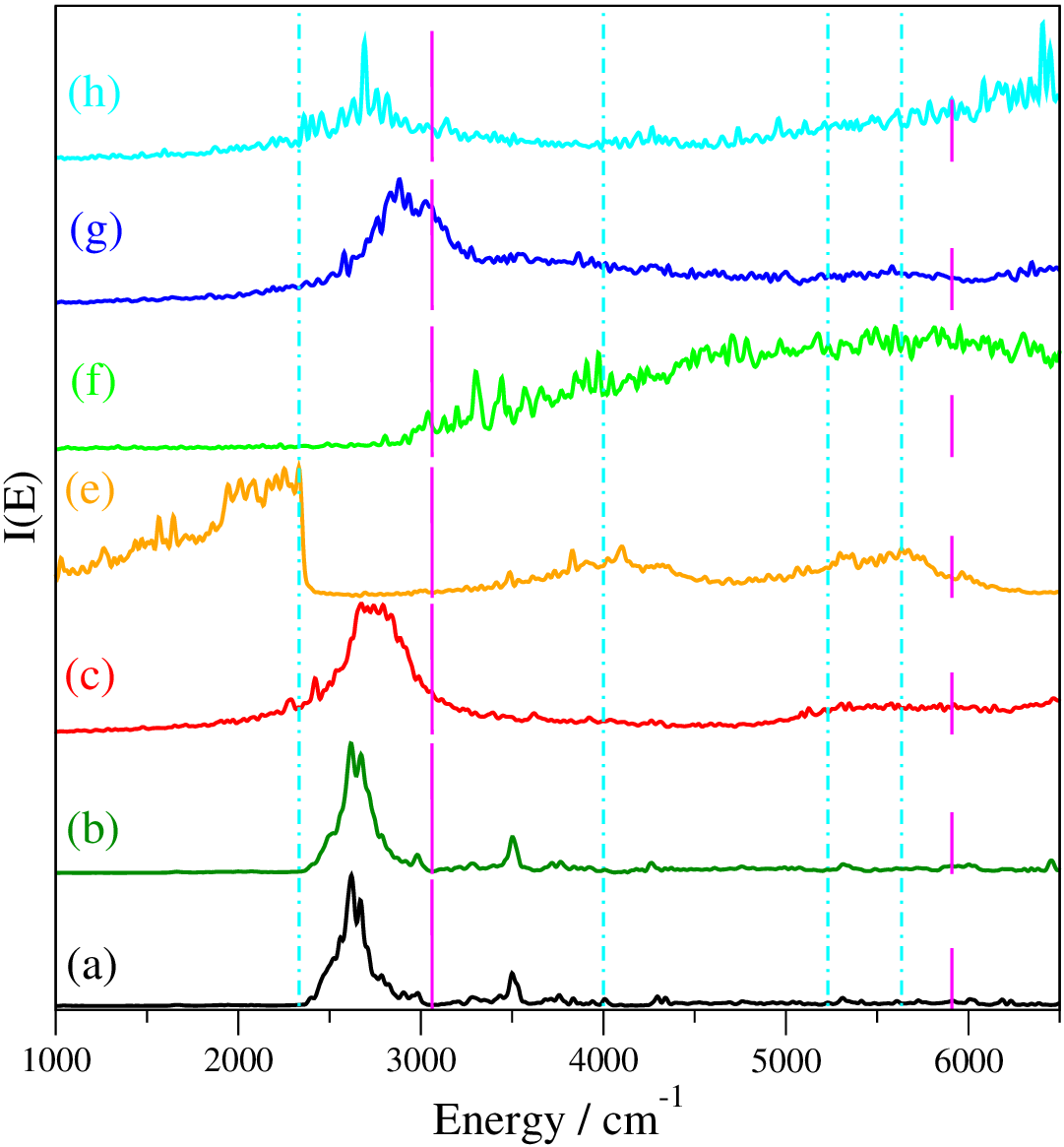}
\par\end{centering}

\caption{\label{fig:2spettri_quartic-strong_TA} The same as in Fig.(\ref{fig:2spettri_quartic_soft_TA})
but with $\lambda$ equals to $2.5\cdot10^{-6}$.}
\end{figure}
 The harmonic approximation is reproducing peaks in harmonic sequence
and the $\mathbf{R}_{t}^{(1)}$ approximation is too noisy. The only
reasonable results are those by Johnson and the new approximations
of Eqs. (\ref{eq:Rt2}) and (\ref{eq:Rt3}). In more details, the
TA-SC-IVR zero point energy (ZPE) is $2620\:\mbox{cm}^{-1}$, $2746\:\mbox{cm}^{-1}$
for the Johnson approximation, $2885\:\mbox{cm}^{-1}$ for Eq.(\ref{eq:Rt2})
and $2688\:\mbox{cm}^{-1}$ for the higher order approximation of
Eq.(\ref{eq:Rt3}). Once again, Eq.(\ref{eq:Rt3}) is more similar
to the original TA-SC-IVR values. However, at any semiclassical level
of calculation, the first fundamental is reproduced.

Overall, the present approximation of Eq.(\ref{eq:Rt3}) is the most
accurate in these model potential energy surface scenarios. We now
turn into real molecules potential energy surfaces.

\subsection{$\mathbf{H_{2}O}$ molecule\label{sec:H2O}}

The water molecule presents strong intermode couplings. In the calculations
presented here, we employ the PES provided by Bowman\cite{Bowman_water}
and Eq.(\ref{eq:separable}). Each trajectory is $1000$ time-step
long with the single time step $10\:\mbox{a.u.}$ long for a total
of $8000$ trajectories both with exact and approximated pre-exponential
factor formulations. Previous calculations\cite{Alex_Mik,Ceotto_MCSCIVR}
showed that phase space Monte Carlo convergence is reached already
with $4000$ trajectories. To better identify each peak, we employ
combinations of antisymmetric coherent states and break down each
spectrum in partial spectra for each irreducible representation of
the $\mbox{C}_{2v}$ point group symmetry, as explained in previous
publications.\cite{Alex_Mik,Ceotto_cursofdimensionality_11} 
\begin{figure}
\begin{centering}
\includegraphics[scale=0.5]{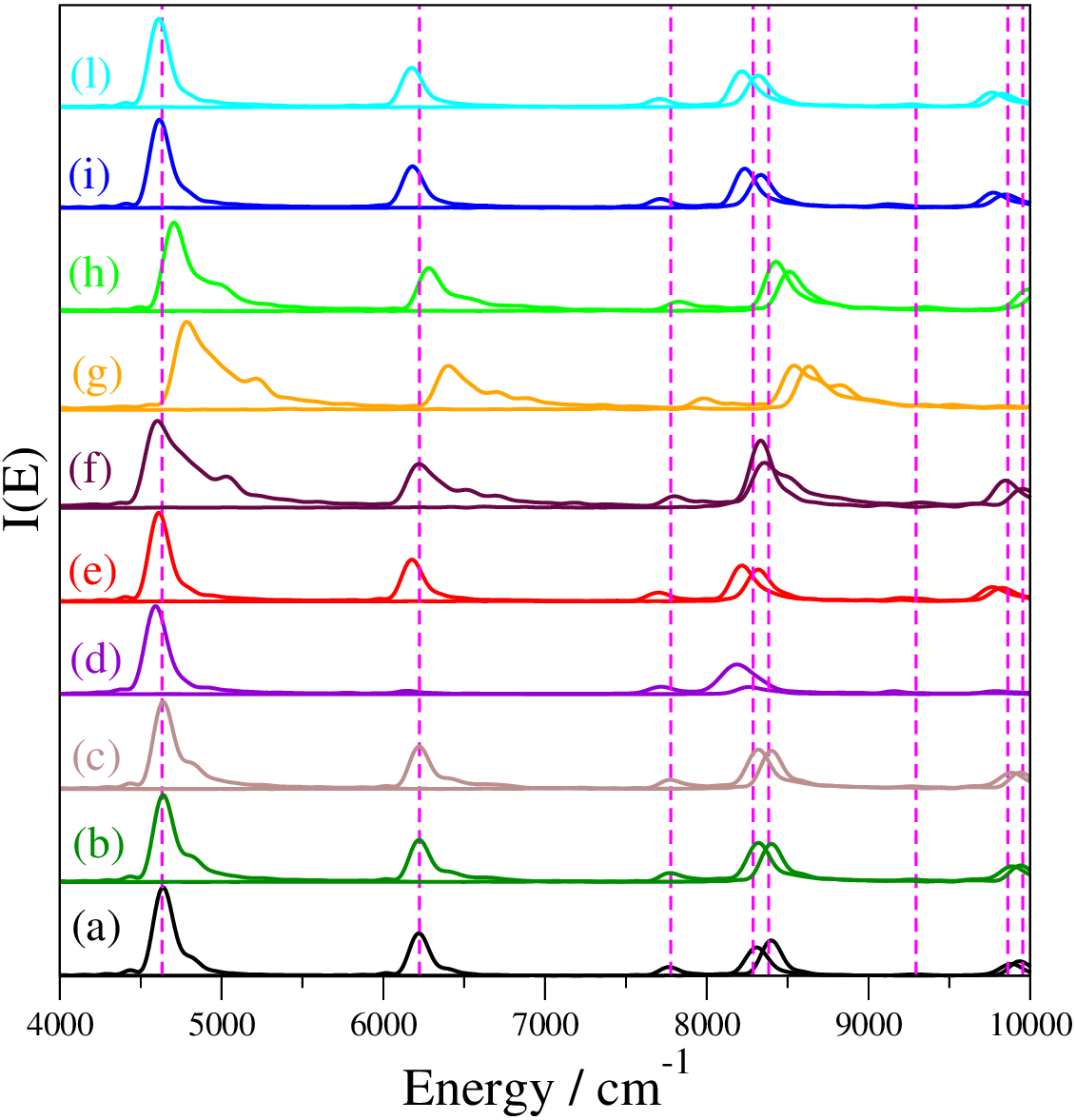}
\par\end{centering}

\centering{}\caption{\label{fig:H2O_spectra}$\mbox{H\ensuremath{_{2}}O}$ spectra. (a)
Black line for the separable SC-IVR (\ref{eq:separable}) spectrum
using the rejection criterium $1-\mbox{det}\left|\mathbf{M}^{T}\left(t\right)\mathbf{M}\left(t\right)\right|>10^{-5}$,
(b) using the ad hoc Kay's rejection method of Eq.(\ref{eq:Kay_criterium}),
(c) brown for the regularization of the monodromy matrix of Eq.(\ref{eq:tamed_monodromy_matrix}),
(d) violet line for the adiabatic approximation (\ref{eq:C_t_adiabatic})
spectrum, (e) red line for the Johnson's approximation (\ref{eq:Johnson_2})
spectrum, (f) maroon line for the PPs approximation (\ref{eq:PPs})
spectrum, (g) orange line for the HO (\ref{eq:prefactor_Harmonic})
approximation spectrum, (h) green line for the $\mathbf{R}_{t}^{(1)}$
approximation (\ref{eq:Miller_approx}) spectrum, (i) blue line for
the spectrum computed using $\mathbf{R}_{t}^{(2)}$ in Eq.(\ref{eq:Rt2}),
and (l) cyan line for the spectrum computed using $\mathbf{R}_{t}^{(3)}$
in Eq.(\ref{eq:Rt3}). The vertical magenta dashed lines represent
the quantum energy levels. $A_{1}$ and $B_{2}$ spectra with the
same color for each approximation. }
\end{figure}
 The spectra with different pre-exponential factor approximations
are reported in Fig.(\ref{fig:H2O_spectra}). For each approximation,
the $A_{1}$ and $B_{2}$ irreducible representation spectra are reported
in Fig.(\ref{fig:H2O_spectra}) with the same color. This figure points
out the major limitations of the harmonic approximation, in particular
for the highest vibrational states. More specifically, the vibrational
level of each state is reported in Table (\ref{tab:H2O_en_lev-1}).
\begin{table}
\centering{}\caption{\selectlanguage{british}%
\label{tab:H2O_en_lev-1}\foreignlanguage{english}{Vibrational energy
levels of $\mbox{H}_{2}\mbox{O}$. Wavenumbers unit. First column
reports the spectroscopic terms, second column reports the exact quantum
mechanical values, third column reports the results computed with
SC-IVR of Eq.(\ref{eq:separable}) using the rejection criterium $1-\mbox{det}\left|\mathbf{M}^{T}\left(t\right)\mathbf{M}\left(t\right)\right|>10^{-5}$,
fourth column SC-IVR calculation using the ad hoc Kay's rejection
method of Eq.(\ref{eq:Kay_criterium}), and the others with the different
pre-exponential factor approximations named as above. In the last
row is reported the Mean Average Error (MAE) of each column.}\selectlanguage{english}%
}
\begin{tabular}{cccccccccccc}
{\scriptsize{}State} & {\scriptsize{}Exact\cite{Bowman_water}} & {\scriptsize{}SC-IVR} & {\scriptsize{}Kay's method} & {\scriptsize{}Regularization} & {\scriptsize{}Adiabatic} & {\scriptsize{}Johnson} & {\scriptsize{}PPs} & {\scriptsize{}HO} & {\scriptsize{}$\mathbf{R}_{t}^{(1)}$} & {\scriptsize{}$\mathbf{R}_{t}^{(2)}$} & {\scriptsize{}$\mathbf{R}_{t}^{(3)}$}\tabularnewline
\hline 
{\scriptsize{}ZPE} & {\scriptsize{}4631.6} & {\scriptsize{}4636} & {\scriptsize{}4640} & {\scriptsize{}4639} & {\scriptsize{}4592} & {\scriptsize{}4612} & {\scriptsize{}4604} & {\scriptsize{}4784} & {\scriptsize{}4704} & {\scriptsize{}4616} & {\scriptsize{}4612}\tabularnewline
\hline 
{\scriptsize{}$A_{1}\left(1_{1}\right)$} & {\scriptsize{}6222.8} & {\scriptsize{}6220} & {\scriptsize{}6220} & {\scriptsize{}6222} & {\scriptsize{}6148} & {\scriptsize{}6176} & {\scriptsize{}6220} & {\scriptsize{}6404} & {\scriptsize{}6280} & {\scriptsize{}6180} & {\scriptsize{}6176}\tabularnewline
\hline 
{\scriptsize{}$A_{1}\left(1_{2}\right)$} & {\scriptsize{}7777.7} & {\scriptsize{}7768} & {\scriptsize{}7772} & {\scriptsize{}7772} & {\scriptsize{}7716} & {\scriptsize{}7704} & {\scriptsize{}7800} & {\scriptsize{}7980} & {\scriptsize{}7828} & {\scriptsize{}7714} & {\scriptsize{}7708}\tabularnewline
\hline 
{\scriptsize{}$A_{1}\left(2_{1}\right)$} & {\scriptsize{}8287} & {\scriptsize{}8308} & {\scriptsize{}8320} & {\scriptsize{}8320} & {\scriptsize{}8188} & {\scriptsize{}8216} & {\scriptsize{}8356} & {\scriptsize{}8540} & {\scriptsize{}8428} & {\scriptsize{}8236} & {\scriptsize{}8218}\tabularnewline
\hline 
{\scriptsize{}$B_{2}\left(3_{1}\right)$} & {\scriptsize{}8382.7} & {\scriptsize{}8400} & {\scriptsize{}8400} & {\scriptsize{}8400} & {\scriptsize{}8400} & {\scriptsize{}8320} & {\scriptsize{}8334} & {\scriptsize{}8632} & {\scriptsize{}8512} & {\scriptsize{}8322} & {\scriptsize{}8319}\tabularnewline
\hline 
{\scriptsize{}$A_{1}\left(1_{3}\right)$} & {\scriptsize{}9294.1} & {\scriptsize{}9286} & {\scriptsize{}9268} & {\scriptsize{}9266} & {\scriptsize{}9156} & {\scriptsize{}9208} & {\scriptsize{}9327} & {\scriptsize{}9510} & {\scriptsize{}9352} & {\scriptsize{}9123} & {\scriptsize{}9264}\tabularnewline
\hline 
{\scriptsize{}$A_{1}\left(1_{1}2_{1}\right)$} & {\scriptsize{}9862.1} & {\scriptsize{}9884} & {\scriptsize{}9888} & {\scriptsize{}9884} & {\scriptsize{}9808} & {\scriptsize{}9764} & {\scriptsize{}9952} & {\scriptsize{}10136} & {\scriptsize{}9988} & {\scriptsize{}9773} & {\scriptsize{}9764}\tabularnewline
\hline 
{\scriptsize{}$B_{2}\left(1_{1}3_{1}\right)$} & {\scriptsize{}9954} & {\scriptsize{}9936} & {\scriptsize{}9940} & {\scriptsize{}9940} & {\scriptsize{}9936} & {\scriptsize{}9828} & {\scriptsize{}9846} & {\scriptsize{}10208} & {\scriptsize{}10056} & {\scriptsize{}9848} & {\scriptsize{}9827}\tabularnewline
\hline 
{\scriptsize{}$A_{1}\left(1_{2}2_{1}\right)$} & {\scriptsize{}11400.5} & {\scriptsize{}11400} & {\scriptsize{}11408} & {\scriptsize{}11409} & {\scriptsize{}11280} & {\scriptsize{}11278} & {\scriptsize{}11609} & {\scriptsize{}11792} & {\scriptsize{}11508} & {\scriptsize{}11294} & {\scriptsize{}11267}\tabularnewline
\hline 
{\scriptsize{}$B_{2}\left(1_{2}3_{1}\right)$} & {\scriptsize{}11490.4} & {\scriptsize{}11440} & {\scriptsize{}11440} & {\scriptsize{}11447} & {\scriptsize{}11440} & {\scriptsize{}11304} & {\scriptsize{}11342} & {\scriptsize{}11780} & {\scriptsize{}11548} & {\scriptsize{}11337} & {\scriptsize{}11305}\tabularnewline
\hline 
{\scriptsize{}$A_{1}\left(2_{2}\right)$} & {\scriptsize{}11833.9} & {\scriptsize{}11876} & {\scriptsize{}11868} & {\scriptsize{}11868} & {\scriptsize{}11660} & {\scriptsize{}11700} & {\scriptsize{}11996} & {\scriptsize{}12176} & {\scriptsize{}12004} & {\scriptsize{}11729} & {\scriptsize{}11704}\tabularnewline
\hline 
{\scriptsize{}$B_{2}\left(2_{1}3_{1}\right)$} & {\scriptsize{}11886} & {\scriptsize{}11918} & {\scriptsize{}11906} & {\scriptsize{}11906} & {\scriptsize{}11920} & {\scriptsize{}11756} & {\scriptsize{}11780} & {\scriptsize{}12272} & {\scriptsize{}12076} & {\scriptsize{}11781} & {\scriptsize{}11760}\tabularnewline
\hline 
{\scriptsize{}$A_{1}\left(3_{2}\right)$} & {\scriptsize{}12069.8} & {\scriptsize{}12060} & {\scriptsize{}12044} & {\scriptsize{}12044} & {\scriptsize{}12164} & {\scriptsize{}11912} & {\scriptsize{}12224} & {\scriptsize{}12408} & {\scriptsize{}12220} & {\scriptsize{}11933} & {\scriptsize{}11900}\tabularnewline
\hline 
{\scriptsize{}$A_{1}\left(1_{1}2_{2}\right)$} & {\scriptsize{}13399.1} & {\scriptsize{}13404} & {\scriptsize{}13412} & {\scriptsize{}13412} & {\scriptsize{}13294} & {\scriptsize{}13212} & {\scriptsize{}13207} & {\scriptsize{}13760} & {\scriptsize{}13536} & {\scriptsize{}13224} & {\scriptsize{}13208}\tabularnewline
\hline 
{\scriptsize{}$B_{2}\left(1_{1}2_{1}3_{1}\right)$} & {\scriptsize{}13443.7} & {\scriptsize{}13452} & {\scriptsize{}13440} & {\scriptsize{}13442} & {\scriptsize{}13452} & {\scriptsize{}13244} & {\scriptsize{}13276} & {\scriptsize{}13824} & {\scriptsize{}13576} & {\scriptsize{}13278} & {\scriptsize{}13254}\tabularnewline
\hline 
{\scriptsize{}$A_{1}\left(1_{1}3_{2}\right)$} & {\scriptsize{}13622} & {\scriptsize{}13560} & {\scriptsize{}13560} & {\scriptsize{}13555} &  & {\scriptsize{}13596} & {\scriptsize{}13582} &  & {\scriptsize{}13712} &  & {\scriptsize{}13674}\tabularnewline
\hline 
{\scriptsize{}MAE} &  & {\scriptsize{}19.6} & {\scriptsize{}21.8} & {\scriptsize{}20.1} & {\scriptsize{}72.6} & {\scriptsize{}108.0} & {\scriptsize{}98.8} & {\scriptsize{}285.8} & {\scriptsize{}110.7} & {\scriptsize{}105.0} & {\scriptsize{}106.3}\tabularnewline
\hline 
\end{tabular}
\end{table}
 For each vibrational state labeled in the first column, one can read
the exact quantum mechanical results in the second column, the separable
SC-IVR ones on the third and fourth and the approximated ones in the
following columns, as labeled in the tables above. From the MAE, it
is clear that the numerical regularization approach of Eq.(\ref{eq:tamed_monodromy_matrix})
is very good with respect to the exact values, showing that the spectroscopic
contribution of the chaotic trajectories is negligible. In fact, the
monodromy matrix is regularized just for $2.1\%$ of the total trajectories,
and Eq.(\ref{eq:tamed_monodromy_matrix}) is applied no more than
$5$ times per trajectory. Instead, $56\%$ of trajectories are rejected
in the standard SC-IVR calculations because of the $\mbox{det\ensuremath{\left[M^{T}M\right]}}$
deviation from unity. This percent difference proves that most of
those chaotic trajectories, that are rejected by the strict criterion
$1-\mbox{det}\left|\mathbf{M}^{T}\left(t\right)\mathbf{M}\left(t\right)\right|>10^{-5}$,
actually do not compromise the accuracy of the calculation. Moreover,
the spectrum obtained using the rejection criterium proposed by Kay
is very similar with the TA-SC-IVR one. From the following columns,
it is evident that the harmonic approximation is the worse one and
that $\mathbf{R}_{t}^{(1)}$, Johnson's, the PPs and the new approximations
$\mathbf{R}_{t}^{(2)}$ and $\mathbf{R}_{t}^{(3)}$ show about the
same accuracy. Once again, the adiabatic approximation is relatively
accurate when Eqs. (\ref{eq:Qdot_Pdot}) can be calculated.

\subsection{$\mathbf{CO_{2}}$ molecule\label{sec:CO2}}

To test the accuracy of the approximations in the case of strong Fermi
resonances, we choose as a test case the carbon dioxide molecule.\cite{Ceotto_1traj,CO2_expt}
We employ Chedin's potential\cite{Chedin pot} and compare with the
exact quantum mechanical results by Vasquez \emph{et al}..\cite{CO2_expt}
Each trajectory is $3000$ time-steps long with a time-step $10\:\mbox{a.u.}$
long. We employ $15000$ trajectories for the phase space integration
both with and without the pre-exponential factor approximations, which
is by far enough for Monte Carlo convergence.
\begin{figure}
\begin{centering}
\includegraphics[scale=0.5]{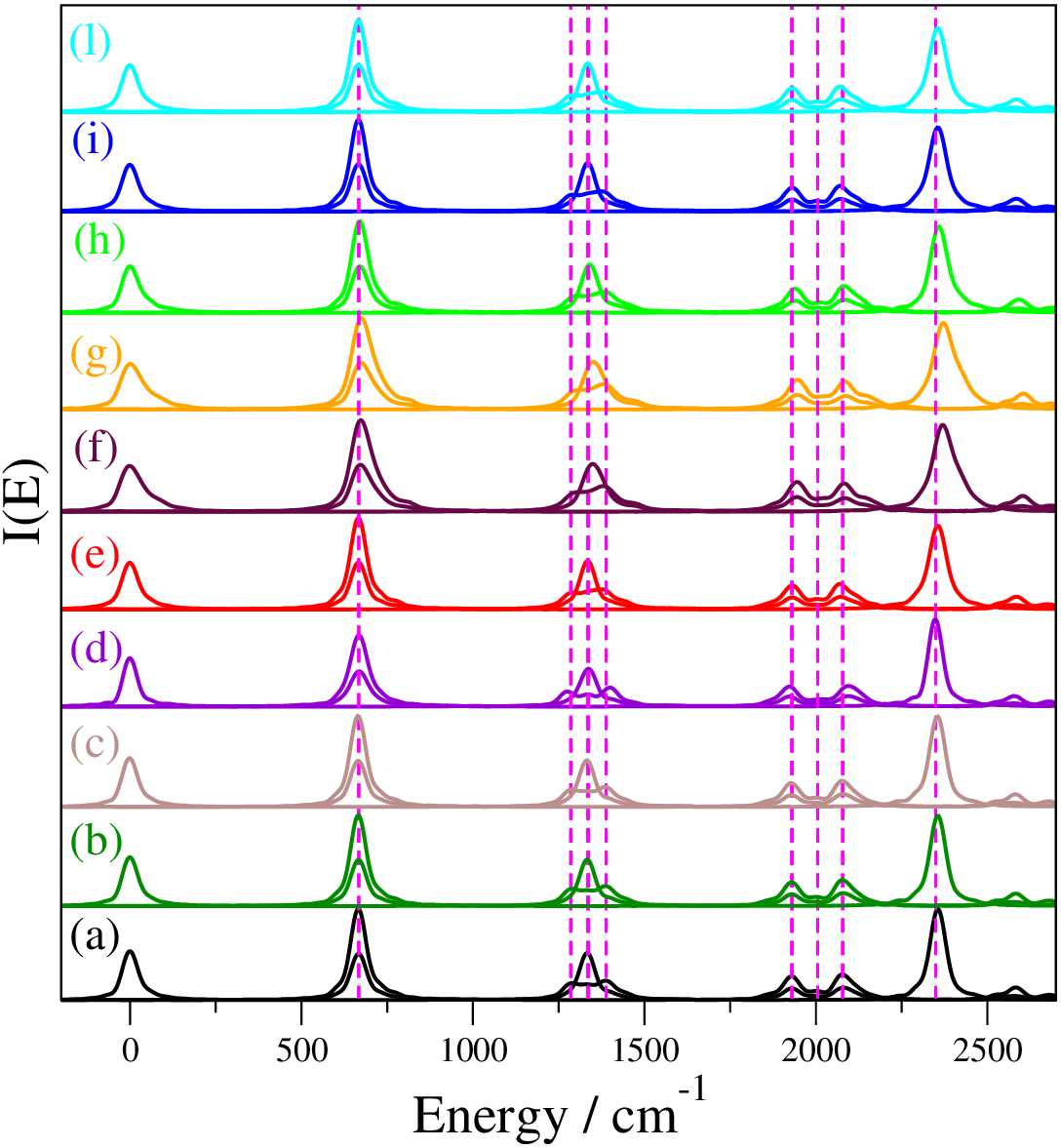}
\par\end{centering}

\caption{\label{fig:3spettri_CO2} The same as in Fig.(\ref{fig:H2O_spectra})
but for the $\mbox{CO\ensuremath{_{2}}}$ molecule. Each approximation
includes the spectra of the $\mbox{A}_{g}$, $\mbox{B}_{1u}$, $\mbox{B}_{2u}$
and $\mbox{B}_{3u}$ irreducible representations of the $\mbox{D}_{2h}$
point group symmetry.\cite{Ceotto_1traj}}
\end{figure}
 Fig.(\ref{fig:3spettri_CO2}) shows a good agreement between all
approximations. Carbon dioxide has higher molecular weight than water
and its dynamics is probably more classical.
\begin{table}
\centering{}\caption{\selectlanguage{british}%
\label{tab:CO2_en_lev}\foreignlanguage{english}{The same as in Table
(\ref{tab:H2O_en_lev-1}) but for $\mbox{CO}_{2}$.}\selectlanguage{english}%
}
{\scriptsize{}}%
\begin{tabular}{cccccccccccc}
{\scriptsize{}State} & {\scriptsize{}Exact\cite{CO2_expt}} & {\scriptsize{}SC-IVR} & {\scriptsize{}Kay's method} & {\scriptsize{}Regularization} & {\scriptsize{}Adiabatic} & {\scriptsize{}Johnson} & {\scriptsize{}PPs} & {\scriptsize{}HO} & {\scriptsize{}$\mathbf{R}_{t}^{(1)}$} & {\scriptsize{}$\mathbf{R}_{t}^{(2)}$} & {\scriptsize{}$\mathbf{R}_{t}^{(3)}$}\tabularnewline
\hline 
{\scriptsize{}$ZPE$} & {\scriptsize{}2536.15} & {\scriptsize{}2535} & {\scriptsize{}2535} & {\scriptsize{}2536} & {\scriptsize{}2531} & {\scriptsize{}2534} & {\scriptsize{}2539} & {\scriptsize{}2564} & {\scriptsize{}2541} & {\scriptsize{}2534} & {\scriptsize{}2534}\tabularnewline
\hline 
{\scriptsize{}$\left(000\right)$} & {\scriptsize{}667.47} & {\scriptsize{}667} & {\scriptsize{}667} & {\scriptsize{}665} & {\scriptsize{}669} & {\scriptsize{}666} & {\scriptsize{}673} & {\scriptsize{}672} & {\scriptsize{}670} & {\scriptsize{}666} & {\scriptsize{}666}\tabularnewline
\hline 
{\scriptsize{}$\left(01^{1}0\right)$} & {\scriptsize{}667.47} & {\scriptsize{}667} & {\scriptsize{}667} & {\scriptsize{}666} & {\scriptsize{}669} & {\scriptsize{}666} & {\scriptsize{}673} & {\scriptsize{}672} & {\scriptsize{}670} & {\scriptsize{}666} & {\scriptsize{}666}\tabularnewline
\hline 
{\scriptsize{}$\left(01^{1}0\right)$} & {\scriptsize{}1285.1} & {\scriptsize{}1290} & {\scriptsize{}1288} & {\scriptsize{}1288} & {\scriptsize{}1275} & {\scriptsize{}1290} & {\scriptsize{}1299} & {\scriptsize{}1297} & {\scriptsize{}1294} & {\scriptsize{}1286} & {\scriptsize{}1291}\tabularnewline
\hline 
{\scriptsize{}$\left(10^{0}0\right)$} & {\scriptsize{}1335.95} & {\scriptsize{}1333} & {\scriptsize{}1332} & {\scriptsize{}1332} & {\scriptsize{}1335} & {\scriptsize{}1334} & {\scriptsize{}1350} & {\scriptsize{}1351} & {\scriptsize{}1341} & {\scriptsize{}1334} & {\scriptsize{}1334}\tabularnewline
\hline 
{\scriptsize{}$\left(02^{2}0\right)$} & {\scriptsize{}1335.95} & {\scriptsize{}1333} & {\scriptsize{}1332} & {\scriptsize{}1334} & {\scriptsize{}1335} & {\scriptsize{}1334} & {\scriptsize{}1350} & {\scriptsize{}1351} & {\scriptsize{}1341} & {\scriptsize{}1334} & {\scriptsize{}1334}\tabularnewline
\hline 
{\scriptsize{}$\left(02^{2}0\right)$} & {\scriptsize{}1387.93} & {\scriptsize{}1388} & {\scriptsize{}1384} & {\scriptsize{}1386} & {\scriptsize{}1400} & {\scriptsize{}1383} & {\scriptsize{}1382} & {\scriptsize{}1393} & {\scriptsize{}1391} & {\scriptsize{}1382} & {\scriptsize{}1374}\tabularnewline
\hline 
{\scriptsize{}$\left(02^{2}0\right)$} & {\scriptsize{}1929.56} & {\scriptsize{}1930} & {\scriptsize{}1928} & {\scriptsize{}1928} & {\scriptsize{}1923} & {\scriptsize{}1933} & {\scriptsize{}1947} & {\scriptsize{}1940} & {\scriptsize{}1940} & {\scriptsize{}1931} & {\scriptsize{}1931}\tabularnewline
\hline 
{\scriptsize{}$\left(11^{1}0\right)$} & {\scriptsize{}1929.56} & {\scriptsize{}1930} & {\scriptsize{}1928} & {\scriptsize{}1929} & {\scriptsize{}1923} & {\scriptsize{}1933} & {\scriptsize{}1947} & {\scriptsize{}1940} & {\scriptsize{}1940} & {\scriptsize{}1931} & {\scriptsize{}1931}\tabularnewline
\hline 
{\scriptsize{}$\left(11^{1}0\right)$} & {\scriptsize{}2005.25} & {\scriptsize{}1997} & {\scriptsize{}2001} & {\scriptsize{}2001} & {\scriptsize{}2015} & {\scriptsize{}2003} & {\scriptsize{}2021} & {\scriptsize{}2021} & {\scriptsize{}2012} & {\scriptsize{}2003} & {\scriptsize{}2003}\tabularnewline
\hline 
{\scriptsize{}$\left(03^{3}0\right)$} & {\scriptsize{}2005.25} & {\scriptsize{}1997} & {\scriptsize{}2001} & {\scriptsize{}2001} & {\scriptsize{}2015} & {\scriptsize{}2003} & {\scriptsize{}2021} & {\scriptsize{}2021} & {\scriptsize{}2012} & {\scriptsize{}2003} & {\scriptsize{}2003}\tabularnewline
\hline 
{\scriptsize{}$\left(03^{3}0\right)$} & {\scriptsize{}2078.15} & {\scriptsize{}2081} & {\scriptsize{}2080} & {\scriptsize{}2077} & {\scriptsize{}2093} & {\scriptsize{}2070} & {\scriptsize{}2083} & {\scriptsize{}2086} & {\scriptsize{}2084} & {\scriptsize{}2071} & {\scriptsize{}2071}\tabularnewline
\hline 
{\scriptsize{}$\left(03^{1}0\right)$} & {\scriptsize{}2078.15} & {\scriptsize{}2081} & {\scriptsize{}2080} & {\scriptsize{}2079} & {\scriptsize{}2093} & {\scriptsize{}2070} & {\scriptsize{}2083} & {\scriptsize{}2084} & {\scriptsize{}2084} & {\scriptsize{}2071} & {\scriptsize{}2071}\tabularnewline
\hline 
{\scriptsize{}$\left(03^{1}1\right)$} & {\scriptsize{}2349.38} & {\scriptsize{}2356} & {\scriptsize{}2355} & {\scriptsize{}2354} & {\scriptsize{}2347} & {\scriptsize{}2356} & {\scriptsize{}2371} & {\scriptsize{}2373} & {\scriptsize{}2359} & {\scriptsize{}2356} & {\scriptsize{}2354}\tabularnewline
\hline 
{\scriptsize{}MAE} &  & {\scriptsize{}3.0} & {\scriptsize{}2.7} & {\scriptsize{}2.1} & {\scriptsize{}6.9} & {\scriptsize{}3.8} & {\scriptsize{}11.4 } & {\scriptsize{}12.4} & {\scriptsize{}6.3} & {\scriptsize{}3.2} & {\scriptsize{}3.9 }\tabularnewline
\hline 
\end{tabular}
\end{table}
 Table(\ref{tab:CO2_en_lev}) reports the values of each vibrational
level for each approximation. In this case, all approximations are
quite accurate, as noted above. The harmonic oscillator approximation
is again the less accurate one, followed by the PPs and $\mathbf{R}_{t}^{(1)}$
ones. Surprisingly, also the adiabatic is not very accurate. The present
approximations ($\mathbf{R}_{t}^{(2)}$ and $\mathbf{R}_{t}^{(3)}$)
and Johnson's one are the most accurate and with almost no difference
with respect to the original SC-IVR integration. The disappointing
performance of the adiabatic approximation is probably due to the
coupling of the $\mbox{CO}_{2}$ modes, which is intermediate between
the fully adiabatic and diabatic regime. The numerical taming approach
of Eq.(\ref{eq:tamed_monodromy_matrix}) is as accurate as the reference
SC-IVR calculation. Their similarity is explained by the small ($0.6\%$)
percentage of trajectory correction using Eq.(\ref{eq:tamed_monodromy_matrix})
with respect to the $14\%$ rejected by looking at the determinant
of the monodromy matrix and 8\% evaluating $\left|C_{t}\left(\mathbf{p}_{0},\mathbf{q}_{0}\right)\right|^{2}$.
The numerical taming is employed no more than $4$ times per trajectory.

\subsection{$\mathbf{CH_{2}O}$ molecule\label{sec:CH2O}}

\begin{figure}
\centering{}\includegraphics[scale=0.5]{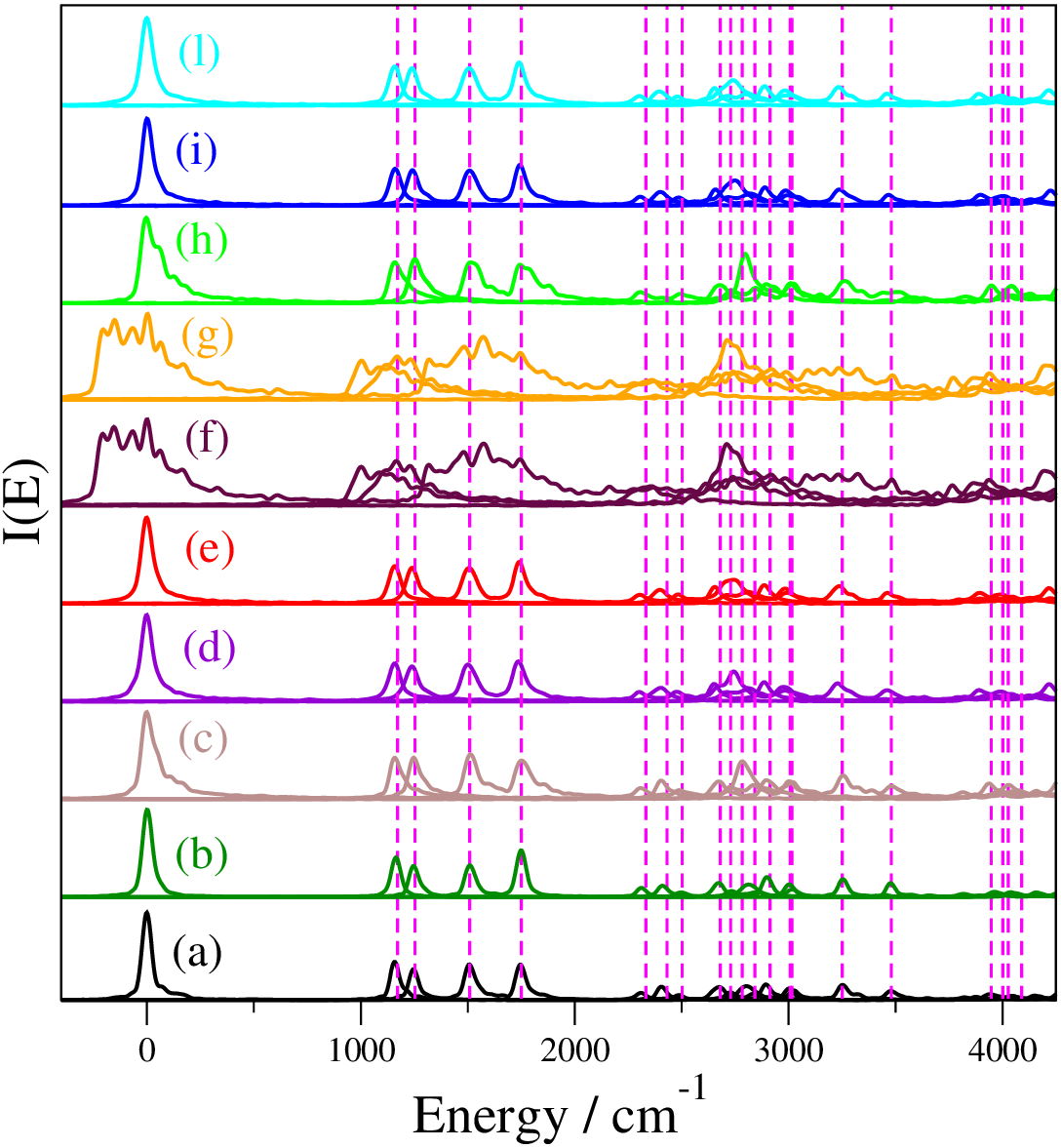}\caption{\label{fig:3spettri_CH2O}The same as in Fig.(\ref{fig:H2O_spectra})
but for the $\mbox{CH\ensuremath{_{2}\mbox{O}}}$ molecule. Each approximation
includes the spectra of the $\mbox{A}_{1}$, $\mbox{A}_{2}$, $\mbox{B}_{1}$
and $\mbox{B}_{2}$ irreducible representations of the $\mbox{C}_{2v}$
point group symmetry.}
\end{figure}

Passing from 3 to 4 atom molecules, we choose to test the pre-exponential
factor approximations with the formaldehyde vibrational spectrum,
since this is a well tested case. Also, $\mbox{CH}_{2}\mbox{O}$ presents
light atoms, as well as strongly coupled dynamics. We employ the PES
designed by Martin \textit{et al}.\cite{Martin_CH2Opot} and we compare
our semiclassical results with the exact quantum mechanical calculations
by Carter \textit{et al}..\cite{Carter_H2CO_exact} We employ $24000$
trajectories for the SC-IVR calculations without pre-exponential factor
approximation (except for the basic one implied by the separable approximation)
and we reject $82.5\%$ with the monodromy matrix determinant criterion
and 85.6\% by using Eq. (\ref{eq:Kay_criterium}). Instead, $8000$
trajectories are used for the approximated and numerically tamed pre-exponential
factor. All trajectories are evolved for $3000$ time-steps with a
time-step $10\:\mbox{a.u.}$ long for all simulations. The point group
symmetry is $\mbox{C}_{2v}$ , and spectra for all four irreducible
representations are reported at each approximation level of accuracy
in Fig.(\ref{fig:3spettri_CH2O}). 

The $\mbox{CH}_{2}\mbox{O}$ spectrum can be divided into a low energy
region, populated by the fundamentals of four vibrational modes, and
an higher energy region, where one can find the fundamentals of the
remaining modes and several overtones. Since the accuracy of each
approximation looks similar in Fig.(\ref{fig:3spettri_CH2O}), we
report in Tables (\ref{tab:CH2O_en_lev}) and (\ref{tab:CH2O_en_lev2})
each vibrational state value. 
\begin{table}
\centering{}\caption{\selectlanguage{british}%
\label{tab:CH2O_en_lev}\foreignlanguage{english}{The same as in Table
(\ref{tab:H2O_en_lev-1}) but for the fundamentals of $\mbox{CH}_{2}\mbox{O}$.}\selectlanguage{english}%
}
\begin{tabular}{cccccccccccc}
{\scriptsize{}simmetry} & {\scriptsize{}Ex.\cite{Carter_H2CO_exact}} & {\scriptsize{}SC-IVR} & {\scriptsize{}Kay's method} & {\scriptsize{}Regularization} & {\scriptsize{}Adiabatic} & {\scriptsize{}Johnson} & {\scriptsize{}PPs} & {\scriptsize{}HO} & {\scriptsize{}$\mathbf{R}_{t}^{(1)}$} & {\scriptsize{}$\mathbf{R}_{t}^{(2)}$} & {\scriptsize{}$\mathbf{R}_{t}^{(3)}$}\tabularnewline
\hline 
{\scriptsize{}$\mbox{ZPE}\left(A_{1}\right)$} &  & {\scriptsize{}5774} & {\scriptsize{}5774} & {\scriptsize{}5780} & {\scriptsize{}5744} & {\scriptsize{}5744} & {\scriptsize{}5932} & {\scriptsize{}6112} & {\scriptsize{}5819} & {\scriptsize{}5744} & {\scriptsize{}5744}\tabularnewline
\hline 
{\scriptsize{}$B_{1}\left(1_{1}\right)$} & {\scriptsize{}1171} & {\scriptsize{}1162} & {\scriptsize{}1162} & {\scriptsize{}1169} & {\scriptsize{}1160} & {\scriptsize{}1159} & {\scriptsize{}1000} & {\scriptsize{}1004} & {\scriptsize{}1159} & {\scriptsize{}1160} & {\scriptsize{}1158}\tabularnewline
\hline 
{\scriptsize{}$B_{2}\left(2_{1}\right)$} & {\scriptsize{}1253} & {\scriptsize{}1245} & {\scriptsize{}1246} & {\scriptsize{}1248} & {\scriptsize{}1240} & {\scriptsize{}1240} & {\scriptsize{}1164} & {\scriptsize{}1168} & {\scriptsize{}1253} & {\scriptsize{}1240} & {\scriptsize{}1240}\tabularnewline
\hline 
{\scriptsize{}$A_{1}\left(3_{1}\right)$} & {\scriptsize{}1509} & {\scriptsize{}1509} & {\scriptsize{}1506} & {\scriptsize{}1513} & {\scriptsize{}1501} & {\scriptsize{}1509} & {\scriptsize{}1573} & {\scriptsize{}1575} & {\scriptsize{}1516} & {\scriptsize{}1509} & {\scriptsize{}1506}\tabularnewline
\hline 
{\scriptsize{}$A_{1}\left(4_{1}\right)$} & {\scriptsize{}1750} & {\scriptsize{}1747} & {\scriptsize{}1745} & {\scriptsize{}1752} & {\scriptsize{}1737} & {\scriptsize{}1743} & {\scriptsize{}1745} & {\scriptsize{}1743} & {\scriptsize{}1745} & {\scriptsize{}1745} & {\scriptsize{}1740}\tabularnewline
\hline 
{\scriptsize{}$A_{1}\left(5_{1}\right)$} & {\scriptsize{}2783} & {\scriptsize{}2810} & {\scriptsize{}2810} & {\scriptsize{}2785} & {\scriptsize{}2745} & {\scriptsize{}2747} & {\scriptsize{}2708} & {\scriptsize{}2711} & {\scriptsize{}2799} & {\scriptsize{}2750} & {\scriptsize{}2741}\tabularnewline
\hline 
{\scriptsize{}$B_{2}\left(6_{1}\right)$} & {\scriptsize{}2842} & {\scriptsize{}2850} & {\scriptsize{}2846} & {\scriptsize{}2836} &  & {\scriptsize{}2801} & {\scriptsize{}2862} & {\scriptsize{}2741} & {\scriptsize{}2846} & {\scriptsize{}2807} & {\scriptsize{}2800}\tabularnewline
\hline 
\end{tabular}
\end{table}
\begin{table}
\centering{}\caption{\selectlanguage{british}%
\label{tab:CH2O_en_lev2}\foreignlanguage{english}{The same as in
Table (\ref{tab:H2O_en_lev-1}) but for the overtones of $\mbox{CH}_{2}\mbox{O}$.}\selectlanguage{english}%
}
\begin{tabular}{cccccccccccc}
{\scriptsize{}State} & {\scriptsize{}Exact\cite{Carter_H2CO_exact}} & {\scriptsize{}SC-IVR} & {\scriptsize{}Kay's method} & {\scriptsize{}Regularization} & {\scriptsize{}Adiabatic} & {\scriptsize{}Johnson} & {\scriptsize{}PPs} & {\scriptsize{}HO} & {\scriptsize{}$\mathbf{R}_{t}^{(1)}$} & {\scriptsize{}$\mathbf{R}_{t}^{(2)}$} & {\scriptsize{}$\mathbf{R}_{t}^{(3)}$}\tabularnewline
\hline 
{\scriptsize{}$A_{1}\left(1_{2}\right)$} & {\scriptsize{}2333} & {\scriptsize{}2310} & {\scriptsize{}2310} & {\scriptsize{}2309} & {\scriptsize{}2302} & {\scriptsize{}2308} & {\scriptsize{}2163} & {\scriptsize{}2453} & {\scriptsize{}2307} & {\scriptsize{}2307} & {\scriptsize{}2304}\tabularnewline
\hline 
{\scriptsize{}$A_{2}\left(1_{1}2_{1}\right)$} & {\scriptsize{}2431} & {\scriptsize{}2410} & {\scriptsize{}2408} & {\scriptsize{}2405} & {\scriptsize{}2403} & {\scriptsize{}2399} & {\scriptsize{}2356} & {\scriptsize{}2360} & {\scriptsize{}2408} & {\scriptsize{}2401} & {\scriptsize{}2396}\tabularnewline
\hline 
{\scriptsize{}$A_{1}\left(2_{2}\right)$} & {\scriptsize{}2502} & {\scriptsize{}2497} & {\scriptsize{}2494} & {\scriptsize{}2489} & {\scriptsize{}2477} & {\scriptsize{}2486} &  & {\scriptsize{}2712} & {\scriptsize{}2495} & {\scriptsize{}2486} & {\scriptsize{}2480}\tabularnewline
\hline 
{\scriptsize{}$B_{1}\left(1_{1}3_{1}\right)$} & {\scriptsize{}2680} & {\scriptsize{}2672} & {\scriptsize{}2670} & {\scriptsize{}2675} & {\scriptsize{}2654} & {\scriptsize{}2656} & {\scriptsize{}2736} &  & {\scriptsize{}2679} & {\scriptsize{}2658} & {\scriptsize{}2654}\tabularnewline
\hline 
{\scriptsize{}$B_{2}\left(2_{1}3_{1}\right)$} & {\scriptsize{}2729} & {\scriptsize{}2731} & {\scriptsize{}2730} & {\scriptsize{}2728} & {\scriptsize{}2800} & {\scriptsize{}2719} & {\scriptsize{}2762} & {\scriptsize{}2761} & {\scriptsize{}2734} & {\scriptsize{}2723} & {\scriptsize{}2716}\tabularnewline
\hline 
{\scriptsize{}$B_{1}\left(1_{1}4_{1}\right)$} & {\scriptsize{}2913} & {\scriptsize{}2898} & {\scriptsize{}2896} & {\scriptsize{}2896} & {\scriptsize{}2886} & {\scriptsize{}2887} & {\scriptsize{}2871} &  & {\scriptsize{}2896} & {\scriptsize{}2888} & {\scriptsize{}2889}\tabularnewline
\hline 
{\scriptsize{}$B_{2}\left(2_{1}4_{1}\right)$} & {\scriptsize{}3007} & {\scriptsize{}3002} & {\scriptsize{}3002} & {\scriptsize{}3002} & {\scriptsize{}2976} & {\scriptsize{}2986} & {\scriptsize{}2946} &  & {\scriptsize{}3010} & {\scriptsize{}2989} & {\scriptsize{}2983}\tabularnewline
\hline 
{\scriptsize{}$A_{1}\left(3_{2}\right)$} & {\scriptsize{}3016} & {\scriptsize{}3018} & {\scriptsize{}3014} & {\scriptsize{}3018} & {\scriptsize{}2986} & {\scriptsize{}2996} & {\scriptsize{}3086} &  & {\scriptsize{}3022} & {\scriptsize{}2993} & {\scriptsize{}3010}\tabularnewline
\hline 
{\scriptsize{}$A_{1}\left(3_{1}4_{1}\right)$} & {\scriptsize{}3250} & {\scriptsize{}3254} & {\scriptsize{}3252} & {\scriptsize{}3256} & {\scriptsize{}3230} & {\scriptsize{}3240} & {\scriptsize{}3157} &  & {\scriptsize{}3263} & {\scriptsize{}3238} & {\scriptsize{}3234}\tabularnewline
\hline 
{\scriptsize{}$A_{1}\left(4_{2}\right)$} & {\scriptsize{}3480} & {\scriptsize{}3476} & {\scriptsize{}3475} & {\scriptsize{}3480} & {\scriptsize{}3462} & {\scriptsize{}3463} & {\scriptsize{}3323} &  & {\scriptsize{}3516} & {\scriptsize{}3468} & {\scriptsize{}3460}\tabularnewline
\hline 
{\scriptsize{}$B_{1}\left(1_{1}5_{1}\right)$} & {\scriptsize{}3947} & {\scriptsize{}3957} & {\scriptsize{}3960} & {\scriptsize{}3937} & {\scriptsize{}3892} & {\scriptsize{}3897} & {\scriptsize{}3864} & {\scriptsize{}3868} & {\scriptsize{}3949} & {\scriptsize{}3897} & {\scriptsize{}3890}\tabularnewline
\hline 
{\scriptsize{}$A_{2}\left(1_{1}6_{1}\right)$} & {\scriptsize{}4001} & {\scriptsize{}3979} & {\scriptsize{}3978} & {\scriptsize{}3974} & {\scriptsize{}3941} & {\scriptsize{}3942} & {\scriptsize{}3858} & {\scriptsize{}3864} & {\scriptsize{}3977} & {\scriptsize{}3945} & {\scriptsize{}3944}\tabularnewline
\hline 
{\scriptsize{}$B_{2}\left(2_{1}5_{1}\right)$} & {\scriptsize{}4027} & {\scriptsize{}4056} & {\scriptsize{}4054} & {\scriptsize{}4029} & {\scriptsize{}3990} & {\scriptsize{}3994} & {\scriptsize{}3934} & {\scriptsize{}3938} & {\scriptsize{}4045} & {\scriptsize{}4010} & {\scriptsize{}3994}\tabularnewline
\hline 
{\scriptsize{}$A_{1}\left(2_{1}6_{1}\right)$} & {\scriptsize{}4089} & {\scriptsize{}4038} & {\scriptsize{}4034} & {\scriptsize{}4043} & {\scriptsize{}4042} & {\scriptsize{}4053} & {\scriptsize{}4196} &  & {\scriptsize{}4074} & {\scriptsize{}4048} & {\scriptsize{}4048}\tabularnewline
\hline 
{\scriptsize{}$A_{1}\left(3_{1}5_{1}\right)$} & {\scriptsize{}4266} & {\scriptsize{}4275} & {\scriptsize{}4273} & {\scriptsize{}4268} & {\scriptsize{}4218} & {\scriptsize{}4225} & {\scriptsize{}4481} & {\scriptsize{}4216} & {\scriptsize{}4281} & {\scriptsize{}4225} & {\scriptsize{}4216}\tabularnewline
\hline 
{\scriptsize{}MAE} &  & {\scriptsize{}12.8} & {\scriptsize{}13.1} & {\scriptsize{}9.9} & {\scriptsize{}31.9} & {\scriptsize{}25.2} & {\scriptsize{}91.1 } & {\scriptsize{}91.9} & {\scriptsize{}12.1} & {\scriptsize{}23.4} & {\scriptsize{}30.2 }\tabularnewline
\hline 
\end{tabular}
\end{table}

For sake of comparison, Table (\ref{tab:CH2O_en_lev}) shows only
the fundamentals excitations and Table (\ref{tab:CH2O_en_lev2}) the
overtones. The MAE reported in the last row of Table (\ref{tab:CH2O_en_lev2})
is calculated over results reported in both tables. For this molecule,
the harmonic approximation is so drastic, that most of the peaks are
missing. As far as the other approximations are concerned, the PPs
is similar to the harmonic one, the adiabatic approximation is a little
bit more accurate, followed by the Johnson one. $\mathbf{R}_{t}^{(2)}$
of Eq.(\ref{eq:Rt2}) and $\mathbf{R}_{t}^{(3)}$ of Eq.(\ref{eq:Rt3})
are quite accurate. In this case also $\mathbf{R}_{t}^{(1)}$ is very
accurate. As far as the numerical regularization is concerned, the
results are very good with respect to the exact values and the ordinary
SC-IVR calculation. A fraction of $20.8\%$ of trajectories has been
tamed and each one no more than 11 times. This percent proves once
again that most of the chaotic trajectories rejected by the determinant
criterion do not jeopardize the accuracy of the spectrum.

\subsection{$\mathbf{CH_{4}}$ and $\mathbf{CH_{2}}\mathbf{D}_{\mathbf{2}}$
molecule\label{sec:CH4_CH2D2}}

In terms of chaotic motion, methane and dideuterated methane are quite
challenging given the nine strongly coupled degrees of freedom and
the light atoms dynamics. We employ the PES by Lee et al.\cite{Lee_CH4_PES}
and compare with the exact quantum energy levels,\cite{Bowman_CH4quantumvalues}
as done in previous semiclassical calculations.\cite[(b)]{Alex_Mik}
We employ $32000$ trajectories for the SC-IVR calculation, out of
which $88.8\%$ and $88.7\%$ are rejected using the monodromy matrix
criterium, while 98.9\% and 97.4\% using the criterium of Kay, respectively
for the $\mbox{CH}_{4}$ and $\mbox{CH}_{2}\mbox{D}_{2}$ molecule.
Instead, $14000$ classical trajectories are used for the approximated
and numerical tamed pre-exponential factor calculations. All trajectories
are made of $3000$ time-steps each, with the same time-step length
as above and for all simulations. In the case of methane, the point
group symmetry is $\mbox{T}_{d}$. 
\begin{figure}
\centering{}\includegraphics[scale=0.45]{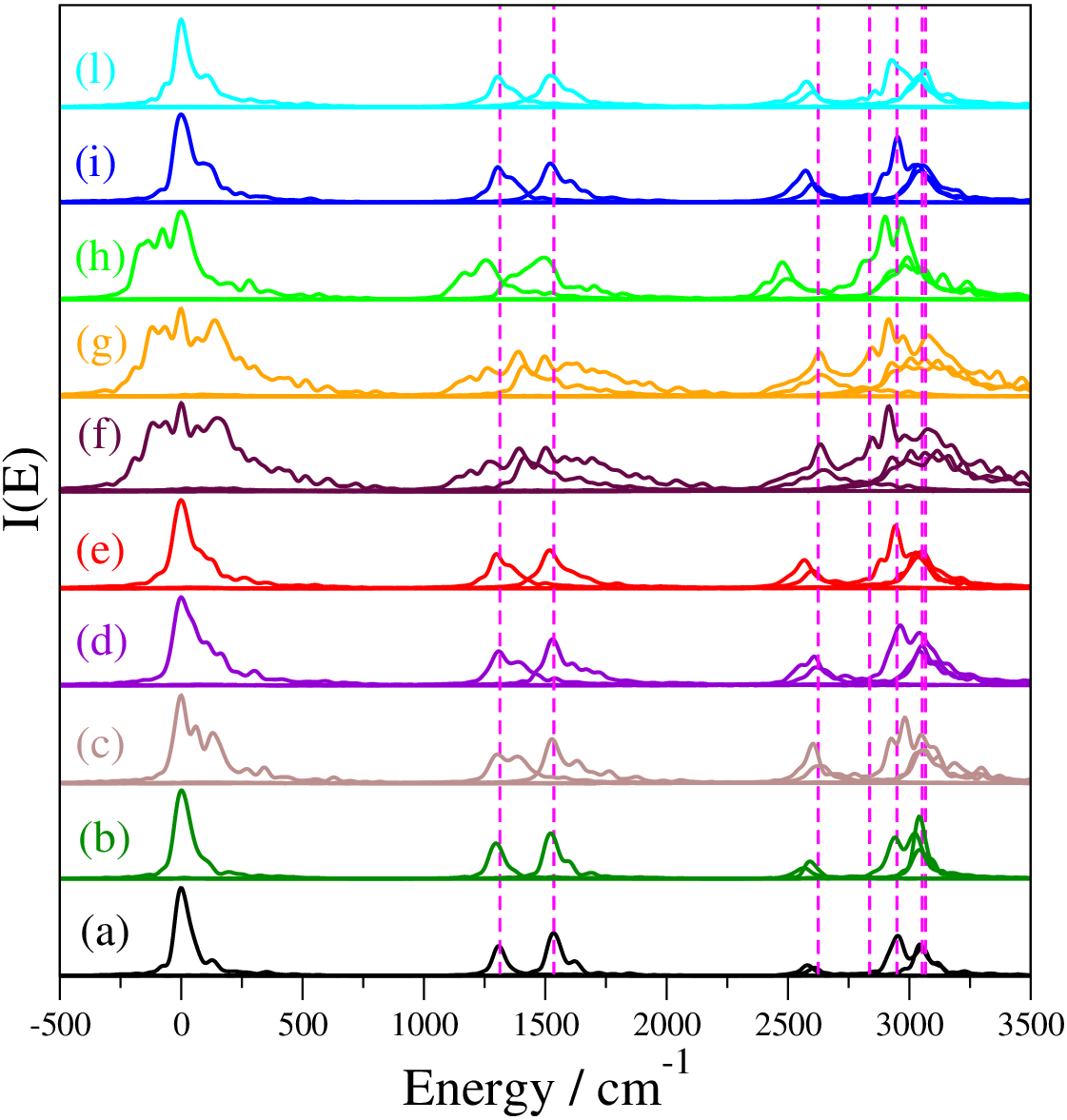}\caption{\label{fig:3spettri_CH4}The same as in Fig.(\ref{fig:H2O_spectra})
but for the $\mbox{CH\ensuremath{{}_{4}}}$ molecule. Each approximation
includes the spectra of the $\mbox{A}_{1}$, $\mbox{E}$, and $\mbox{T}_{2}$
irreducible representations of the $\mbox{T}_{d}$ point group of
symmetry.}
\end{figure}
The spectrum of each irreducible representation is reported in Fig.(\ref{fig:3spettri_CH4})
with the same color code as above and for different approximations.
\begin{table}
\centering{}\caption{\selectlanguage{british}%
\label{tab:CH4_en_lev}\foreignlanguage{english}{The same as in Table
(\ref{tab:H2O_en_lev-1}) but for $\mbox{CH}_{4}$.}\selectlanguage{english}%
}
\begin{tabular}{cccccccccccc}
{\scriptsize{}State} & {\scriptsize{}Exact\cite{Bowman_CH4quantumvalues}} & {\scriptsize{}SC-IVR} & {\scriptsize{}Kay's method} & {\scriptsize{}Regularization} & {\scriptsize{}Adiabatic} & {\scriptsize{}Johnson} & {\scriptsize{}PPs} & {\scriptsize{}HO} & {\scriptsize{}$\mathbf{R}_{t}^{(1)}$} & {\scriptsize{}$\mathbf{R}_{t}^{(2)}$} & {\scriptsize{}$\mathbf{R}_{t}^{(3)}$}\tabularnewline
\hline 
{\scriptsize{}$\mbox{ZPE}\left(A_{1}\right)$} & {\scriptsize{}9707} & {\scriptsize{}9708} & {\scriptsize{}9708} & {\scriptsize{}9704} & {\scriptsize{}9669} & {\scriptsize{}9657} & {\scriptsize{}9846 } & {\scriptsize{}10124} & {\scriptsize{}9941} & {\scriptsize{}9659} & {\scriptsize{}9652 }\tabularnewline
\hline 
{\scriptsize{}$T_{2}\left(1_{1}\right)$} & {\scriptsize{}1313} & {\scriptsize{}1296} & {\scriptsize{}1297} & {\scriptsize{}1304} & {\scriptsize{}1309} & {\scriptsize{}1300} & {\scriptsize{}1390} & {\scriptsize{}1390} & {\scriptsize{}1257} & {\scriptsize{}1305} & {\scriptsize{}1304}\tabularnewline
\hline 
{\scriptsize{}$E\left(2_{1}\right)$} & {\scriptsize{}1535} & {\scriptsize{}1524} & {\scriptsize{}1524} & {\scriptsize{}1528} & {\scriptsize{}1531} & {\scriptsize{}1518} & {\scriptsize{}1500} & {\scriptsize{}1497} & {\scriptsize{}1496} & {\scriptsize{}1522} & {\scriptsize{}1520}\tabularnewline
\hline 
{\scriptsize{}$T_{2}\left(1_{2}\right)$} & {\scriptsize{}2624} & {\scriptsize{}2596} & {\scriptsize{}2593} & {\scriptsize{}2636} & {\scriptsize{}2616} & {\scriptsize{}2601} & {\scriptsize{}2646} & {\scriptsize{}2636} & {\scriptsize{}2497} & {\scriptsize{}2605} & {\scriptsize{}2600}\tabularnewline
\hline 
{\scriptsize{}$T_{2}\left(1_{1}2_{1}\right)$} & {\scriptsize{}2836} & {\scriptsize{}2820} & {\scriptsize{}2821} & {\scriptsize{}2832} & {\scriptsize{}1309} & {\scriptsize{}2818} & {\scriptsize{}2890} & {\scriptsize{}2887} & {\scriptsize{}2753} & {\scriptsize{}2827} & {\scriptsize{}2824}\tabularnewline
\hline 
{\scriptsize{}$T_{1}\left(1_{1}2_{1}\right)$} & {\scriptsize{}2836} & {\scriptsize{}2820} & {\scriptsize{}2821} & {\scriptsize{}2832} & {\scriptsize{}1309} & {\scriptsize{}2818} & {\scriptsize{}2890} & {\scriptsize{}2887} & {\scriptsize{}2753} & {\scriptsize{}2827} & {\scriptsize{}2824}\tabularnewline
\hline 
{\scriptsize{}$A\left(3_{1}\right)$} & {\scriptsize{}2949} & {\scriptsize{}2942} & {\scriptsize{}2942} & {\scriptsize{}2982} & {\scriptsize{}2963} & {\scriptsize{}2944} & {\scriptsize{}2914} & {\scriptsize{}2916} & {\scriptsize{}2936} & {\scriptsize{}2951} & {\scriptsize{}2928}\tabularnewline
\hline 
{\scriptsize{}$E\left(2_{2}\right)$} & {\scriptsize{}3067} & {\scriptsize{}3040} & {\scriptsize{}3042} & {\scriptsize{}3062} & {\scriptsize{}3052} & {\scriptsize{}3028} & {\scriptsize{}3065} & {\scriptsize{}3066} & {\scriptsize{}2993} & {\scriptsize{}3035} & {\scriptsize{}3044}\tabularnewline
\hline 
{\scriptsize{}$T_{2}\left(4_{1}\right)$} & {\scriptsize{}3053} & {\scriptsize{}3038} & {\scriptsize{}3040} & {\scriptsize{}3052} & {\scriptsize{}3044} & {\scriptsize{}3037} & {\scriptsize{}3092} & {\scriptsize{}3069} & {\scriptsize{}2983} & {\scriptsize{}3041} & {\scriptsize{}3044}\tabularnewline
\hline 
{\scriptsize{}MAE} &  & {\scriptsize{}15.3} & {\scriptsize{}16.6} & {\scriptsize{}8.7} & {\scriptsize{}7.8} & {\scriptsize{}18.6} & {\scriptsize{}39.8} & {\scriptsize{}34.9} & {\scriptsize{}68.1} & {\scriptsize{}13.0} & {\scriptsize{}15.6 }\tabularnewline
\hline 
\end{tabular}
\end{table}
 Table (\ref{tab:CH4_en_lev}) shows the low lying energy levels.
These can be compared to the exact ones reported, as before, in the
second column. The fifth column reports the regularization results,
where $37.6\%$ of the trajectories experienced a monodromy matrix
regularization for no more than $21$ times. This was enough to not
reject any trajectory and reproduced the quantum mechanical results
quite accurately. The PPs approximation is very similar to the harmonic
one. Overall, $\mathbf{R}_{t}^{(2)}$ and $\mathbf{R}_{t}^{(3)}$
are offering the most accurate pre-exponential factor approximation,
a part from the adiabatic and the regularization ones that imply the
integration of the equation of motion of the monodromy matrix elements.

\begin{figure}
\centering{}\includegraphics[scale=0.45]{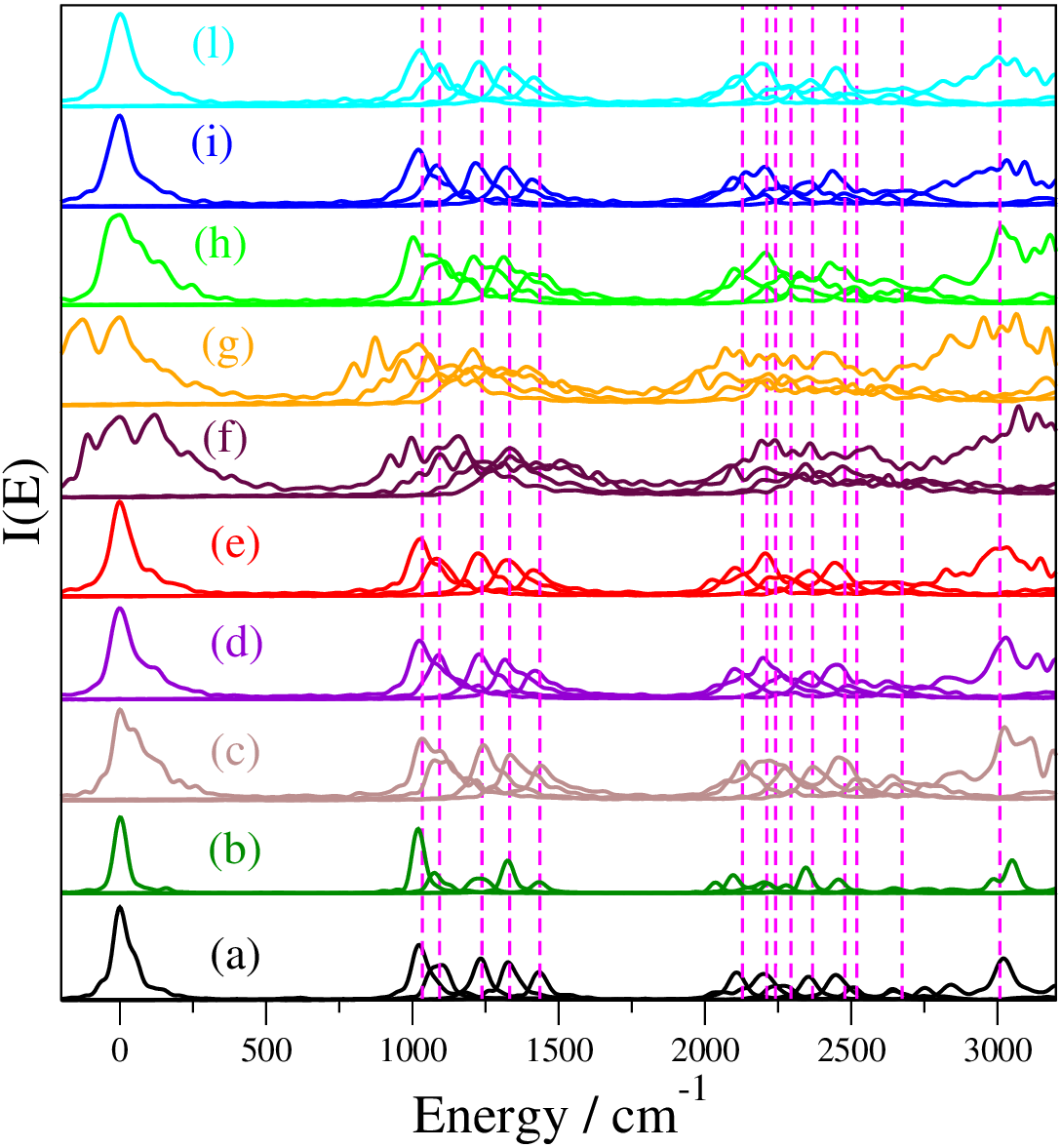}\caption{\label{fig:3spettri_CH2D2}The same as in Fig.(\ref{fig:H2O_spectra})
but for the $\mbox{CH\ensuremath{{}_{2}}D\ensuremath{_{2}}}$ molecule.
Each approximation includes the spectra of the $\mbox{A}_{1}$, $\mbox{A}_{2}$,
$\mbox{B}_{1}$ and $\mbox{B}_{2}$ irreducible representations of
the $\mbox{C}_{2v}$ point group of symmetry.}
\end{figure}

The point group symmetry for $\mbox{C}\mbox{H}_{2}\mbox{D}_{2}$ is
$\mbox{C}_{2v}$ and each irreducible representation is reported in
Fig.(\ref{fig:3spettri_CH2D2}).
 As in previous figures, Fig.(\ref{fig:3spettri_CH2D2}) reports the
results for each approximation. From this figure, results are quite
similar, except for the highest vibrational levels.
\begin{table}
\centering{}\caption{\selectlanguage{british}%
\label{tab:CH2D2_en_lev}\foreignlanguage{english}{The same as in
Table (\ref{tab:H2O_en_lev-1}) but for $\mbox{CH}_{2}\mbox{D}_{2}$.}\selectlanguage{english}%
}
\begin{tabular}{cccccccccccc}
{\scriptsize{}simmetry} & {\scriptsize{}Ex.\cite{Bowman_CH4quantumvalues}} & {\scriptsize{}SC-IVR} & {\scriptsize{}Kay's method} & {\scriptsize{}Regularization} & {\scriptsize{}Adiabatic} & {\scriptsize{}Johnson} & {\scriptsize{}PPs} & {\scriptsize{}HO} & {\scriptsize{}$\mathbf{R}_{t}^{(2)}$} & {\scriptsize{}$\mathbf{R}_{t}^{(2)}$} & {\scriptsize{}$\mathbf{R}_{t}^{(3)}$}\tabularnewline
\hline 
{\scriptsize{}$\mbox{ZPE}\left(A_{1}\right)$} & {\scriptsize{}8443} & {\scriptsize{}8442} & {\scriptsize{}8438} & {\scriptsize{}8440} & {\scriptsize{}8410} & {\scriptsize{}8401} & {\scriptsize{}8510} & {\scriptsize{}8860} & {\scriptsize{}8508} & {\scriptsize{}8408} & {\scriptsize{}8404}\tabularnewline
\hline 
{\scriptsize{}$A_{1}\left(1_{1}\right)$} & {\scriptsize{}1034} & {\scriptsize{}1019} & {\scriptsize{}1018} & {\scriptsize{}1035} & {\scriptsize{}1026} & {\scriptsize{}1027} & {\scriptsize{}997} & {\scriptsize{}875} & {\scriptsize{}1003} & {\scriptsize{}1021} & {\scriptsize{}1025}\tabularnewline
\hline 
{\scriptsize{}$B_{2}\left(2_{1}\right)$} & {\scriptsize{}1093} & {\scriptsize{}1078} & {\scriptsize{}1074} & {\scriptsize{}1076} & {\scriptsize{}1092} & {\scriptsize{}1084} & {\scriptsize{}1185} & {\scriptsize{}1056} & {\scriptsize{}1108} & {\scriptsize{}1086} & {\scriptsize{}1092}\tabularnewline
\hline 
{\scriptsize{}$B_{1}\left(3_{1}\right)$} & {\scriptsize{}1238} & {\scriptsize{}1240} & {\scriptsize{}1224} & {\scriptsize{}1244} & {\scriptsize{}1228} & {\scriptsize{}1225} & {\scriptsize{}1335} & {\scriptsize{}1208} & {\scriptsize{}1208} & {\scriptsize{}1216} & {\scriptsize{}1228}\tabularnewline
\hline 
{\scriptsize{}$A_{2}\left(4_{1}\right)$} & {\scriptsize{}1332} & {\scriptsize{}1326} & {\scriptsize{}1324} & {\scriptsize{}1334} & {\scriptsize{}1316} & {\scriptsize{}1323} & {\scriptsize{}1425} & {\scriptsize{}1306} & {\scriptsize{}1312} & {\scriptsize{}1321} & {\scriptsize{}1315}\tabularnewline
\hline 
{\scriptsize{}$A_{1}\left(5_{1}\right)$} & {\scriptsize{}1436} & {\scriptsize{}1431} & {\scriptsize{}1431} & {\scriptsize{}1432} & {\scriptsize{}1421} & {\scriptsize{}1414} &  &  & {\scriptsize{}1409} & {\scriptsize{}1409} & {\scriptsize{}1413}\tabularnewline
\hline 
{\scriptsize{}$B_{2}\left(1_{1}2_{1}\right)$} & {\scriptsize{}2128} & {\scriptsize{}2098} & {\scriptsize{}2094} & {\scriptsize{}2128} & {\scriptsize{}2104} & {\scriptsize{}2105} & {\scriptsize{}2098} & {\scriptsize{}2068} & {\scriptsize{}2103} & {\scriptsize{}2101} & {\scriptsize{}2111}\tabularnewline
\hline 
{\scriptsize{}$A_{1}\left(6_{1}\right)$} & {\scriptsize{}2211} & {\scriptsize{}2203} & {\scriptsize{}2202} & {\scriptsize{}2220} & {\scriptsize{}2200} & {\scriptsize{}2207} & {\scriptsize{}2194} &  & {\scriptsize{}2207} & {\scriptsize{}2205} & {\scriptsize{}2192}\tabularnewline
\hline 
{\scriptsize{}$B_{1}\left(1_{1}2_{1}\right)$} & {\scriptsize{}2242} & {\scriptsize{}2222} & {\scriptsize{}2214} &  &  & {\scriptsize{}2224} & {\scriptsize{}2344} & {\scriptsize{}2217} &  & {\scriptsize{}2218} & {\scriptsize{}2212}\tabularnewline
\hline 
{\scriptsize{}$B_{1}\left(7_{1}\right)$} & {\scriptsize{}2294} & {\scriptsize{}2270} & {\scriptsize{}2276} & {\scriptsize{}2273} & {\scriptsize{}2267} & {\scriptsize{}2273} &  &  & {\scriptsize{}2265} & {\scriptsize{}2269} & {\scriptsize{}2288}\tabularnewline
\hline 
{\scriptsize{}$A_{2}\left(1_{1}4_{1}\right)$} & {\scriptsize{}2368} & {\scriptsize{}2349} & {\scriptsize{}2342} & {\scriptsize{}2370} & {\scriptsize{}2359} & {\scriptsize{}2360} & {\scriptsize{}2474} & {\scriptsize{}2273} & {\scriptsize{}2325} & {\scriptsize{}2358} & {\scriptsize{}2360}\tabularnewline
\hline 
{\scriptsize{}$A_{1}\left(1_{1}5_{1}\right)$} & {\scriptsize{}2474} & {\scriptsize{}2465} & {\scriptsize{}2455} & {\scriptsize{}2457} & {\scriptsize{}2459} & {\scriptsize{}2444} & {\scriptsize{}2359} & {\scriptsize{}2413} & {\scriptsize{}2428} & {\scriptsize{}2437} & {\scriptsize{}2448}\tabularnewline
\hline 
{\scriptsize{}$B_{2}\left(2_{1}5_{1}\right)$} & {\scriptsize{}2519} & {\scriptsize{}2510} & {\scriptsize{}2518} & {\scriptsize{}2512} & {\scriptsize{}2491} & {\scriptsize{}2497} & {\scriptsize{}2592} & {\scriptsize{}2456} & {\scriptsize{}2513} & {\scriptsize{}2477} & {\scriptsize{}2494}\tabularnewline
\hline 
{\scriptsize{}$B_{1}\left(3_{1}5_{1}\right)$} & {\scriptsize{}2674} & {\scriptsize{}2658} & {\scriptsize{}2647} & {\scriptsize{}2650} & {\scriptsize{}2631} & {\scriptsize{}2635} & {\scriptsize{}2742} & {\scriptsize{}2624} & {\scriptsize{}2656} & {\scriptsize{}2627} & {\scriptsize{}2631}\tabularnewline
\hline 
{\scriptsize{}$A_{2}\left(4_{1}5_{1}\right)$} & {\scriptsize{}2769} & {\scriptsize{}2764} & {\scriptsize{}2762} & {\scriptsize{}2754} & {\scriptsize{}2741} & {\scriptsize{}2745} &  &  & {\scriptsize{}2748} & {\scriptsize{}2743} & {\scriptsize{}2748}\tabularnewline
\hline 
{\scriptsize{}$A_{1}\left(8_{1}\right)$} & {\scriptsize{}3008} & {\scriptsize{}3044} & {\scriptsize{}3048} & {\scriptsize{}3024} & {\scriptsize{}3032} & {\scriptsize{}3033} & {\scriptsize{}3074} & {\scriptsize{}3065} & {\scriptsize{}3014} & {\scriptsize{}3035} & {\scriptsize{}3000}\tabularnewline
\hline 
{\scriptsize{}MAE} &  & {\scriptsize{}14.6} & {\scriptsize{}18.1} & {\scriptsize{}9.7} & {\scriptsize{}18.5} & {\scriptsize{}18.3 } & {\scriptsize{}74.7 } & {\scriptsize{}60.3 } & {\scriptsize{}22.9} & {\scriptsize{}23.4 } & {\scriptsize{}17.5}\tabularnewline
\hline 
\end{tabular}
\end{table}
 A more detailed view is provided by Table (\ref{tab:CH2D2_en_lev}),
where $44.1\%$ of the $14000$ trajectories have been regularized
for no more than $19$ times each. The PPs is confirming to be about
as accurate as the harmonic one, and the adiabatic approximation is
a quite accurate one. The Jonhson approximation is also quite accurate.
The $\mathbf{R}_{t}^{(3)}$ approximation of Eq.(\ref{eq:Rt3}) is
overall more accurate than $\mathbf{R}_{t}^{(2)}$ and $\mathbf{R}_{t}^{(1)}$.
An harmonic approximation of the pre-exponential factor would be too
brutal in this case and some of the peak signals are missing.

\section{Conclusions\label{sec:Conclusions}}

The series of calculations reported above show the importance of the
semiclassical pre-exponential factor of Eq.(\ref{eq:C_t_prefactor})
to properly account for the quantum mechanical effects of the semiclassical
propagator. Unfortunately, the semiclassical calculation of the pre-exponential
factor of classical trajectories for chaotic systems is hampered by
numerical issues, as already known and once more demonstrated here
on several model systems. To bypass this numerical \emph{empasse},
we recall and present possible approximations to the pre-exponential
factor in SC-IVR dynamics. These approximations are motivated either
by analytical considerations or by numerical regularizations. Each
approximation is presented, derived and then applied separately to
both model systems with an artificial amount of chaos and real systems
of growing dimensionality and complexity. The accuracy of each approximation
has been tested with the Herman-Kluk and the time-averaging SC-IVR
methods versus the number of rejected trajectories, which is an empirical
measure of the amount of chaos as well as respect to the established
ad-hoc method of Kay.\cite{Kay_101} The numerical regularization
is quite accurate but it can not be applied \emph{a priori} for any
system since it implies the calculation of the monodromy matrix. The
regularization results are very similar to the original SC-IVR ones,
since the chaotic trajectories are not counting in the regularized
monodromy matrix. The pre-exponential factor analytical approximations,
which are $\mathbf{R}_{t}^{(2)}$ in Eq.(\ref{eq:Rt2}) and $\mathbf{R}_{t}^{(3)}$
in Eq.(\ref{eq:Rt3}), are quite accurate compared to both the exact
values and the SC-IVR ones, and we suggest them for semiclassical
simulations of systems when the integration of the monodromy matrix
and its regularization are not possible.
\begin{acknowledgments}
We acknowledges support from the European Research Council (ERC) under
the European Union\textquoteright s Horizon 2020 research and innovation
programme (grant agreement No {[}647107{]} \textendash{} SEMICOMPLEX
\textendash{} ERC-2014-CoG). M.C. acknowledges also the CINECA and
the Regione Lombardia award under the LISA initiative (grant SURGREEN)
for the availability of high performance computing resources. Dr.
Riccardo Conte and Prof. Dmitry Shalashilin are warmly thanked for
useful discussions.\end{acknowledgments}


\begin{thebibliography}{10}
\bibitem{Miller_avd_74}W. H. Miller, Adv. Chem. Phys. \textbf{25},
69 (1974).

\bibitem{Miller_PNAS}W. H. Miller, Proc. Natl. Acad. Sci. U.S.A.
\textbf{102}, 6660 (2005).

\bibitem{Kay_review}K. G. Kay, \emph{Annu. Rev. Phys. Chem.}, 2005,
\textbf{56}, 255.

\bibitem{Miller_IVR}W. H. Miller, J. Chem. Phys. \textbf{53}, 3578
(1970); W. H. Miller and T. F. George, J. Chem. Phys. \textbf{56},
5668 (1972).

\bibitem{Heller_frozengaussian}E. J. Heller, J. Chem. Phys. \textbf{62},
1544 (1975); E. J. Heller, J. Chem. Phys. \textbf{75}, 2923 (1981).

\bibitem{Herman_Kluk_2} M. F. Herman, J. Chem. Phys. \textbf{85},
2069 (1986); E. Kluk, M. F. Herman and H. L. Davis, J. Chem. Phys.
\textbf{84}, 326 (1986).

\bibitem{Kay_100}K. G. Kay, J. Chem. Phys. \textbf{100}, 4377 (1994);
K. G. Kay, J. Chem. Phys. \textbf{100}, 4432 (1994).

\bibitem{Miller_includingQuantumEffects}W. H. Miller, J. Chem. Phys.
\textbf{125}, 132305 (2006).

\bibitem{Kay_HKderivation_06}W. H. Miller, J. Chem. Phys. \textbf{125},
132305 (2006); K. G. Kay, Chem. Phys. \textbf{322}, 3 (2006).

\bibitem{Kay_IVRcorrelationfunctions_04}T. Sklarz and K. G. Kay,
J. Chem. Phys. \textbf{120}, 2606 (2004).

\bibitem{Kay_collinearHe_07}C. Harabati and K. G. Kay, J. Chem. Phys.
\textbf{127}, 084104 (2007).

\bibitem{Kay_tunnellingHigerorderHK} G. Hochman and K. G. Kay,\emph{
}J. Chem. Phys. \textbf{130}, 061104 (2009).

\bibitem{Herman_review}M. F. Herman, Annu. Rev. Phys. Chem. \textbf{45},
83 (1994).

\bibitem{Makri_review}N. Makri, Annu. Rev. Phys. Chem. \textbf{50},
167 (1999).

\bibitem{Pollak_prefactorfree_04} (a) S. Zhang and E. Pollak, J.
Chem. Phys. \textbf{121}, 3384 (2004); (b) S. Zhang and E. Pollak,
J. Chem. Theory Comput. \textbf{1}, 345 (2005).

\bibitem{Pollak_gaussianpropagator_06}J. Shao and E. Pollak, J. Chem.
Phys. \textbf{125}, 133502 (2006).

\bibitem{Pollak_autocorrelation} (a) E. Pollak and E. Martin-Fierro,
J. Chem. Phys. \textbf{126}, 164107 (2007); (b) E. Martin-Fierro and
E. Pollak, J. Chem. Phys. \textbf{125}, 164104 (2006).

\bibitem{Manolopoulos} (a) A. R. Walton, D. E. Manolopoulos, Mol.
Phys. \textbf{87}, 961 (1996); (b) A. R: Walton, D. E. Manolopoulos,
Chem. Phys. Lett\emph{.} \textbf{244}, 448 (1995); (c) M. L. Brewer,
J. S. Hulme, D. E. Manolopoulos, J. Chem. Phys\emph{.} \textbf{106},
4832 (1997).

\bibitem{Coker}S. Bonella, D. Montemayor , and D. F. Coker, Proc.
Natl. Am. Soc\emph{.} \textbf{102}, 6715 (2005). 

\bibitem{Coker2}S. Bonella and D. F. Coker, J. Chem. Phys. \textbf{118},
4370 (2003).

\bibitem{Grossmann} (a) C. Harabati, J. M. Rost, and F. Grossmann,
J. Chem. Phys. \textbf{120}, 26 (2004); (b) F. Grossmann, Comments
At. Mol. Phys. \textbf{34}, 243 (1999).

\bibitem{DeAguiar_SC_IVR_11}T. F. Viscondi and M. A. M. de Aguiar,
J. Chem. Phys. \textbf{134}, 234105 (2011).

\bibitem{Roy}B. B. Issack and P. N. Roy, J. Chem. Phys. \textbf{127},
054105 (2007).

\bibitem{Charu_electrontranssferSCIVR_11}C. Venkataraman, J. Chem.
Phys. \textbf{135}, 204503 (2001).

\bibitem{Nakamura_PCCP}H. Nakamura, S. Nanbu, Y. Teranishic, and
A. Ohtab, Phys. Chem. Chem. Phys. \textbf{18}, 11972 (2016)

\bibitem{Nanbu_nonadaFrozenGauss}Al. D. Kondorskiy and S. Nanbu,
J. Chem. Phys. \textbf{143}, 114103 (2015)

\bibitem{Ceotto_mixedSC}M. Buchholz, F. Grossmann, and M. Ceotto,
J. Chem. Phys. \textbf{144}, 094102 (2016)

\bibitem{Grossmann_SCBoseHubbardmodel_16}S. Ray, P. Ostmann, L. Simon,
F. Grossmann, and W.T. Strunz, J. Phys. A: Math. Theor. \textbf{49},
165303 (2016)

\bibitem{Ceotto_MCSCIVR} M. Ceotto, S. Atahan, G. F. Tantardini,
and A. Aspuru-Guzik,\emph{ }J. Chem. Phys. \textbf{130}, 234113, (2009).

\bibitem{Ceotto_1traj}M. Ceotto, S. Atahan, S. Shim, G. F. Tantardini,
and A. Aspuru-Guzik, Phys. Chem. Chem. Phys\emph{.} \textbf{11}, 3861
(2009).

\bibitem{Ceotto_cursofdimensionality_11}M. Ceotto, G. F. Tantardini,
and A. Aspuru-Guzik, J. Chem. Phys. \textbf{135}, 214108 (2011).

\bibitem{Ceotto_acceleratedSCIVR}M. Ceotto, Y. Zhuang, and W. L.
Hase, J. Chem. Phys. \textbf{138}, 054116 (2013).

\bibitem{Ceotto_NH3}R. Conte, A. Aspuru-Guzik, and M. Ceotto, \emph{J.
Phys. Chem. Lett.} \textbf{4}, 3407 (2013).

\bibitem{Jorge} J. Tatchen and E. Pollak, \emph{J. Chem. Phys.},
2009, \textbf{130}, 041103.

\bibitem{Pollak_CH2OInternalconversion_13}R. Ianconescu, J. Tatchen,
and E. Pollak, J. Chem. Phys \textbf{139}, 154311 (2013).

\bibitem{Roy_AbinitioSCIVR}S. Y. Y. Wong, D. M. Benoit, M. Lewerenz,
A. Brown, and P.-N. Roy, J. Chem. Phys. \textbf{134}, 094110 (2011).

\bibitem{Hase_abinitioMDformaldehyde_94}W. Chen, W. L. Hase, H. B.
Schlegel, Chem. Phys. Lett. \textbf{228}, 436 (1994).

\bibitem{Hase_BOMDHessianIntegrator_99}J. M. Millam, V. Bakken, W.
Chen, W. L. Hase, H. B. Schlegel, J. Chem. Phys. \textbf{111}, 3800
(1999).

\bibitem{Hase_reviewBODirectDynamics_03} L. Sun and W. L. Hase, Rev.
Comput. Chem. \textbf{19}, 79 (2003).

\bibitem{Jiri_intjquantchem} T. Zimmermann, J. Ruppen, B. Li, and
J. Vaní\v{c}ek, Int. J. Quantum Chem. \textbf{110}, 2426 (2010).

\bibitem{Martinez_AIMS}M. Ben-Nun, T. J. Martinez,\emph{ }Adv. Chem.
Phys\emph{.} \textbf{121}, 439 (2002); B. G. Levine, J. D. Coe, A.
M. Virshup, and Todd J. Martinez, Chem. Phys. \textbf{347}, 3 (2008);
J. D. Coe, B. G. Levine, and T. J. Martinez J. Phys. Chem. \textbf{111},
11302 (2007).

\bibitem{Martins_AIMD_91}R. M. Wentzcovitch and J. L. Martins, Solid
State Commun. \textbf{78}, 831 (1991).

\bibitem{Marx_book}D. Marx adn J. Hutter, \emph{Modern Methods and
Algorithms of Quantum Chemistry}, edited by J. Grotendorst (John von
Neumann Institute for Computing, Julich, Germany, 2000), 2nd ed.

\bibitem{Hase_ScienceSN2}L. Sun, K. Song, and W. L. Hase, Science
\textbf{296}, 875 (2002).

\bibitem{Miller308_Wigner_98}H. Wang, X. Sun, and W. H. Miller, J.
Chem. Phys. \textbf{108}, 9726 (1998).

\bibitem{Miller312_FaradayDisc_98}W. H. Miller, Faraday Disc. Chem.
Soc. \textbf{110}, 1 (1998).

\bibitem{Miller320_LSCIVRgeneralization_99}W. H. Miller, J. Phys.
Chem. A \textbf{103}, 9384 (1999).

\bibitem{Miller_Liu} (a) J. Liu and W. H. Miller, J. Chem. Phys.
\textbf{125}, 224104 (2006); (b)\emph{ ibidem} \textbf{126}, 234110
(2007); (c)\emph{ ibidem} \textbf{127}, 114506 (2007); (d)\emph{ ibidem}
\textbf{128}, 144511 (2008). 

\bibitem{Geva} (a) I. Navrotskaya and E. Geva, J. Phys. Chem. A \textbf{111},
460 (2007); (b) B. K. Ka, Q. Shi, and E. Geva, J. Phys. Chem. A \textbf{109},
5527 (2005); (c) F. X. Vazquez, S. Talapatra, and E. Geva, J. Phys.
Chem. A \textbf{115}, 9775 (2011).

\bibitem{Koda_IVRWigner}S. Koda, J. Chem. Phys. \textbf{143}, 244110
(2015)

\bibitem{Koda_Mix_SC_Wigner}S. Koda, J. Chem. Phys. \textbf{144},
154108 (2016)

\bibitem{Pollak_Petersen_interaction}J. Petersen and E. Pollak, J.
Chem. Phys. \textbf{143}, 224114 (2015)

\bibitem{MillerFBSCIVR} (a) X. Sun and W. H. Miller, J. Chem. Phys.
\textbf{110}, 6635 (1999); (b) H. Wang, M. Thoss, K. Sorge, R. Gelabert,
X. Gimenez and W. H. Miller, J. Chem. Phys. \textbf{114}, 2562 (2001);
(c) R. Gelabert, X. Gimenez, M. Thoss, H. Wang and W. H. Miller, J.
Chem. Phys. \textbf{114}, 2572 (2001); (d) M. Thoss, H. Wang and W.
H. Miller, J. Chem. Phys. \textbf{114}, 9220 (2001);

\bibitem{Miller340_GeneralizedFilinov_01}H. Wang, D. E. Manolopoulos
and W. H. Miller, J. Chem. Phys. \textbf{115}, 6317 (2001).

\bibitem{Makri_FBSCIVR} (a) K. Thompson and N. Makri, Phys. Rev.
E \textbf{59}, R4729 (1999); (b) J. Shao and N. Makri, J. Phys. Chem.
A \textbf{103}, 7753, 9479 (1999).

\bibitem{Conte_Pollak10}R. Conte and E. Pollak Phys. Rev. E 81, 036704
(2010).

\bibitem{Conte_Pollak12}R. Conte and E. Pollak J. Chem. Phys. 136,
094101 (2012).

\bibitem{Takatsuka_eigenstates_05} (a) H. Ushiyama and K. Takatsuka,
J. Chem. Phys. 122, 224112 (2005); (b) S. Takahashi and K. Takatsuka,
J. Chem. Phys. 127, 084112 (2007)

\bibitem{Filinov_filter} (a) V. S. Filinov, Nucl. Phys. B \textbf{271},
717 (1986); (b) N. Makri and W. H. Miller, Chem. Phys. Lett. \textbf{139},
10 (1987); (c) J. D. Doll, D. L. Freeman, and T. L. Beck, Adv. Chem.
Phys. \textbf{78}, 61 (1994); (d) S. M. Anderson, D. Neuhauser, and
R. Baer, J. Chem. Phys. \textbf{118}, 9103 (2003).

\bibitem{Alex_Mik}A. L. Kaledin and W. H. Miller, J. Chem. Phys.
\textbf{118}, 7174 (2003); A. L. Kaledin and W. H. Miller, J. Chem.
Phys. \textbf{119}, 3078 (2003).

\bibitem{Ceotto_Zhang_JCTC} Y. Zhuang, M. R. Siebert, W. L. Hase,
K. G. Kay, and M. Ceotto, J. Chem. Theory and Comput \textbf{9}, 54
(2013).

\bibitem{Ceotto_GPU} D. Tamascelli, F. S. Dambrosio, R. Conte, and
M. Ceotto, J. Chem. Phys. \textbf{140}, 174109 (2014).

\bibitem{Ceotto_david}M. Ceotto, D. dell'Angelo, and G. F. Tantardini,
J. Chem. Phys. \textbf{133}, 054701 (2010).

\bibitem{Ceotto_eigenfunctions}M. Ceotto, S. Valleau, G. F. Tantardini,
and A. Aspuru-Guzik, J. Chem Phys. \textbf{134}, 234103 (2011).

\bibitem{Feynman_Hibbs} R.P. Feynman and A.R. Hibbs, \emph{Quantum
Mechanics and Path Integrals} (McGraw-Hill Companies, 1965).

\bibitem{Berry_Mount} M. V. Berry and K. E. Mount, Semiclassical
approximations in wave mechanics, Rep. Prog. Phys. 35, 315 (1972).

\bibitem{Goldstein}H. Goldstein, Classical Mechanics, 2nd ed. Addison-Wesley,
New York, 1988

\bibitem{WangThossReviewSCIVR}M. Thoss and H. Wang, \emph{Annu. Rev.
Phys. Chem.}, 2004, \textbf{55}, 299.

\bibitem{signproblem}D. Thirumalai, B.J. Berne, Annu. Rev. Phys.
Chem. \textbf{37}, 401 (1986) ;C.H. Mak, D. Chandler, Phys. Rev. A
\textbf{41}, 5709 (1990); J.D. Doll, D.L. Freeman, T.L. Beck, Adv.
Chem. Phys. \textbf{78}, 61 (1994); C.H. Mak, R. Egger, H. Weber-Gottschick,
Phys. Rev. Lett. \textbf{81}, 4533 (1998).

\bibitem{Kay_101}K.G. Kay, J. Chem. Phys. \textbf{101}, 2250, 1994.

\bibitem{Pollak_pertrubation} J. Ankerhold, M. Saltzer, E. Pollak,
J. Chem. Phys. \textbf{116}, 5925 (2002); E. Pollak, J. Shao, J. Phys.
Chem. A \textbf{107}, 7112 (2003); S. Zhang, E. Pollak, J. Chem. Phys.
\textbf{119}, 11058 (2003); S. Zhang, E. Pollak, Phys. Rev. Lett.
\textbf{91}, 190201 (2003); S. Zhang, E. Pollak, Phys. Rev. Lett.
\textbf{93}, 140401 (2004).

\bibitem{Miller327_logderivative_00}R. Gelabert, X. Gimenez, M. Thoss,
H. Wang and W. H. Miller, J. Phys. Chem. A \textbf{104}, 10321-10327
(2000).

\bibitem{Pollak_renormalizationFrozenGaussian_11}J. Tatchen, E. Pollak,
G. Tao, and W. H. Miller, J. Chem. Phys. \textbf{134}, 134104 (2011).

\bibitem{Wang_Dmatrix}H. Wang, D. E. Manolopoulos, and W. H. Miller,
J. Chem. Phys. \textbf{115}, 6317 (2001).

\bibitem{Miller318_prefactoradiabaticGuallar_99}V. Guallar, V. S.
Batista and W. H. Miller, J. Chem. Phys. \textbf{110}, 9922 (1999).

\bibitem{Miller328_prefactorapprox_00}V. Guallar, V. S. Batista and
W. H. Miller, J. Chem. Phys. \textbf{113}, 9510 (2000).

\bibitem{Roy_ZPEaccurate_05}B. B. Issack and P.N. Roy, J. Chem. Phys.
\textbf{123}, 084103 (2005) .

\bibitem{Roy_singeltrajectory_07}B. B. Issack and P.-N. Roy, J. Chem.
Phys. \textbf{12}7, 144306 (2007).

\bibitem{Roy_excitedstatesconstrained_07}B. B. Issack and P.-N. Roy,
J. Chem. Phys. \textbf{126}, 024111 (2007). 

\bibitem{Roy_hydrogenbonded_07}B. B. Issack and P.-N. Roy, J. Chem.
Phys. \textbf{127}, 054105 (2007)

\bibitem{Calvo}M. P. Calvo and J. M. Sanz-Serna, SIAM J. Sci. Comput.
\textbf{14}, 936 (1993)

\bibitem{Gray_Manolopoulos}D. E. Manolopoulos and S. K. Gray, J.
Chem. Phys. \textbf{102}, 9214 (1995)

\bibitem{Brewer_99}M.L. Brewer, J. Chem. Phys.\textbf{ 111}, 6168
(1999).

\bibitem{Colbert_DVR} D. Colbert and W. H. Miller, J. Chem. Phys.
\textbf{96}, 1982 (1992).

\bibitem{Pollak_chaoticquartic_89}B. Eckhardt, G. Hose, E. Pollak,
Phys. Rev A \textbf{39}, 3776 (1989).

\bibitem{Bowman_water}J.M. Bowman, A. Wierzbicki, J. Zuniga, Chem.
Phys. Lett. \textbf{150}, 269 (1988).

\bibitem{CO2_expt}J. Vazquez, M.E. Harding, J.F. Stanton, and J.
Gauss, J. Chem. Theory Comput. \textbf{7}, 1428 (2011).

\bibitem{Chedin pot}A. Chedin, J. Mol. Spectrosc. \textbf{76}, 430
(1979).

\bibitem{Martin_CH2Opot}J.M.L. Martin, T.J. Lee, and P.R. Taylor,
J. Mol. Spectrosc. \textbf{160}, 105 (1993).

\bibitem{Carter_H2CO_exact}S. Carter, N. Pinnavaia, and N.C. Handy,
Chem. Phys. Lett. \textbf{240}, 400 (1995).

\bibitem{Lee_CH4_PES}T.J. Lee, J.M.L. Martin, P.R. Taylor, An accurate
ab initio quartic force field and vibrational frequencies for CH4
and isotopomers, J. Chem. Phys. \textbf{102}, 254 (1995).

\bibitem{Bowman_CH4quantumvalues}S. Carter S., H.M. Shnider, J.M.
Bowman, J. Chem. Phys. \textbf{110}, 8417 (1999).\end{thebibliography}
\end{document}